\documentclass[a4paper,11pt]{article}
\pdfoutput=1 % if your are submitting a pdflatex (i.e. if you have
\usepackage{jheppub} % for details on the use of the package, please
  \usepackage{etoolbox}% http://ctan.org/pkg/etoolbox
  \patchcmd{\maketitle}{\@fpheader}{}{}{}

\usepackage{empheq} %used to have a bigger box in LL subsection

\usepackage[T1]{fontenc} % if needed
\usepackage[utf8]{inputenc} % Input
\usepackage{microtype} % Subliminal refinements towards typographical perfection
\usepackage{mdframed} % Framed boxes \begin{mdframed}
\usepackage{graphicx}
\usepackage{todonotes} % TODOS \todo{}
\usepackage{tensor}
\usepackage{soul}
\usepackage{amsmath}
\usepackage{amsfonts}
\usepackage{amssymb}
\usepackage{stmaryrd}
\usepackage[utf8]{inputenc}
\usepackage{mathtools}
\usepackage{mathrsfs}
\usepackage{mleftright} \mleftright
\usepackage{tensor} % Tensor with \indices{^ab_cd}
\usepackage{braket} % Braket works \bra \ket \braket{ | }
\usepackage{mathrsfs} % Nice fonts for Lagrangian, Hamiltonian, Path intgralmeasure
\usepackage{bbold}
\usepackage{color}
\usepackage{braket}
\usepackage{slashed}
\usepackage{hyperref}

\usepackage{comment}
\usepackage{cancel}
\usepackage{array, makecell} %

\usepackage{adjustbox}

\newcommand{\Rob}[1]{{ \textcolor{blue}{{#1}}}}
\newcommand{\Diego}[1]{{ \textcolor{orange}{{#1}}}}

\DeclareMathAlphabet{\mathfs}{U}{rsfs}{m}{n}                     %

\newcommand{\inter}{{\lrcorner}}
\newcommand{\mfs}[1]{\mathfs {#1}}                               %
\newcommand{\sL}{{\mfs L}}
\newcommand{\Lie}{\sL}

\newcommand{\n}{\nonumber}
\newcommand{\be}{\nopagebreak[3]\begin{equation}}
\newcommand{\ee}{\end{equation}}
\newcommand{\bee}{\nopagebreak[3]\begin{equation*}}
\newcommand{\eee}{\end{equation*}}
\newcommand{\ba}{\nopagebreak[3]\begin{eqnarray}}
\newcommand{\ea}{\end{eqnarray}}
\newcommand{\baa}{\nopagebreak[3]\begin{eqnarray*}}
\newcommand{\eaa}{\end{eqnarray*}}

\usepackage{lipsum}% For 'Lorem Ipsum' dummy text
\usepackage[pscoord]{eso-pic}% The zero point of the coordinate systemis the lower left corner of the page (the default).

\title{Surface Charges Toolkit for Gravity}

\author[1]{Ernesto Frodden,}
\author[2,3,4]{Diego Hidalgo}

\affiliation[1]{Instituto de Sistemas Complejos de Valpara\'iso (ISCV), Subida Artiller\'ia 470, Valpara\'iso, Chile}
\affiliation[2]{Centro de Estudios Cient\'ificos (CECs), Av. Arturo Prat 514, Valdivia, Chile}
\affiliation[3]{Departamento de F\'isica, Universidad de Concepci\'on, Casilla 160-C, Concepci\'on, Chile}
\affiliation[4]{Instituto de Ciencias F\'isicas y Matem\'aticas, Universidad Austral de Chile, Edificio Emilio Pugin, cuarto piso, Campus Isla Teja, Valdivia, Chile}

\emailAdd{efrodden@gmail.com}
\emailAdd{dihidalgot@gmail.com}

\abstract{These notes provide a detailed catalog of surface charge formulas for different classes of gravity theories. The present catalog reviews and extends the existing literature on the topic. 
Part of the focus is on reviewing the method to compute quasi-local surface charges for gauge theories in order to clarify conceptual issues and their range of applicability. Many surface charge formulas for gravity theories are expressed in metric, tetrads-connection, Chern-Simons connection, and even BF variables. For most of them the language of differential forms is exploited and contrasted with the more popular metric components language. The gravity theory is coupled with matter fields as scalar, Maxwell, Skyrme, Yang-Mills, and spinors. Furthermore, three examples with ready-to-download notebook codes, show the method in full action. Several new results are highlighted through the notes.
}
\begin{document}

\maketitle
\flushbottom

%--------------------------------------------------------
\section{Introduction}
\label{sec:introduction}
%--------------------------------------------------------

In physicist's eyes one of the most beautiful results of mathematical-physics is the deep connection between global symmetries and conserved quantities elegantly established by Emmy Noether one century ago \cite{Noether:1918zz,Tavel:1971zz}. She taught us, as a particular consequence, that the celebrated principle of energy conservation in a given theory is a consequence of the symmetry under time translations of that theory. Her general result is known as the First Noether Theorem (for a recent review see \cite{Banados:2016zim}). 

In the same paper, written in 1918,  Noether showed also that applying the same strategy for gauge theories we can not obtain conserved quantities, but instead, she showed another interesting result: For each independent gauge symmetry, a relation among field equations holds (off-shell relations), the so-called Noether identities (\emph{e.g.} Bianchi identity for gravity). This is the content of the Second Noether Theorem.   

As you may appreciate after the great work done by Noether, an important question remained unanswered: Is there still a connection between charges and global symmetries in the case of gauge theories? As physics developed through the last century it became clear that all fundamental theories in physics are in fact gauge theories. Therefore, it was crucial to have an answer. 

First Noether Theorem is still partially\footnote{Because of the gauge symmetry, think of a connection field, there is already an ambiguity on the field symmetry transformation that is inherited in the computation of the stress-energy tensor through this method. See the discussion around Belinfante tensor.} useful for gauge theories if the theory, such as Yang-Mills theory, is defined on a fix background spacetime, such as Minkowski spacetime. The global symmetries of spacetime are translated into Noether charges, usually packed into a stress-energy tensor.
However, by unfreezing spacetime and considering a more realistic theory coupled to General Relativity, itself a gauge theory of spacetime, hope is lost. Most attempts of directly using First Noether Theorem to define charges out of global symmetries run into troubles or are condemned to ambiguity. 

To deal with gauge theories and specially with General Relativity several successful methods where developed during the last fifty years. The most popular are, by name of authors, Abbott-Deser-Tekin \cite{Abbott:1981ff,Deser:2002jk}, Regge-Teitelboim \cite{Regge:1974zd}, Iyer-Wald \cite{Iyer:1994ys,Wald:1999wa}, Barnich-Brandt \cite{Barnich:2001jy,Barnich:2000zw}, and Torre-Anderson \cite{Anderson:1996sc,Torre:1997cd} method. Their developments were sometimes independent and cross inseminated. Nevertheless, all of them have common features. One is their relying, directly or indirectly, on the structure of the phase space, more specifically on the {\it symplectic structure}. Another common feature is that in all of them the obtained charges are expressed as closed surface integrals (or $(D-2)$-surface integrals for $D$-dimensional spacetimes). The equivalence and connection among some of them have been established in the literature\footnote{See for instance a recent review about the equivalence among the Iyer-Wald and Barnich-Brandt procedure \cite{Ruzziconi:2019pzd}. Analogously, in \cite{Ding:2019dhy} was shown that the of-shell Abbot-Deser-Tekin formalism is related to the Iyer-Wald and Barnich-Brandt-Comp\'ere formalism.}. However, a systematic study of their connection is still missing. 

%\footnote{\Ernesto{For the interested reader here we present a diagram showing stablished connection and references (disclaimer, it is not exhaustive and it may well be that the connection have been established before than the cited reference, the diagram may be thought as a guide for a deeper study). [CITE: 1) IW with RT at the end of the paper? 2) BB with RT in the Compere-thesis asymptotic charges paper 3) TA with RT in their asymptotic charges paper 4) ADT with IW and BB in recent paper by Chinese (older) 5) IW and BB in Compere-thesis and also studied in our paper with Diego]}}

In these notes we revisit and explore the method to compute surface charges for several gravity theories. We follow in general terms the Iyer-Wald (IW) symplectic method but being well aware about the close equivalent Barnich-Brandt symplectic method, and also, being well aware that the method differs from the original Noether proposal. One main difference of the approach presented here is the emphasis on the quasi-local nature of the formulas for charges, but of course, they might be used at spacetime asymptotic regions too\footnote{It is worth to note here that the quasi-local treatment for the charge conservation ensures, given a spacetime with exact symmetries, the independence of the radius. This contrast with some asymptotic approaches where $r\to\infty$ is required to get rid of terms appearing on the specific computation of charge formula even if the solution is everywhere specified, \emph{e.g.} a black hole solution.}. To achieve clarity our terminology and notation slightly differ from the original IW treatment but the core logic is the same. For a complementary approach see the lecture \cite{Compere:2019qed}.

Our purpose with these notes is to make a self-contained systematic review and to widen the applicability of this symplectic method to compute charges. We treat several gravity theories with metric variable but a special emphasis is on gravity theories written in the {\it differential form language}. Due to its geometric nature and elegance the use of forms deserves a closer attention in the light of recent developments in physics. Some conceptual issues regarding gauge symmetries are transparent in this language. Further, equivalent formulas for charges usually get more compact expressions and become more tractable. An indication of this are the several new results that we encounter when analyzing the surface charges of the theories. 

The procedure to compute charges for a given solution within a gauge theory can be ordered in five simple steps 1) Identify the fields infinitesimal gauge transformations, 2) Obtain the general surface charge density formula for the theory, 3) Identify the parameters solving the exact symmetry condition (\emph{e.g.} generalized Killing equation), 4) Compute the surface charge integral with those parameters (to get a differential charge), and 5) Integrate the differential charge on phase space. In the present notes we extensively explore the first and second steps for different gravity theories. In order to further pursue the rest of the steps a particular family solution of field equations have to be chosen.  In the final Section \ref{examples} three familiy solutions are picked as examples to show the complete scheme in full detail for three different gravity formulations.

There are multiple ways to use this toolkit. Mainly it can be useful for readers interested in any of the exhibited theories. Check the index.  Also in the following Table it is summarized all the formulas found for the surface charges density corresponding to each theory. For purposes of comparison we have grouped theories written with metric variable in Section \ref{metric} and with a differential form language in Section \ref{forms}. The Table summarizes the main results of both sections. In each respective subsection the main formula is also highlighted in a square box and accompanied with explanation and comments. It is worth to stress, that for most theories considered, there is an associated appendix with extra comments and a step-by-step calculations that leads to the corresponding surface charge formula. 

In this toolkit there is also scope for readers unaware but interested in being introduced to the surface charge method itself. Simple examples are presented in great detail with pedagogical purposes. This is the case of the pure electromagnetic theory treated already in the next introductory Subsection \ref{emwarmup} or the Subsection \ref{CSsection} devoted to Chern-Simons theory in $2+1$. Those readers can also follow the general method to compute surface charges for an arbitrary theory presented in full detail in Section \ref{IW}.

Experts on the surface charge method can also find interesting new results in Section \ref{forms}. We bring to their attention 1) The connection with the Barnich-Brandt method to compute surface charge density and the use of new expressions of the contracting homotopy operator adapted to Einstein-Cartan or Chern-Simons theories in Eqs. \eqref{EChomo} and \eqref{CShomo} of the appendices, respectively, 2) The unexpected results on simple theories with torsion, see Subsection \ref{ECtorsion}, where it is shown that contorsion disappears from the formulas, or 3) The compact expression for surface charges one gets by assuming an asymptotic constant curvature in the Einstein-Cartan theory with cosmological constant, see Subsection \ref{ECads}. Also the general overview and the easy comparison of formulas in the metric and differential form languages for the different or equivalent theories is useful for a deeper analysis of the surface charge method itself.

\newpage 

\renewcommand{\arraystretch}{1.58}
\begin{center}
\resizebox{0.89\textwidth}{!}{
\begin{adjustbox}{rotate=270}
\begin{tabular}{|c|c|c|}
\hline
{\bf Theory} & {\bf Action} & {\bf Surface charge density}\\
\hline 

\makecell{Einstein-Hilbert (EH)} &$\displaystyle  S^{\scriptscriptstyle (EH)}=\frac{\kappa}{2}\int_{\scriptscriptstyle \mathcal{M}}d^{\scriptscriptstyle D}x \sqrt{-g} \left(R-2\Lambda\right) $& $\displaystyle \mathring{k}^{\mu\nu}_\xi=\sqrt{-g}\kappa\left(\xi^{[\nu}\nabla_\sigma \delta g^{\mu]\sigma}-\xi^{[\nu}\nabla^{\mu]} \delta g+\xi_\sigma\nabla^{[\mu} \delta g^{\nu]\sigma}-\frac{1}{2}\delta g\nabla^{[\nu}\xi^{\mu]}+\delta g^{\sigma[\nu}\nabla_\sigma\xi^{\mu]}\right)$\\

\makecell{EH-Maxwell}  & $\displaystyle S^{\scriptscriptstyle (EH)}- \frac{1}{4}\int_{\scriptscriptstyle \mathcal{M}}d^{{\scriptscriptstyle D}}x \sqrt{-g}F_{\mu\nu}F^{\mu\nu}$ & $\displaystyle \mathring{k}^{\mu\nu}_{\xi}+\sqrt{-g}\left[\lambda\left(\delta F^{\mu\nu}-\frac{1}{2}\delta g F^{\mu\nu}\right)-\delta A_\alpha \left(\xi^{\alpha}F^{\mu\nu}+2\xi^{[\mu}F^{\nu]\alpha}\right)\right]$ \\

\makecell{EH-$\Phi$} &$ \displaystyle S^{\scriptscriptstyle(EH)}-\frac{1}{2}\int_{\scriptscriptstyle \mathcal{M}}d^{\scriptscriptstyle D}x\sqrt{-g}\left(\partial^\mu\Phi\partial_\mu\Phi-\zeta_{\scriptscriptstyle D}R\Phi^2\right)$ & $\displaystyle \mathring{k}^{\mu\nu}_{\xi}\left(1+\frac{\zeta_{\scriptscriptstyle D}}{\kappa} \Phi^2\right)+\zeta_{\scriptscriptstyle D} \sqrt{-g}\xi^{[\mu}\left(\delta g^{\nu]\alpha}\nabla_\alpha \Phi^2-\nabla^{\nu]}\delta \Phi^2-\frac{2}{\zeta_{\scriptscriptstyle D}}\delta\Phi\nabla^{\nu]}\Phi\right)$\\

\makecell{EH-Skyrme} & $\displaystyle S^{\scriptscriptstyle(EH)}+\frac{K}{4} \int_{\scriptscriptstyle \mathcal{M}} d^{\scriptscriptstyle D}x \sqrt{-g}  \left< R^\mu R_\mu+\frac{\lambda}{8}F_{\mu\nu}F^{\mu\nu} \right> $   &  $\displaystyle \mathring{k}^{\mu\nu}_{\xi}+K\sqrt{-g}\xi^{[\mu}\left<\left(R^{\nu]}+\frac{\lambda}{4}[R_\sigma,F^{\nu]\sigma}]\right)U^{-1}\delta U\right> $ \\

\makecell{Lanczos-Lovelock} &$\displaystyle \int_{\scriptscriptstyle \mathcal{M}}d^{\scriptscriptstyle D}x \sqrt{-g}\sum_{m=0}^{{\scriptscriptstyle [(D-1)/2]}}c_mL_m$& $\displaystyle 

\sqrt{-g} \left( \delta g P\indices{^\mu^\nu^\alpha_\beta} \nabla_{\alpha} \xi^{\beta} -2\delta P\indices{^\mu^\nu^\alpha_\beta}\nabla_\alpha \xi^\beta  +2{P^{\mu \nu \alpha}}_{\beta}  \xi_\gamma\nabla_\alpha \delta g^{\gamma \beta} - 4\xi^{[ \mu} {{{{P^{\nu ]}}_{\alpha}}}_{\beta}}^{\gamma} \nabla_\gamma \delta g^{\alpha \beta}\right)$
\vspace{0.1cm}\\

\hline 

\makecell{Einstein-Cartan (EC)} & $\displaystyle S^{\scriptscriptstyle(EC)} =\kappa'\int_{\scriptscriptstyle \mathcal{M}} \varepsilon_{abcd} \left( R^{ab}e^c e^d \pm \frac{1}{2\ell^2} e^a e^b e^c e^d      \right)$ &  $\displaystyle\mathring{k}_\epsilon=-\kappa' \varepsilon_{abcd}\left(\lambda^{ab}\delta (e^ce^d)-\delta \omega^{ab}\xi\inter (e^ce^d)\right)$\\

\makecell{EC-Maxwell} & $\displaystyle S^{\scriptscriptstyle(EC)}+\alpha  \int_{\scriptscriptstyle \mathcal{M}}  F \star F $ &  $\displaystyle\mathring{k}_\epsilon   -2\alpha \left(  \lambda  \delta\star  F - \delta A \xi \inter \star F\right)$\\

\makecell{EC-Yang-Mills} & $\displaystyle S^{\scriptscriptstyle (EC)}+\alpha_{\scriptscriptstyle YM}\int_{\scriptscriptstyle \mathcal{M}} \left<   F \star F  \right>$ &  $\displaystyle \mathring{k}_\epsilon    -2 \alpha_{\scriptscriptstyle YM}  \left(   \lambda^i \delta \star F^i - \delta A^i \xi \inter \star F^i   \right)$\\

\makecell{Torsional $(2+1)$-EC}& $\displaystyle \int_{\scriptscriptstyle \mathcal{M}} \left(\varepsilon_{abc}e^aR^{bc}+ \beta e_aT^a\right)$ &  $-\varepsilon_{abc}(\tilde \lambda^{ab}\delta e^c-\delta\tilde \omega^{ab} \xi\inter e^c)$ \\

\makecell{EC-Dirac}& $\displaystyle S^{\scriptscriptstyle (EC)} -\frac{i}{3} \alpha_\psi \int_{\scriptscriptstyle \mathcal{M}} \varepsilon_{abcd} e^a e^b  e^c \left(\bar\psi \gamma^d\gamma_5 d_\omega \psi+\overline{d_\omega\psi}\gamma^d\gamma_5\psi \right)$ &  $\displaystyle -\kappa'\varepsilon_{abcd}\left(\tilde\lambda^{ab}\delta(e^ce^d)-\delta\tilde\omega^{ab}\xi\inter(e^ce^d)\right)$ \\

\makecell{Lovelock-Cartan}& $\displaystyle \int_{\scriptscriptstyle \mathcal{M}}\sum_{p=0}^{\scriptscriptstyle [D/2]}\kappa^{\scriptscriptstyle D}_p\varepsilon_{a_1\cdots a_D}R^{a_1a_2}\cdots R^{a_{2p-1}a_{2p}}e^{a_{2p+1}}\cdots e^{a_D}$ &  $\displaystyle -\sum_{p=1}^{\scriptscriptstyle \left[D/2\right]}p\kappa^{\scriptscriptstyle D}_p\,\varepsilon_{a_1\cdots a_D}\left(\lambda^{a_1a_2}\delta-\delta\omega^{a_1a_2}\xi\inter\right)R^{a_3a_4}\cdots R^{a_{2p-1}a_{2p}} e^{a_{2p+1}}\cdots e^{a_D}$\\

\makecell{$(2+1)$-Chern-Simons} & $\displaystyle \displaystyle \kappa_{\scriptscriptstyle CS} \int_{\scriptscriptstyle \mathcal{M}} \text{Tr}\left(A\wedge dA+\frac{2}{3}A\wedge A\wedge A\right)$ &  $\displaystyle 2\kappa_{\scriptscriptstyle CS} \text{Tr} \left( \lambda \delta A \right)$\\

\makecell{$(2n+1)$-Chern-Simons} & $\displaystyle \kappa_n \int_{\scriptscriptstyle \mathcal{M}} \int_{0}^{1} dt   \left< A \wedge  F_{t}^{n} \right>$ &  $\displaystyle n\kappa_n  \left< \lambda \delta A \wedge F^{n-1} \right>$\\

\makecell{BF Gravity}& $\displaystyle \int_{\scriptscriptstyle \mathcal{M}} \left(B^i F^i\pm\frac{1}{2\ell^2}B^iB^i+\frac{1}{2}\chi^{ij}B^iB^j\right)$ & $\displaystyle  -\delta B^i \lambda^i + \delta A^i \xi \inter B^i$ \\

\makecell{BF-like Jackiw-Teitelboim}& $\displaystyle \int_{\scriptscriptstyle \mathcal{M}} \left(    B_a (de^a +{\varepsilon^a}_b \omega e^b  ) +\tilde B (d\omega + \varepsilon_{ab} e^a e^b)    \right)$& $\displaystyle -\delta \tilde B \lambda - \delta B^a \lambda_a$
\vspace{0.1cm}
\\

\hline

\end{tabular}
\end{adjustbox}
}

\end{center}

\newpage

%----------------------------------------------
\subsection{Electromagnetic warm up}
\label{emwarmup}
%----------------------------------------------

To appreciate the relevance of surface charge method we start by naively applying the First Noether Theorem for the gauge symmetry of pure Maxwell theory. Then, we stress the problem due to the gauge symmetry of the theory and immediately use the surface charge method to save the day. A reader familiar with the method may skip this subsection.

The Lagrangian is $L[A]=F\star F$, where the two-form field strength is the exterior derivative of the one-form connection, $F=dA$, and $\star$ is the Hodge product.\footnote{In components this corresponds to the usual Lagrangian density $\sqrt{-g}F_{\mu\nu}F^{\mu\nu}$. If you are unfamiliar with this notation check \eqref{Hodgespacetime} in convention's Appendix \ref{conventions} and translate each equation.} The theory has a $U(1)$ gauge symmetry, $A\to A'=A+d\Lambda$, with the infinitesimal version $\delta_\lambda A=-d\lambda$. The infinitesimal parameter $\lambda$ is a spacetime function. The variation of the Lagrangian is 
\ba
\delta L=E_A\delta A+d\Theta(\delta A)=-2 (d\star F) \delta A +2d(\delta A\star F).
\ea  
If we restrict the variation to an infinitesimal gauge symmetry, off-shell, we get 
\be
0=2(d\star F)d\lambda-2d(d\lambda \star F)=-2d((d\star F)\lambda+d\lambda \star F),
\label{EMidentity}
\ee
because trivially $-2d(d\star F)\lambda=0$ due to $d^2=0$. We may be tempted to interpret the previous equation as a conservation law, $dJ_\lambda=0$, for the current 
\be
J_\lambda\equiv -2 (d\star F)\lambda- 2d\lambda \star F=-2d(\lambda \star F),
\ee
however, the second equality tells us that the current is trivially conserved, that is, even without imposing the equation of motion. In the differential form parlance a current built from a gauge symmetry is always an exact form and therefore a closed form, this is another form of the old Second Noether Theorem. Therefore, the would-be Noether current $J_\lambda$ is trivial and thus is physically meaningless to define a charge with it. In fact \eqref{EMidentity} is an off-shell identity by virtue of the Noether identity $N_\lambda=-2d(d\star F)\lambda=0$ which here, in the case of electromagnetism, is a mere consequence of $d^2=0$. This analysis is in agreement with the expected well-known result that gauge symmetries do not produce charges.

Now, a subset of gauge symmetries in special cases may become global symmetries (also called exact symmetries). In the case of pure electromagnetism this stands for solving $\delta_\lambda A=-d\lambda=0$ which has the solution $\lambda= \lambda_0=cte$\footnote{\label{footnoteEM} Just for electromagnetism this solution does not depend on the fields and thus the rest of the analysis is quite general. In the case of general relativity the analogous equation is the Killing equation which is a property of certain symmetric spacetime, or for the case of Yang-Mills theory $\delta_\lambda A^i=-d_A \lambda^i=0$ also depends on the fields an there is no a general solution.}. Now, still the previous analysis will not change its triviality because it was completely general and this is just a particular choice of $\lambda$. This is important to stress because a naive use of $J_{\lambda_0}$ as defined before will actually produce here the right formula for the electric charge, however, the logic is misleading and that mistake will hit back in other gauge theories.

To get a sensitive charge from the exact symmetry in gauge theories, $\delta_{\lambda_0}A=0$, we should not follow Noether's approach but use another strategy. Consider the two-form on phase space known as the symplectic structure density
\be
\Omega(\delta_1,\delta_2)\equiv \delta_1\Theta(\delta_2)-\delta_2\Theta(\delta_1)=2\delta_1 A\star d\delta_2 A-2\delta_2 A\star d\delta_1 A,
\ee
evaluated on an infinitesimal gauge direction $\delta_2A\to \delta_\lambda A=-d\lambda$ (and $\delta_1A\to \delta A$)
\ba
\Omega(\delta,\delta_\lambda)=2d\lambda\star d(\delta A)=2d(\lambda\star d\delta A)-2\lambda d\star d(\delta A)=2d(\lambda \star \delta F),
\label{sympEM}
\ea
 where we used that $d^2=0$. Further, we used that the field is on-shell and that $\delta A$ satisfies the linearized equation of motion $\delta (d\star F)=d\star \delta F=d\star d(\delta A)=0$. At this level to consider $\lambda=\lambda_0$ truly makes a sensitive difference because the l.h.s. of \eqref{sympEM} simply vanishes, because of $\delta_{\lambda_0}A=0$, and we get a {\it conservation law}
\be
2d(\lambda_0\star \delta F)=0,
\label{conservedk}
\ee
this suggest us to define a (varied) charge, named {\it surface charge}, as
\be
\delta Q_{\lambda_0}=2\lambda_0\oint_{\scriptscriptstyle S} \star \delta F, 
\ee
this is meaningful information we can extract from the exact symmetry assumption. The key point was to establish a conservation law for the {\it surface charge density} $k_{\lambda_0}\equiv2\lambda_0\star\delta F$. Note that the expression is a variation on phase space, however, because electromagnetism is simple enough the integration is trivial and, taking the integration constant zero, we can just drop the $\delta$ symbols
\be
Q_{\lambda_0}=2\lambda_0\oint_{\scriptscriptstyle S} \star F,
\label{cargae}
\ee 
this is the well-known formula for the electric charge obtained as a consequence of an exact symmetry of the theory. 

Some remarks are in order: 
\begin{itemize}
\item Note that \eqref{conservedk} is a quasi-local expression, the integration is over any close two-surface $S$. The same holds in the general case.

\item To get a non-vanishing charge $Q$ the source for the charge should be enclosed by the surface $S$, fields sourcing the charge could be in principle included in the Lagrangian but because the analysis leading to \eqref{conservedk} is fully local, we do not need to know the field description of the source as far as they are non-zero just in a compact region. 

\item As stated in footnote \ref{footnoteEM}, electromagnetism is a very special case because we have a conserved charge without knowing the details of the field. In general cases charges are obtained only when the exact symmetry condition holds and that usually depends on particular solutions of the field equations, \emph{e.g.} symmetric spacetimes. 
 
\end{itemize}

%----------------------------------------------
\section{Derivation of Surface Charges}
\label{IW}
%----------------------------------------------

In this section, we aim at a pedagogical step-by-step derivation of the surface charges following in general terms the Iyer-Wald \cite{Iyer:1994ys} approach. 

Let $\boldsymbol{L}[\phi]$ be a diffeomorphism-invariant Lagrangian and $\phi$ be a collection of dynamical fields. The first variation of the Lagrangian can always be written as
\be
\delta \boldsymbol{L} = \boldsymbol{E}(\phi) \delta \phi + d \boldsymbol{\Theta}(\phi, \delta \phi),
\ee
where $\boldsymbol{E}(\phi)$ are the equations of motion of the Lagrangian theory and they locally depend on the dynamical fields and their derivatives, while $\boldsymbol{\Theta}(\phi, \delta \phi)$ locally depends on the dynamical fields $\phi$, their variations $\delta \phi$ and derivatives. The letter $d$ stands for exterior derivative. The boundary term $\boldsymbol{\Theta}(\phi, \delta \phi)$ is linear in the variations $\delta \phi$; it is a $(D-1)$-form and is called the symplectic potential form. It suffers from two types of ambiguities: The first ambiguity arises from adding an exact $D$-form to the Lagrangian top-form; the second one arises from adding an exact $(D-1)$-form to $\boldsymbol{\Theta}$.

In the following we write the formulas in differential form language using bold letters for them. In parallel we also express the formulas in the more conventional language, without using differential form, with ordinary letters.   

Let $\xi=\xi^\mu\partial_\mu$ be an arbitrary vector field and let us compute the first variation of the Lagrangian with respect to $\xi$. Since $\delta_{\xi} \boldsymbol{L} = \Lie_{\xi} \boldsymbol{L}$ and $\boldsymbol{L}$ is a top-form, by using the Cartan's magic formula\footnote{For an arbitrary form $\omega$ and a vector field $\xi$ the Lie derivative is $\Lie_\xi\omega=d(i_\xi\omega)+i_\xi (d\omega)$. For the inner operation we use either the $i_\xi$ notation (here) or the $\xi\inter$ notation (later).}, one has
\be \label{variationL}
\partial_\mu\left(\xi^\mu L\right)=E(\phi)\delta_\xi \phi+\partial_\mu \Theta^\mu(\phi, \delta_\xi \phi), \qquad \left[d\left(i_{\xi} \boldsymbol{L}\right) = \boldsymbol{E}(\phi) \delta_{\xi} \phi + d \boldsymbol{\Theta}(\phi, \delta_{\xi} \phi)\right].
\ee
Here $L$ is the Lagrangian density (\emph{e.g.}, in General Relativity $L[g_{\mu\nu}] = \sqrt{-g} R$).

We now assume that the field transformations, $\delta_{\xi} \phi$, are linear in $\xi$. If so, one can always write the first term in the variation of the Lagrangian density along $\xi$ as
\be
E(\phi)\delta_\xi \phi = \partial_\mu S^\mu_\xi + N_\xi, \qquad \left[\boldsymbol{E}(\phi) \delta_{\xi} \phi = d \boldsymbol{S}_{\xi} + \boldsymbol{N}_{\xi}\right],
\label{noetheride}
\ee
where $N_\xi$ collects all terms linear in the vector field. Now by replacing Eq.~\eqref{noetheride} into Eq.~\eqref{variationL} one can arrange terms in a total derivative as
\be \label{totd}
\partial_\mu\left(\Theta^\mu(\phi, \delta_{\xi} \phi)  - \xi^\mu L + S^\mu_\xi \right) =  N_\xi, \qquad \left[d\left(\boldsymbol{\Theta}(\phi, \delta_{\xi} \phi) - i_{\xi} \boldsymbol{L} +\boldsymbol{S}_{\xi} \right)=\boldsymbol{N}_\xi \right].
\ee
We obtained an equation of the form $\partial_\mu J^\mu_\xi =N_\xi$ (in forms $d \boldsymbol{J}_\xi =\boldsymbol{N}_\xi$). Because $\xi$ is an arbitrary spacetime function, the very structure of this equation implies that $N_\xi=0$. This result is the Second Noether Theorem and $N_\xi=0$ are known as the Noether identities. There is one of them for each independent arbitrary function $\xi$ (\emph{e.g.} in four dimensional GR the four independent diffeomorphisms imply the four component Bianchi identity).

Therefore, one can defines a current vector, $J^{\mu}_{\xi} = \Theta^\mu - \xi^\mu L + S^\mu_\xi$, which is identically conserved, $\partial_\mu J^\mu_\xi=0$, even without the use of the equations of motion. In this sense, one says that $J^{\mu}_{\xi}$ is trivially conserved. When the equations of motion are used, $J^{\mu}_{\xi}$ becomes the sometimes called Noether current $J^{\mu}_{\xi} \approx \Theta^\mu - \xi^\mu L$, which due to the previous analysis is also trivially conserved.

At this stage, we can evoke the Poincar\'e's lemma to the identically conserved current $J^{\mu}_{\xi}$. Therefore, there must locally exist a codimension two-form sometimes called Noether potential, $\widetilde Q_{\xi}^{\mu\nu}$, such that $J^{\mu}_{\xi} = \partial_{\nu} \widetilde Q_{\xi}^{\mu\nu}$. Notice that $\widetilde Q_{\xi}^{\mu\nu}$ is ambiguous up to a closed codimension-two form.

Because of this trivial conservation, charges should not be defined by using $J^\mu_\xi$. In order to define charges for gauge theories one has to rely on the symplectic structures of the theory. A way to define the symplectic structure current is
\be
\Omega^{\mu}(\phi, \delta_1, \delta_2) \equiv \delta_{1}\Theta^{\mu}(\phi, \delta_2 \phi) - \delta_{2}\Theta^{\mu}(\phi, \delta_1\phi) - \Theta^{\mu}(\phi, [\delta_1, \delta_2]\phi]),
\ee
where the last term ensures linearity on the variations.

Now, consider the arbitrary variation of the current $J^{\mu}_{\xi}$, we keep the vector field fix, \emph{i.e.} $\delta \xi^\mu=0$, then
\be
\delta J^{\mu}_{\xi} =\delta \Theta^\mu (\phi, \delta_\xi \phi)-\xi^\mu\delta L+\delta S^{\mu}_\xi=\partial_\nu \delta \widetilde  Q^{\mu\nu}_\xi,
\qquad \left[\delta\boldsymbol{J}_\xi=d(\delta\boldsymbol{\widetilde Q}_{\xi}) \right].
\label{laos}
\ee
Evaluating the symplectic current on the gauge symmetry, $\delta_2=\delta_{\xi}$ and $\delta_1=\delta$, one obtains
\ba
\Omega^{\mu}( \delta\phi, \delta_{\xi}\phi) &\equiv& \delta \Theta^\mu( \delta_\xi \phi)-\delta_\xi \Theta^\mu( \delta \phi)\n\\
&=&\xi^\mu\left(E\, \delta \phi +\partial_\nu \Theta^\nu( \delta \phi)\right)-\delta S^\mu_\xi+\partial_\nu \delta\widetilde  Q^{\mu\nu}_\xi-2\partial_\nu\left(\xi^{[\nu} \Theta^{\mu]} ( \delta \phi)\right)-\xi^\mu\partial_\nu \Theta^\nu( \delta \phi) \n\\
&=&\xi^\mu E\, \delta \phi -\delta S^{\mu}_\xi+\partial_\nu \left(\delta \widetilde  Q^{\mu\nu}_\xi+2\xi^{[\mu} \Theta^{\nu]} ( \delta \phi)\right),
\ea
 in second line we replaced $\delta\Theta^\mu(\phi, \delta_\xi\phi)$ from Eq.~\eqref{laos}, we expressed the variation of the Lagrangian $\delta L$, and we used the Lie derivative of the tensor density $\Theta^{\mu}$ as $\delta_{\xi} \Theta^{\mu} = \Lie_\xi\Theta^\mu=2\partial_\nu(\xi^{[\nu} \Theta^{\mu]})+\xi^\mu\partial_\nu \Theta^\nu$. 

To be consistent, with differential forms we have\footnote{In the following, we are assuming that variations commute, \emph{i.e.}, $[\delta, \delta_{\xi}]\phi = 0$ and, therefore, $\boldsymbol{\Theta}(\phi, [\delta, \delta_{\xi}]\phi]) =0$. This is true if $\xi$ is assumed to be fix on the phase space as we do here. But if besides $\xi$ the theory has more gauge symmetry parameters involved, \emph{e.g.} a collection $\epsilon=(\xi,\lambda^{ab}, \lambda^i,\dots)$, then the term $\boldsymbol{\Theta}(\phi, [\delta, \delta_{\epsilon}]\phi)$ can always be decomposed as  (in analogy to Eq.~\eqref{noetheride})
\be
\boldsymbol{\Theta}[\phi, [\delta, \delta_{\epsilon}]\phi] = d\boldsymbol{B}_{\delta\epsilon} + \bold{C}_{\delta\epsilon},
\ee
where $\bold{C}_{\delta\epsilon}$ collects all terms linear in the variation of the symmetry parameter and it vanishes on-shell,  $\bold{C}_{\delta\epsilon}\approx 0$. The term $\boldsymbol{B}_{\delta\epsilon}$ will affect the surface charge density as far as some of the symmetry parameter could be field dependent, $\delta\epsilon=(\delta\xi,\delta\lambda^{ab},\delta\lambda^i,\dots)\neq 0$, still with $\delta\xi=0$.}
\ba
\bf\Omega( \delta\phi, \delta_{\xi}\phi) &\equiv& \delta \bf\Theta( \delta_\xi \phi)-\delta_\xi \bf\Theta( \delta \phi)\n\\
&=&i_\xi\left({\bf E}\delta \phi +d{\bf\Theta}( \delta \phi)\right)-\delta {\bf S}_\xi+d (\delta{\bf \widetilde  Q}_\xi)-d(i_\xi{\bf \Theta}( \delta \phi))-i_\xi (d {\bf \Theta}( \delta \phi)) \n\\
&=&i_\xi{\bf E}\delta \phi -\delta {\bf S}_\xi+d\left(\delta {\bf \widetilde  Q}_\xi-i_\xi{\bf \Theta}( \delta \phi)\right).
\ea
Taking a field configuration $\phi$ satisfying the equations of motion and such that their variation $\delta \phi$ satisfy the linearized equations of motion, one gets that the symplectic current reads as
\be
\Omega^\mu(\phi, \delta \phi,\delta_\xi \phi) \approx \partial_\nu \left(\delta \widetilde  Q^{\mu\nu}_{\xi}+2\xi^{[\mu}\Theta^{\nu]}(\phi, \delta \phi)\right),
\qquad \left[\boldsymbol{\Omega}(\phi, \delta \phi,\delta_\xi \phi)=d\boldsymbol{k }_{\xi} \right].
\ee
The final step consists in defining, up to a total derivative, the {\it surface charge density}
\be
k_\xi^{\mu\nu}(\phi, \delta \phi) \equiv \delta\widetilde  Q^{\mu\nu}_{\xi}+2\xi^{[\mu}\Theta^{\nu]}(\phi, \delta \phi), \qquad \left[\boldsymbol{k}_{\xi}(\phi, \delta \phi) \equiv \delta \boldsymbol{\widetilde  Q}_{\xi} - i_{\xi} \boldsymbol{\Theta}(\phi, \delta \phi) \right].
\ee
In the special case of an exact symmetry, \emph{i.e.} specific $\xi$ function satisfying $\delta_{\xi} \phi =0$, it turns out that the surface charge density is conserved on-shell: $\partial_\mu k_\xi^{\mu\nu}\approx0$ (in the language of forms, $d\boldsymbol{k}_{\xi} \approx 0$). The exactness of the surface charge density guarantees that the value of its integration on a closed surface is independent of the choice of the surface. By integrating it over any closed surface $S$ (\emph{e.g.}, a $(D-2)$-dimensional sphere), one obtains the {\it surface charge}
\begin{equation}
\slashed{\delta} Q_{\xi}(\phi, \delta \phi) =  \frac{1}{2(D-2)!} \oint_{S} k_\xi^{\mu\nu} \varepsilon_{\mu\nu \alpha_3\dots\alpha_{D}}dx^{\alpha_3}\wedge \dots \wedge dx^{\alpha_D} = \oint_{S} \boldsymbol{k}_{\xi}(\phi, \delta \phi),
\label{scintegral}
\end{equation}
naturally on both languages one obtains the same value for the surface charge. Note that this quantity is a differential or one-form on the phase space.
The symbol $\slashed{\delta}$ emphasizes that the surface charge is not necessarily an exact differential on the phase space of the solutions. In other words the function $Q_\xi$ may not exists. A sufficient condition for its existence is $\displaystyle \delta \left(\oint_{\scriptscriptstyle S} \boldsymbol{ k}_\xi \right)=0$. If the condition holds, then, after an integration on the phase space it is possible to obtain the finite surface charge $Q_\xi$.

%--------------------------------------------------------
\section{Gravity Theories with Metric Variable}
\label{metric}
%--------------------------------------------------------
In this section, we aim at those gravity theories, in the metric formalism, whose action reads
\be
S[g_{\mu\nu}, \phi] = S^{{\scriptscriptstyle(EH)}}[g_{\mu\nu}] + S^{(m)}[g_{\mu\nu}, \phi],
\ee
where $ S^{\scriptscriptstyle (EH)}[g_{\mu\nu}]$ is the Einstein-Hilbert (EH) action describing General Relativity (GR) and $S^{(m)}[g_{\mu\nu}, \phi]$ is the action describing the dynamics of the matter sector.

To achieve this program, we first study in details the Einstein-Hilbert action and how surface charges are derived in General Relativity.

%----------------------------------------------
\subsection{Einstein-Hilbert-$\Lambda$ action}
%----------------------------------------------

We start by considering the Einstein-Hilbert (EH) action in $D$-dimensional spacetime with an additional cosmological constant term $\Lambda$ 
\be
S^{\scriptscriptstyle (EH)}[g_{\mu\nu}]=\frac{\kappa}{2}\int_{\cal M}dx^{\scriptscriptstyle D}\sqrt{-g} \left(R-2\Lambda\right),
\label{EHaction0}
\ee
where $\kappa = c^4/(8\pi G)$, $c$ the speed of light and $G$ the Newton's constant.

The variation of the metric field under an infinitesimal diffeomorphism amounts for its Lie derivative generated by the vector field $\xi$:  $\delta_\xi g_{\mu\nu} = \Lie_{\xi}g_{\mu\nu} = \nabla_\mu\xi_\nu+\nabla_\nu\xi_\mu$. This corresponds to the gauge symmetry of GR\footnote{We take the perspective of GR as a gauge theory in the sense that for arbitrary vector field the Lie derivative of the metric is both 1) a local transformation and 2) a symmetry of the action (\emph{i.e.} the varied action after replacement of the symmetry becomes a boundary term).}. Therefore, as we know, to compute charges we should use the surface charge method.

The surface charge density for GR is derived in full detail in the Appendix~\ref{gr}, the results is the following expression\footnote{See also \cite{Compere:2019qed} or \cite{Hajian:2015eha}. Note the overall sign difference. It is due to our conventions on the variation symbol $\delta$. We assume it respects $\delta g^{\mu\nu}=-g^{\mu\alpha}g^{\nu\beta}\delta g_{\alpha\beta}$.}
\be
\boxed{\mathring{k}^{\mu\nu}_\xi=\sqrt{-g}\kappa\left(\xi^{[\nu}\nabla_\sigma \delta g^{\mu]\sigma}-\xi^{[\nu}\nabla^{\mu]} \delta g+\xi_\sigma\nabla^{[\mu} \delta g^{\nu]\sigma}-\frac{1}{2}\delta g\nabla^{[\nu}\xi^{\mu]}+\delta g^{\sigma[\nu}\nabla_\sigma\xi^{\mu]}\right)}
\label{metricGR}
\ee
where  $\delta g=g_{\alpha\beta}\delta g^{\alpha\beta}$. The surface charge density does not depend on the cosmological constant term. Moreover it is linear in the symmetry vector field $\xi$ and in the variation of the metric field $\delta g^{\mu\nu}$. Note also that $\mathring{k}^{\mu\nu}_\xi$ is anti-symmetric: An indication that the differential form language may be appropriated here. 

If we assume the condition that $\xi$ is an exact symmetry for the metric field, \emph{i.e.}, $\xi$ is a Killing vector
\be
\Lie_\xi g_{\mu\nu}=\nabla_\mu \xi_\nu+\nabla_\nu \xi_\mu=0,
\label{killingeq}
\ee
the surface charge density satisfies a conservation, $\partial_\mu \mathring{k}^{\mu\nu}_\xi=0$, and therefore it is suitable to define a surface charge through the integral expression \eqref{scintegral}. 

%----------------------------------------------
\subsection{Einstein-Hilbert-Maxwell action}
%----------------------------------------------

Let us consider the Einstein-Hilbert-Maxwell action. The theory is described by the following action
\be
S[g_{\mu\nu}, A_\mu]= S^{{\scriptscriptstyle(EH)}}[g_{\mu\nu}] - \frac{1}{4}\int_{\scriptscriptstyle \cal M} d^{{\scriptscriptstyle D}}x \sqrt{-g}F_{\mu\nu}F^{\mu\nu},
\ee
where $A_{\mu}$ is the electromagnetic gauge potential and $F_{\mu\nu} = \partial_{\mu}A_{\nu} - \partial_{\nu}A_{\mu}$ is the field strength.

The infinitesimal gauge transformations for this theory are   
\ba
\label{symmEM0}
\delta_\xi g_{\mu\nu}&=&\Lie_\xi g_{\mu\nu}=\nabla_\mu \xi_\nu+\nabla_\nu \xi_\mu, \\
\delta_{\epsilon} A_\mu&=&\delta_{(\xi, \lambda)} A_\mu=\Lie_\xi A_\mu-\nabla_\mu\lambda'=\xi^\nu F_{\nu\mu}-\nabla_\mu \lambda,
\label{symmEM}
\ea
where in addition to the diffeomorphisms on the $A_\mu$ field we must consider the $U(1)$ gauge symmetry of the electromagnetism. Note that we use $\lambda=\lambda'-\xi^\mu A_\mu$, which is the prescription for the so-called {\it improved gauge transformations}. We use $\epsilon=(\xi,\lambda)$ to pack all gauge symmetry parameters. For convenience we use this prescription all along this toolkit each time there is a gauge transformation acting on connections. The advantage is that the transformations become explicitly covariant (here invariant) as far as the gauge parameter $\lambda$ transforms in a covariant way (here invariant).
To undo the prescription, a simple replacement of $\lambda$ in the final formulas is enough.

The surface charge density for Einstein-Hilbert-Maxwell theory is\footnote{This formula should be compared with the results in \cite{Compere:2007az}, (4.22) in \cite{Compere:2009dp}, (4.13) in \cite{Hajian:2015xlp}, or recently (51) in \cite{Ding:2019dhy}. In the explicit formulas of \cite{Compere:2009dp} and \cite{Hajian:2015xlp}, an extra term $2F^{\mu\rho} \delta {g_{\rho\sigma}} g^{\sigma\nu}$ appears, due to their different definition of the variation of the field strength with the indices up: $\delta F^{\mu\nu}\equiv g^{\mu\sigma}g^{\nu\rho}\delta F_{\sigma\rho}$. We thank Kamal Hajian for the clarification of this point through \cite{Ghodrati:2016vvf}.} (see Appendix \ref{appEHM})
\be
\boxed{k_\epsilon^{\mu\nu}=\mathring{k}^{\mu\nu}_{\xi}+\sqrt{-g}\left[\lambda\left(\delta F^{\mu\nu}-\frac{1}{2}\delta g F^{\mu\nu}\right)-\delta A_\alpha \left(\xi^{\alpha}F^{\mu\nu}+2\xi^{[\mu}F^{\nu]\alpha}\right)\right]}
\label{SCem}
\ee
where again $\delta g \equiv g_{\mu\nu}\delta g^{\mu\nu}$. 

In order to define a conserved surface charge the exact symmetry conditions must be satisfied. The conditions stand for equating the infinitesimal gauge transformations \eqref{symmEM0} and \eqref{symmEM} to zero and solve for the parameters $\epsilon=(\xi,\lambda)$. As we know for, pure gravity this is the Killing condition on $\xi$, but here we also need to solve $\lambda$ in terms of $\xi$. For a given Killing field this $\lambda(\xi)$ is the one needed for computing the charge. Note, that even if there is no Killing field at all, still $\xi=0$ and $\lambda=\lambda_0$ is a general solution. This is the origin of the electric charge in curved spacetimes. 

Check the example of the electrically charged and rotating $2+1$ black hole in \ref{exBTZ1} to see how this works exactly.

%----------------------------------------------
\subsection{Einstein-Hilbert-$\Lambda$ action with a conformally coupled scalar field}
\label{EHscalar}
%----------------------------------------------

Consider gravity field interacting with a scalar field. A particularly simple model is to add, a so-called, conformally coupled scalar field, $\Phi$. 
The theory is described by the Lagrangian action (see for instance \cite{Martinez:1996gn}) 
\be
S[g_{\mu\nu},\Phi]=\frac{\kappa}{2}\int_{\scriptscriptstyle \cal M}d^{\scriptscriptstyle D}x\sqrt{-g}\left[R-2\Lambda-\frac{1}{\kappa}\left(\partial^\mu\Phi\partial_\mu\Phi-\zeta_{\scriptscriptstyle D}R\Phi^2\right)\right].
\label{scalarL}
\ee
Although the theory does not possess a conformal symmetry, in this model the equation of motion for the scalar field  (a generalization of the Klein-Gordon equation given by $(\square +\zeta_{\scriptscriptstyle D}R)\Phi=0 $) is conformally invariant\footnote{The conformal transformations are $\Phi\to \Omega^{1-\frac{D}{2}}(x)\Phi$ and $g_{\mu\nu}\to \Omega^2(x)g_{\mu\nu}$ for $\Omega(x)$ an arbitrary spacetime function.} for the specific values $\zeta_{\scriptscriptstyle D}=\displaystyle \frac{D-2}{4(D-1)}$.

The infinitesimal diffeomorphism on the metric field is the usual \eqref{symmEM0}. There is no a new gauge symmetry in this theory, then, for the scalar field the transformation is simply
\ba
\delta_\xi\Phi=\Lie_\xi\Phi&=&\xi^\mu\partial_\mu \Phi.
\ea
The surface charge density is (see the details in the Appendix \ref{appEHscalar} )
\be
\boxed{k^{\mu\nu}_{\xi}=\mathring{k}^{\mu\nu}_{\xi}\left(1+\frac{\zeta_{\scriptscriptstyle D}}{\kappa} \Phi^2\right)+\zeta_{\scriptscriptstyle D} \sqrt{-g}\xi^{[\mu}\left(\delta g^{\nu]\alpha}\nabla_\alpha \Phi^2-\nabla^{\nu]}\delta \Phi^2-\frac{2}{\zeta_{\scriptscriptstyle D}}\delta\Phi\nabla^{\nu]}\Phi\right)}
\label{metricGRPhi}
\ee
which is expressed in terms of pure GR surface charge density (\ref{metricGR}). Again to have a conserved charge one must use a Killing vector field, \eqref{killingeq}, but also the exact symmetry equation on the scalar field must hold:   $\xi^\mu\partial_\mu \Phi=0$. In words, there must be spacetime directions along which the scalar field is constant.

%----------------------------------------------
\subsection{Einstein-Hilbert-Skyrme action}
%----------------------------------------------

Here we consider gravity coupled to a Skyrme field, denoted $U(x^\mu)$, which is a $SU(2)$ group valued field on spacetime. The Skyrme model, in flat spacetime, is an effective field theory describing either nuclear and particle physics in the low energy regime of QCD (see the review \cite{Ma:2016npf}). Currently, the research on the gravitational Skyrme model is in development. Notable results come form the side of black hole physics (see for instance \cite{Droz:91,Dvali:2016mur}) and particular self-gravitating Skyrme solutions (see \cite{Canfora:16}).

The Einstein-Skyrme action is
\be
S[g_{\mu\nu},U]=\int_{\scriptscriptstyle \cal M} d^{\scriptscriptstyle D}x \sqrt{-g}\left(\frac{\kappa}{2}\left( R-\Lambda \right)+\frac{K}{4}\left<R^\mu R_\mu+\frac{\lambda}{8}F_{\mu\nu}F^{\mu\nu}\right>\right),
\label{skyrmeaction}
\ee
with $R_\mu=U^{-1} \partial_\mu U$, therefore $R_\mu$ is valued in the $SU(2)$ algebra  ($R_\mu={R_\mu}^{j}\tau_j$,  we use the generators $\tau_j=-i\sigma_j$  with $\sigma_j$ the Pauli matrices),  and $F_{\mu\nu}=[R_\mu,R_\nu]$ with $[\cdot,\cdot]$ the algebra commutator. The  $K$ and $\lambda$ are positive coupling constants;  and the $\left<\cdot \right>$ denotes the trace on the algebra elements.

Analogously with the previous theory there is no new gauge symmetry. Then, we must consider just the infinitesimal diffeomorphism transformations on the fields 
\ba
\label{skyrmesym0}
\delta_\xi g_{\mu\nu}&=&\Lie_\xi g_{\mu\nu}=\nabla_\mu \xi_\nu+\nabla_\nu \xi_\mu\\
\delta_\xi U&=&\Lie_\xi U=\xi^\mu\partial_\mu U.
\label{skyrmesym}
\ea
In the Appendix \ref{appEHS} it is worked out the derivation of the surface charge density. The result is
\be
\boxed{k_\xi^{\mu\nu}=\mathring{k}^{\mu\nu}_{\xi}+K\sqrt{-g}\xi^{[\mu}\left<\left(R^{\nu]}+\frac{\lambda}{4}[R_\sigma,F^{\nu]\sigma}]\right)U^{-1}\delta U\right>}
\label{SCskyrme}
\ee
Again, to have the conservation law, $\partial_\mu k^{\mu\nu}_\xi\approx 0$, the exactness symmetry condition must hold. That is, equating  \eqref{skyrmesym0} and \eqref{skyrmesym} to zero and solving for $\xi$ (have a Killing vector). 

As an application, this is the formula one should use, within this formalism,  to compute the mass/energy of a spherically symmetric black holes in the presence of a Skyrme field. 

%----------------------------------------------
\subsection{Lanczos-Lovelock action}
%----------------------------------------------

Gravity theories can be extended to higher and lower spacetime dimensions. With the criteria of having second order field equations for the metric Lanczos in 1938 \cite{Lanczos:1938sf} and Lovelock in 1971 \cite{Lovelock:1971yv} wrote the most general expression for the Lagrangian in an arbitrary dimension. In addition to the cosmological constant and the Einstein-Hilbert term, the Lanczos-Lovelock (LL) Lagrangian includes a series of higher power curvature terms that depend on the dimension. For a recent review we suggest \cite{Padmanabhan:2013xyr}. The Lanczos-Lovelock (LL) action is
\be
S[g_{\mu\nu}]=\int_{\scriptscriptstyle \cal M}\sqrt{-g}\sum_{m=0}^{{\scriptscriptstyle [(D-1)/2]}}c_mL_m ,
\ee
with $c_m$ arbitrary constants, here $[\cdot]$ is the integer part function, and each Lagrangian term in the sum is
\be
L_m=\frac{1}{2^m}\,\delta^{\alpha_1\beta_1}_{\gamma_1\sigma_1}\cdots ^{\alpha_m\beta_m}_{\gamma_m\sigma_m} R\indices{_{\alpha_1}_{\beta_{1}}^{\gamma_1\sigma_1}}\cdots R\indices{_{\alpha_m}_{\beta_{m}}^{\gamma_m\sigma_m}},
\ee
where the symbol $\delta_{\cdots}^{\cdots}$ are the totally anti-symmetric Kronecker's deltas (for instance $\delta^{\alpha\beta}_{\mu\nu}=\delta^{\alpha}_\mu\delta^\beta_\nu-\delta^{\alpha}_\nu\delta^\beta_\mu$). The $L_{m=0}=1$ is defined to have the cosmological constant term. The Einstein-Hilbert-$\Lambda$ action \eqref{EHaction0},  is recovered with $c_{m=0}=-\kappa\Lambda$, $c_1=\kappa/2$, and $c_2=0$.

The only field in this theory is the metric, thus the infinitesimal gauge transformation is $\delta_\xi g_{\mu\nu}=\Lie_\xi g_{\mu\nu}$. With the gauge transformation identified we can compute the surface charge density from the general formula where the only gauge symmetry is diffeomorphism, remember, it reads 
\be \label{LLquasifinal}
k^{\mu\nu}_\xi=\delta \widetilde Q^{\mu\nu}_\xi+2\xi^{[\mu}\Theta^{\nu]}(\delta).
\ee
It is straightforward to compute the specific quantities for LL Lagrangian, they can also be read from \cite{Padmanabhan:2013xyr} or \cite{Bueno:2016ypa}, the result
\ba
\widetilde Q^{\mu\nu}_\xi&=&-2\sqrt{-g}P^{\mu\nu\alpha\beta}\nabla_\alpha \xi_\beta,\\
\Theta^\nu(\delta)&=&2\sqrt{-g}P^{\nu\alpha\beta\gamma}\nabla_\gamma \delta g_{\alpha\beta},
\ea
with
\be
P^{\alpha\beta\gamma\delta}\equiv \frac{\partial\ \ \ \ \ }{\partial R_{\alpha\beta\gamma\delta}}\left(\sum_{m=0}^{\scriptscriptstyle [(D-1)/2]}c_mL_m\right)=\sum_{m=1}^{\scriptscriptstyle [(D-1)/2]}\frac{c_m}{2^m}P_{(m)}^{\alpha\beta\gamma\delta},
\ee
each term in the sum is
\be\label{PmLovelock}
P_{(m)}^{\alpha\beta\gamma\delta}\equiv \delta^{\alpha_1\beta_1}_{\gamma_1\sigma_1}\cdots ^{\alpha_{m-1}\beta_{m-1}\alpha\beta}_{\gamma_{m-1}\sigma_{m-1}\rho\lambda} g^{\rho\gamma}g^{\lambda\delta} R\indices{_{\alpha_1}_{\beta_{1}}^{\gamma_1\sigma_1}}\cdots R\indices{_{\alpha_{m-1}}_{\beta_{m-1}}^{\gamma_{m-1}\sigma_{m-1}}}.
\ee
Notice that by its definition $P^{\alpha\beta\gamma\delta}$ inherits the symmetries of the Riemann tensor $P^{\alpha\beta\gamma\delta}=-P^{\beta\alpha\gamma\delta}=-P^{\alpha\beta\delta\gamma}=P^{\gamma\delta\alpha\beta}$, and that due to the Bianchi identity we also have $\nabla_\alpha P^{\alpha\beta\gamma\delta}=0$.
 
Now we manipulate each term in \eqref{LLquasifinal} separately
\ba\label{LL1}
 \delta \widetilde{Q}^{\mu \nu} & =  & -2\delta \left[\sqrt{-g}P\indices{^\mu^\nu^\alpha_\beta}\nabla_\alpha \xi^\beta\right]\n, \\
 & =  & \sqrt{-g} \left[   \delta g P\indices{^\mu^\nu^\alpha_\beta} \nabla_{\alpha} \xi^{\beta} - 2  \delta P\indices{^\mu^\nu^\alpha_\beta} \nabla_{\alpha} \xi^{\beta }+2P\indices{^\mu^\nu^\alpha_\beta} \nabla_\alpha \delta g^{\beta\gamma} \xi_\gamma  \right],
\ea 
where in the first line we made a slight change of indices that will save us of one extra term in the final expression. For the second line, remember  $\delta \sqrt{-g}=\frac{1}{2}\sqrt{-g}g^{\sigma\rho}\delta g_{\sigma\rho}=-\frac{1}{2}\sqrt{-g}g_{\sigma\rho}\delta g^{\sigma\rho}=-\frac{1}{2}\sqrt{-g}\delta g$. To compute the third term we use $\delta(\nabla_\alpha \xi^\lambda)=\delta\Gamma^\lambda_{\alpha\beta}\xi^{\beta}$, and the formula $\delta {\Gamma^{\lambda}}_{\alpha \beta} = \frac{1}{2} g^{\lambda \sigma} \left(  \nabla_\beta \delta g_{\alpha \sigma}+ \nabla_\alpha \delta g_{\beta \sigma} \right. $ 
$\left.- \nabla_\sigma \delta g_{\alpha \beta}  \right)$. 

The second term in \eqref{LLquasifinal}, reads
\ba\label{LL2}
2 \xi^{[ \mu} \Theta^{\nu ]} (\delta) & = & 4 \sqrt{-g}\xi^{[\mu} P^{\nu] \alpha\beta \gamma} \nabla_{\gamma} \delta g_{\alpha \beta}=4 \sqrt{-g} \xi^{[ \mu} {{{P^{\nu ]}}_{\alpha}}^{\gamma}}_{\beta} \nabla_\gamma \delta g^{\alpha \beta} ,
 \ea
where we used  $P^{\mu \nu \alpha \beta}=-P^{\mu \nu  \beta \alpha}$ and $\delta g_{\alpha \beta} = - g_{\alpha\alpha'}\delta g^{\alpha' \beta'}g_{\beta'\beta}$. Plugging back \eqref{LL1} and \eqref{LL2} into \eqref{LLquasifinal}, the final result is\footnote{Our result coincides exactly with the surface charge formula in \cite{Bueno:2016ypa} found for general higher gravity theories. Notice that for Lovelock-Lanczos gravity we have $\nabla_{\alpha}P^{\alpha\beta\mu\nu}=0$.}
\ba
\label{scLL}
\boxed{
k^{\mu\nu}_{\xi}=
\sqrt{-g} \left( \delta g P\indices{^\mu^\nu^\alpha_\beta} \nabla_{\alpha} \xi^{\beta} -2\delta P\indices{^\mu^\nu^\alpha_\beta}\nabla_\alpha \xi^\beta  +2{P^{\mu \nu \alpha}}_{\beta}  \xi_\gamma\nabla_\alpha \delta g^{\gamma \beta} - 4\xi^{[ \mu} {{{{P^{\nu ]}}_{\alpha}}}_{\beta}}^{\gamma} \nabla_\gamma \delta g^{\alpha \beta}\right) 
}\n\\
\ea
Analogously to General Relativity, the conservation law $\partial_\mu k^{\mu \nu} \approx 0$ is fulfilled when the Killing equation holds, $\Lie_\xi g_{\mu\nu}=0$ (exact symmetry condition). A short calculation shows that, as expected, in four dimensions the replacement of $\displaystyle P^{\mu\nu\alpha\beta}=\frac{\kappa}{2}g^{[\mu|\alpha}g^{\nu]\beta}$ produces exactly the formula for $\mathring k^{\mu\nu}_\xi$ given by \eqref{metricGR}.

We finish the section with an aside comment. Let us establish the connection of the formula just found and the black hole entropy for LL theories. The observation made in \cite{Iyer:1994ys}, is that eternal black holes possess a {\it bifurcated horizon surface} where the horizon generator, which is proportional to a Killing vector, vanishes  there, $\left.\tilde\xi\!\mid \right._{bf}=0$. Over that surface, the formula \eqref{LLquasifinal} is trivially integrable and the so-called Noether potential, $\widetilde Q^{\mu\nu}_{\tilde\xi}$, is enough to compute the full associated charge. This analysis can be generalized to LL theories, see \cite{Jacobson:1993xs} and \cite{Banados:1993qp}. A further analysis allows for the obtained quantity to be interpreted as the entropy of the black hole. Here we stress that although the formula simplifies on the bifurcated horizon, one is also able to compute the same value for the entropy outside the horizon, on an {\it arbitrary}  closed surface containing the black hole singularity, by using the full formula \eqref{scLL}\footnote{However,  as in the standard case an extra input should be considered to fix the normalization of the horizon generator. The usual argument is based on a normalization of the Killing field at the asymptotic region. For a quasi-local argument based on the first law of black hole mechanics and the black hole area see \cite{Frodden:2017qwh}.}. The reason is that the full formula is protected by the conservation while the Noether potential is not.

%\begin{multline} 
%k^{\mu\nu}_{\xi}=  
%\sqrt{-g} \left( \delta g P^{\mu \nu \alpha \beta} \nabla_{\alpha} \xi_{\beta} -2\delta P^{\mu\nu\alpha\beta}\nabla_\alpha \xi_\beta+2 P\indices{^\mu^\nu^\alpha_\beta}\delta g^{\beta\gamma} \nabla_{\alpha} \xi_{\gamma} \right. \\+2{P^{\mu \nu \alpha}}_{\beta}  \xi_\gamma\nabla_\alpha \delta g^{\gamma \beta}+ \left. 4\xi^{[ \mu} {{{P^{\nu ]}}_{\alpha}}^{\gamma}}_{\beta} \nabla_\gamma \delta g^{\alpha \beta}\right).\n
%\end{multline}

%--------------------------------------------------------
\section{Gravity Theories in Differential Form Language}
\label{forms}

Theories of gravity are mainly studied in the metric formalism, where the metric tensor is the dynamical variable describing the gravitational field. 
However, they can also be recast using the language of differential forms. In the most common alternative, known as the \emph{Cartan formulation} of gravity, the metric tensor is traded for tetrads and spin connections as dynamical variables in the action principle. 

In this section we also consider in the form language general Chern-Simons (CS) theories in arbitrary dimensions and the BF theories. To recover gravity in the CS formulation the tetrad must be encoded in the algebra valued connection field. In the BF formulation, instead, the metric is directly traded by a two-form field and an one-form connection.

Coming back to the Cartan formulation of gravity, though less preferred in the literature\footnote{One recent exception is the work \cite{Oliveri:2019gvm}, where boundary charges for the Holst Lagrangian in tetrad-connection variables are analyzed.}, it has undoubtedly advantages to provide an explicit coordinate-invariant description, to describe the coupling with fermionic matter fields, and to study dynamical and non-dynamical torsion.

In the following we start by presenting theories of gravity in tetrad formalism. The reader might appreciate the comparison between the two above-mentioned formalism, and the benefits and disadvantages in either formulations. We aim at presenting the surface charge formulas for the Einstein-Cartan theory coupled to electromagnetic and matter fields. For the equivalent theories the formulas in this section are of course equivalent to those derived in the previous section with the metric formalism. We comment about some features, highlighted by the use of differential forms, by deepening on direct consequences of the formulas themselves. 

For example, in the case of a pure gravity theory and for asymptotically (anti)-de Sitter spacetimes, we show that the corresponding surface charge evaluated in the asymptotic region gets a very compact form, (\ref{altas-tekin}), as recently noted in \cite{Altas:2018pkl, Altas:2018zjr}.
A new result, as far as the authors' knowledge concerns, is about torsional gravity theories: We show two relevant examples where the charges are unaffected by the presence of non-dynamical torsion fields.

\subsection*{Preliminaries on Einstein-Cartan formalism}
%----------------------------------------------

Here we introduce the basics of the Einstein-Cartan formulation of gravity. For a complete review we suggest \cite{Eguchi:1980jx}. In order to write Einstein-Cartan with forms the metric field should be replaced by a vielbein field, $e\indices{^a_\mu}$ satisfying $g_{\mu\nu}=e\indices{^a_\mu}e\indices{^b_\nu}\eta_{ab}$. This variable exhibits a local Lorentz transformation as a gauge symmetry, which is nothing but the symmetries of the tangent space of the manifold at each spacetime point. In four dimensions, with internal indices $a,b,c,\dots =0,1,2,3$, the vielbein are in fact the coordinate components of four one-form fields, the coframe field, $e^a=e\indices{^a_\mu}dx^\mu$, which are at the same time an orthogonal basis at each point of the cotangent space. %In the previous simple formula is already clear the elegance of the language, because of the contraction of the spacetime indices, differential forms always leave the coordinate system implicit. No reference to field coordinate-system-components is needed. 

%This slightly more abstract approach has advantages when dealing with problematic coordinate systems.\footnote{A classic example is the Schwarzschild metric in Schwarzschild coordinates at the event horizon, while metric components blows up at the horizon the physics is completely smooth there. To fully realize it through a better choice of coordinate systems took decades. In contrast, the vielbein one-form field is smooth from the beginning because its existence relies on the basic properties of smoothness of the manifold, and thus, nothing can blows up at the horizon.} 

%Another advantage of using differential forms appears when dealing with complicated formal field expressions, which is exactly the case here for surface charges. Certainly this fact deserves to be further exploited, particularly in recent literature concerning asymptotic analysis of symmetries.\footnote{For instance it is clear that defining spacetime families by requiring specific tailings for the metric components is not a coordinate invariant way to do it. Still it is an efficient an concrete approach but we stress that caution should be taken: How coordinate dependent the conclusions are? And more specific: How much of the so-called asymptotic symmetries groups are just a trivial subset of the group of coordinate diffeomorphism at the asymptotic regions? These questions remain unclear in most of the BMS-related works.}

The elegance of the differential form language for gravity is clear when one adopts the first-order formulation. To have first-order equations of motion, besides $e^a$, we introduce another one-form field called the spin connection field $\omega\indices{^a^b}=-\omega^{ba}=\omega\indices{^a^b_\mu}dx^\mu$. Then, we can define the covariant exterior derivative, $d_\omega(\cdot)=d(\cdot)+\omega(\cdot)$, which acting on a vielbein defines the torsion field $T^a\equiv d_\omega e^a\equiv de^a+\omega\indices{^a_b}\wedge e^b$, and with it we can also define the spacetime curvature two-form $R\indices{^a^b}\equiv d\omega\indices{^a^b}+\omega\indices{^a_c}\wedge \omega\indices{^c^b}$.

%----------------------------------------------
\subsection{Einstein-Cartan-$\Lambda$ }
%----------------------------------------------

The Einstein-Hilbert action with cosmological constant \eqref{EHaction0} is equivalent to the so-called Einstein-Cartan action with cosmological constant 
\be 
S[e^a,\omega^{ab}]=\kappa'\int_{\scriptscriptstyle \cal M} \varepsilon_{abcd} \left( R^{ab}e^c e^d \pm \frac{1}{2\ell^2} e^a e^b e^c e^d      \right),
\label{EC4D}
\ee
where the wedge product among forms is left implicit, for instance $e^ae^b=e^a\wedge e^b =-e^b\wedge e^a$. The constants are related to the old ones by $\kappa'=\kappa/4= c^4/(32\pi G)$ and $\ell^2 = \frac{3}{|\Lambda|}$. The $\pm$ signs correspond to negative and positive cosmological constants, respectively. As expected from the metric analysis, the surface charges will not depend on the cosmological constant in the Einstein-Cartan formalism either.

In the metric formalism, spacetime symmetries or isometries are encoded in the Killing equation. The general wording used to refer to it on arbitrary fields is the \emph{exact symmetry condition}. As we showed before, it is not always just a Lie derivative because when a local gauge symmetry is present and a gauge transformation easily spoils the symmetry condition. For the vielbein and connection variables the exact symmetry condition that is gauge invariant is\footnote{We use the notation $\xi\inter = i_\xi$ for the interior product, for instance $\xi\inter e^a=\xi^\mu e\indices{^a_\mu}$.}
\ba
\delta_\epsilon e^a&=&d_\omega ( \xi\inter e^a)+\xi\inter (d_\omega e^a)+\lambda\indices{^a_b}e^b=0
\label{symconde}
\\
\delta_\epsilon \omega^{ab}&=&\xi\inter R^{ab}-d_\omega \lambda^{ab}=0,
\label{symcondomega}
\ea
which in fact can be understood as a Lie derivative on forms (Cartan magic formula: $\Lie_\xi e^a=d(\xi\inter e^a)+\xi\inter (de^a)$) plus a specific infinitesimal Lorentz transformation, $\delta_{\lambda'}e^a={\lambda'}\indices{^a_b}e^b$ generated by the field dependent parameter ${\lambda'}\indices{^a^b}=\lambda^{ab}+\xi\inter \omega^{ab}$. We group both parameters in $\epsilon=(\xi,\lambda\indices{^{ab}})$, such that $\delta_\epsilon\equiv\Lie_\xi+\delta_{\lambda+\xi\inter \omega}$. This combination of infinitesimal transformation is just a convenient prescription, sometimes called improved transformation, and it has the advantage of being homogeneous under local Lorentz transformations, $\delta_\epsilon (\Lambda\indices{^a_b} e^b)=\Lambda\indices{^a_b}\delta_\epsilon e^b$, which is crucial to keep the local Lorentz gauge symmetry explicitly free while imposing the exact symmetry. If we do not do this a local Lorentz transformation would change the Killing equation and one has to keep track of the extra piece in all formulas. 
A simple analysis shows that the exact symmetry condition $\delta_\epsilon e^a=0$ implies the usual Killing equation on the metric field
\be
\Lie_\xi g = \Lie_\xi e^a\otimes e_a+ e^a\otimes \Lie_\xi e_a=\delta_\epsilon e^a\otimes e_a+ e^a\otimes \delta_\epsilon e_a=0,
\ee
where we used that $g=g_{\mu\nu} (dx^\mu\otimes dx^\nu)=\eta_{ab}(e^a\otimes e^b)$ with $\otimes$ the symmetric tensor product, and the anti-symmetry of $\lambda^{ab}$ such that $\delta_\lambda e^a\otimes e_a =\lambda^{ab}\, e_b\otimes e_a=0$.
Note also that if the connection can be expressed in terms of the vielbein, $\omega^{ab}(e)$, the condition $\delta_\epsilon\omega^{ab}=0$ is a trivial consequence of $\delta_\epsilon e^a=0$. And note also that the equation is linear and thus it is straightforward to solve for the parameter $\lambda^{ab}$; in fact $\lambda^{ab}=e^a\inter( d_\omega (\xi\inter e^b))$ which is equivalent to $\lambda^{ab}=e^{a\mu}e^{b\nu}\nabla_{\mu}\xi_\nu$, where the anti-symmetry is an explicit consequence of the Killing equation $\nabla_{(\mu}\xi_{\nu)}=0$.

Now, we consider the surface charge density. For the reader interested in the details we refer to the Appendix \ref{appendixECL}, where the surface charge density for this theory is worked out step by step. The final result is simply (see \cite{Barnich:2016rwk,Frodden:2017qwh})
\be
\boxed{\mathring{k}_\epsilon=-\kappa' \varepsilon_{abcd}\left(\lambda^{ab}\delta (e^ce^d)-\delta \omega^{ab}\xi\inter (e^ce^d)\right)}
\label{scEC}
\ee
or equivalently
\be
\mathring{k}_\epsilon=-2\kappa' \varepsilon_{abcd}\left(\lambda^{ab}\delta e^c-\delta \omega^{ab} \xi\inter e^c\right)e^d.
\label{scEC2}
\ee
Thus, as mentioned earlier, it does not depend explicitly on the cosmological constant. In particular, if there is no a cosmological constant term in the action, the formula for the surface charge density is the same one.

The previous formula can be used to define charges in wherever spacetime region the exact symmetry condition holds. If the spacetime is assumed to have an exact symmetry with parameters $\epsilon$ defined on a limited region, in that region the surface charge density will satisfy $d\mathring{k}_\epsilon=0$ and therefore a surface charge may be defined. In particular for exact solutions with exact symmetries, like simple black hole solutions, the surface charge defines charges quasi-locally: an asymptotic analysis is not needed, and the charge is defined at any two-surface enclosing the black hole. This makes surface charge a useful tool to compute the charges in spacetimes with complicated asymptotic structures.

As a final comment, the surface charges are \emph{insensitive} to the addition of topological terms (Euler/Gauss-Bonnet, Nieh-Yan, Pontryagin, etc.) or any boundary term in the Lagrangian. This may not be completely clear at first sight, because in some cases the surface charge density formula gets modified. Yet, it can be shown explicitly that those modifications disappear after the spacetime integration is done. The proof of this statement can be found in \cite{Frodden:2017qwh}. It also can be understood as an indirect consequence of the equivalence (for exact symmetries) between the Barnich-Brandt and Iyer-Wald symplectic methods. The former uses directly the equation of motions to define surface charges.

%----------------------------------------------
\subsubsection{Asymptotically (Anti)-de Sitter Spacetimes}
\label{ECads}
%----------------------------------------------

Now, that we have the simple formula at hand, we may consider go to particular situations. Depending of the physical problem we are interested in, further conditions on the fields or even restriction on the gauge symmetries, might be imposed. With them, there is a chance for the formula to become even simpler. %In the following we show that this is exactly the case if we assume an asymptotic region with a constant value for the curvature.

Let us assume that at the spacetime boundary, $\partial{\cal M}$, the curvature is constant, namely
\be
\left.R^{ab}\right|_{\partial\cal M}=\left.\pm\frac{1}{\ell^2}e^ae^b \right|_{\partial\cal M},
\label{alascond}
\ee
where the plus sign is for de Sitter (positive curvature) and the minus sign is for anti-de Sitter (negative curvature). This is nothing but the equivalent of the constant curvature condition $R_{\mu\nu}=\pm |\Lambda|g_{\mu\nu}$ in form language. To simplify the notation, we define the (anti)-de Sitter curvature as $\bar R^{ab}=R^{ab}\mp \frac{1}{\ell^2}e^ae^b$. With the asymptotic condition at hand, the surface charge density for pure gravity with cosmological constant \eqref{scEC}, may be evaluated at the asymptotic region. There, it gets the simpler expression\footnote{To explicitly prove (\ref{altas-tekin}) note that in the asymptotic region we have 
$$\varepsilon_{abcd}\delta R^{ab}\lambda^{cd}=\varepsilon_{abcd} (d_\omega \delta\omega^{ab})\lambda^{cd}=\varepsilon_{abcd} \delta\omega^{ab}d_\omega\lambda^{cd}+d(\cdot)=\varepsilon_{abcd} \delta\omega^{ab}\xi\inter R^{cd}+d(\cdot)=\pm\frac{1}{\ell^2}\varepsilon_{abcd} \delta\omega^{ab}\xi\inter (e^ce^d)+d(\cdot),$$ 
where in the last two equalities we used the exact symmetry condition and the asymptotic condition. Then, we should remember that surface charge densities are defined as equivalence classes that are unaffected by adding exact forms, see (\ref{scintegral}), then we can disregard any $d(\cdot)$ term in the formula. The previous equation allows us to trade the second term in \eqref{scEC} and we obtain the desired result.
%The proof is a straightforward calculation using the exact symmetry condition  $\delta_\epsilon\omega =0$ (or $d_\omega\lambda^{cd}=\xi\inter R^{cd}$).
} 

\be
\boxed{
\left.{\mathring k}_\epsilon\right|_{\partial \cal M}= \pm \ell^2 \kappa'\varepsilon_{abcd}\delta \bar R^{ab}\lambda^{cd}
}
\label{altas-tekin}
\ee 
%{\color{red} 2)  compare it with et al  \cite{Aros:1999id}}\\
this beautiful formula is remarkable because in that compact expression is encoded the same information that one would compute in the asymptotic region using all the terms appearing in (\ref{metricGR}) for the metric formalism. This expression is the equivalent one in differential forms language of the result recently obtained by Altas and Tekin \cite{Altas:2018pkl,Altas:2018zjr}, and of course it strongly uses a non-vanishing cosmological constant. 

\begin{comment}
\Rob{this paragraph is not clear at all!}
As an extra comment, it is interesting to note that the formula does not depend explicitly on the vector field $\xi$, but of course it depends implicitly through the parameter $\lambda^{ab}=\lambda^{ab}(\xi)$ that solves the exact symmetry condition (\ref{symconde}). This apparent balance, between $\xi$ and $\lambda^{ab}$, smoothly lead us to consider how fundamental is a vector field to implement spacetime isometries. When using the metric field we use the Killing equation applied on the metric field, but as far as we use a different variable with a different gauge symmetry, as $e^a$ here, we need to generalize the Killing equation considering the gauge parameter. Then, an interesting question is how far can we go in that direction. Or more precisely, are there variables describing gravity such that the isometries, or here the exact symmetry condition, can be implemented on them without any reference to Killing fields? We do not have the answer. %{\color{red}We discuss more on this when we work out the surface charges for BF theories.}  
\end{comment}

%----------------------------------------------
\subsection{Einstein-Cartan-Maxwell} 
%----------------------------------------------

The electromagnetic field is described by the one-form potential $A$ and the field strength is simply the exterior derivative of the potential, $F=dA$. The Einstein-Cartan action coupled to the electromagnetic field is 
\be 
S[e^a,\omega^{ab},A]=\int_{\scriptscriptstyle \cal M} \left( \kappa' \varepsilon_{abcd}R^{ab} e^ce^d  +\alpha F \star F \right),
\label{ECM}
\ee
with $\alpha=-1/2$. The coupling with the vielbein field in the second term is through the Hodge star $\star$. Explicit components on the frame field are $F=\frac{1}{2}F_{ab}e^ae^b$, and the Hodge dual is $\star F=\frac{1}{4}\varepsilon_{abcd}F^{ab}e^ce^d$ (see Appendix \ref{conventions} for conventions). 

To impose the exact symmetry conditions we still should impose the equations $\delta_\epsilon e^a=0$ and $ \delta_\epsilon\omega^{ab}=0$ as in (\ref{symconde}) and (\ref{symcondomega}), but now we have the field $A$ with its own extra gauge symmetry. The corresponding infinitesimal gauge transformation is $\delta_{\lambda'} A=-d\lambda'$. Therefore, besides the two previous conditions, we should add a third exact symmetry condition directly on the potential, namely    
\be
\delta_\epsilon A =  \xi \inter F - d\lambda =0.
\label{exactMAX}
\ee
Once more for the symmetry condition, we use the improved transformation that is the composition of a Lie derivative and a particular $U(1)$ gauge transformation with parameter $\lambda'=\lambda+\xi\inter A$; explicitly $\delta_\epsilon A=\Lie_\xi A+\delta_{(\lambda+\xi\inter A)}A=d(\xi\inter A)+\xi\inter dA-d(\lambda+\xi\inter A)$. In this set up we group together all the parameters as $\epsilon=(\xi,\lambda^{ab},\lambda)$ although $A$ does not transform with local Lorentz and $e^a$ or $\omega^{ab}$ do not change with $U(1)$ gauge transformation.

Then, the surface charge density is the sum of $\mathring{k}_\epsilon$, from (\ref{scEC}), and an extra electromagnetic piece 
\be
\boxed{k^{\scriptscriptstyle  ECM}_\epsilon=\mathring{k}_\epsilon   -2\alpha \left(  \lambda\, \delta\star  F - \delta A\, \xi \inter \star F\right)}
\ee
The derivation is presented in Appendix \ref{ECYM} for a general Yang-Mills theory. The surface charge for gravity coupled with extra fields is then given by the surface charge of the pure gravity plus the contributions from the additional fields. The reason for this is that these extra fields enter the boundary term $\Theta(\delta)$ in a linear way. An exception to this structure is the conformally coupled scalar field that we studied in the metric formalism.

\begin{comment}
\Rob{To remove}
This structure in pieces of the surface charge density for gravity coupled with extra fields is a constant. The reason is that in most cases the extra fields only enters in the calculation at the boundary term $\Theta(\delta)$ at the variation of the Lagrangian. An exception to this is the conformally coupled scalar field we studied in the metric formalism.
\end{comment}
%----------------------------------------------
\subsubsection{Einstein-Cartan-Yang-Mills}
%----------------------------------------------

The previous case of Einstein-Cartan-Maxwell theory is a particular case of the more general derivation of surface charges for Yang-Mills theories (YM). For completeness, we include the result here and work out the details of the derivation in the Appendix \ref{ECYM}.
This class of gauge theories involves a gauge symmetry described by a semi-simple Lie group $G$, usually the $SU(N)$ group. They are also described by a fundamental one-form gauge connection now valued on the algebra generators of the group, $A={A^i}\tau_i$, the indices $i,j,k,...= 1,2,..., \text{dim}(G)$. The field strength is an algebra valued two-form, $F=F^i\tau_i$, defined as $F=dA+ A\wedge A$, and the covariant derivative is $d_A (\cdot)=d (\cdot)+ [A,(\cdot)]$  with $d$ the exterior derivative in spacetime and $[\cdot, \cdot]$ the algebra commutator. The Einstein-Cartan-Yang-Mills action is
\be \label{YMaction}
S[e^a,\omega^{ab},A]=\int_{\scriptscriptstyle \cal M} \left( \kappa' \varepsilon_{abcd}R^{ab} e^ce^c  +\alpha_{\scriptscriptstyle YM} \big< F \star F \big> \right),
\ee
where $\star$ is again the Hodge operator, the bracket $\big< \cdot \big>$ denotes the trace acting on the group generators $\big< F \star F \big>= \text{Tr}(\tau_i \tau_j) F^i \star F^j $, and $ \alpha_{\scriptscriptstyle YM} $ the coupling constant. %\Ernesto{Interesting results of this YM coupled to gravity theories are generalizations of the Dirac magnetic monopole or new analytic solutions (e.g gravitating Merons), in both cases it is useful to have reliable formulas to compute the associated charges.}
%\Rob{I would remove this comment here. It is not necessary and distract the reader.}

The action \eqref{YMaction} is invariant under diffeomorphisms, local Lorentz transformation, and now also under the non-abelian gauge transformations with the infinitesimal realization $\delta_{\lambda'}A^i=-d_A\lambda'^i=-d\lambda'^i-[A,\lambda']^i$. We group all the associated infinitesimal parameters in $\epsilon=(\xi, \lambda^{ab}, \lambda^i)$. Again we should impose an extra exact symmetry condition
\be
\delta_\epsilon A^i =  \xi \inter F^i - d_A\lambda^i = 0,
\ee
which is the trivial generalization of \eqref{exactMAX} also with the improved prescription. Then, the surface charge density for gravity coupled to a YM theory is (Appendix \ref{ECYM}) 
\be
\boxed{  k^{\scriptscriptstyle YM}_\epsilon=\mathring{k}_\epsilon    -2 \alpha_{\scriptscriptstyle YM}  \big<  \lambda \delta \star F - \delta A\xi \inter \star F   \big>}
\ee
There is an interesting pattern on the structure of the surface charge densities: The contribution from the YM field is structurally similar to the pure gravity one. In fact, even for pure gravity, when considering a Macdowell-Mansouri action \cite{MacDowell:1977jt}, that is, a Lagrangian of the form $\varepsilon_{abcd}\bar R^{ab} \bar R^{cd}$ (with $\bar R^{ab}$ the (anti)-de Sitter field strength as defined after (\ref{alascond})), the surface charge density resembles even more this structure too  (check it in Eq. (3.22) of \cite{Frodden:2017qwh}). As we show later, the same is true for BF theories. This constant pattern is explained by the similarity of the form of the Lagrangians. 
%\Rob{I think it comes from the fact that the Lagrangian functional has the same structure. Nothing surprising.}

It would be also possible to couple Einstein-Cartan theory with scalar fields or Skyrme fields. However, we will not proceed along this direction. We prefer taking advantage of the differential form approach to discuss gravity theories with non-dynamical torsion.

%----------------------------------------------
\subsection{Einstein-Cartan with Torsion: Two Examples}
\label{ECtorsion}
%----------------------------------------------

In the Einstein-Cartan theory, as a first order formulation, the connection $\omega^{ab}$ is an independent variable. Therefore, in this frame the torsion does not vanish in general. It is enough to have a source in the corresponding field equation to turn on the torsion field. While being a natural option to consider for the geometry of real spacetimes, so far, the torsion seems an elusive feature of physical spacetime and there is no experimental indication for it at the moment. Still, it is not ruled out and thus for the sake of generality it is worth to study.

Here, in particular, we wonder how is that surface charges get modified with the presence of torsion in spacetime? How the torsion affects the spacetime charges? We do not have a general answer. But by studying two particular and quite different theories, the conclusion for both of them is that torsion field does not affect the general formula for the charges.    

The role of torsion in the following charge formulas is analogous with the role played by the cosmological constant term: Although present in gravity theories, explicitly modifying the field equations and its respective solutions, it does not appear as a direct contributing term into surface charges.

The first theory we consider is a simple pure gravity example in $(2+1)$-spacetime where torsion is sourced without adding any extra fields. In fact the term, first introduced by Mielke and Baekler in \cite{Mielke:1991nn}, is built out of the same gravity fields. In a second example, now in $(3+1)$-dimensional spacetime, we consider the well-known Einstein-Cartan-Dirac theory where torsion is sourced by Dirac spinors.

For both theories we find the remarkable fact that torsion is explicitly absent from the charge formula (this result may be contrasted with the recent work \cite{DeLorenzo:2018odq}). It remains as an open problem to specify under which conditions the torsion does not affect the charges.

%----------------------------------------------
\subsubsection{Einstein-Cartan in $(2+1)$-dimensions plus a Torsion Term}
%----------------------------------------------

Consider the gravitational action in three dimensions
\be
S[e^a,\omega^{ab}]=\int_{\scriptscriptstyle \cal M}\left(\varepsilon_{abc}e^aR^{bc}+ \beta e_aT^a\right),
\ee
for simplicity we set the overall parameter to one and introduce $\beta$ as the coupling constant for the new term, where $T^a=d_\omega e^a$. As matter of fact $e_aT^a$ can not be written as a boundary term. Furthermore, it produces a source for torsion, as can be checked in the equations of motion \eqref{varLtorsion}. This action can be seen as a sector of the more general Mielke-Baekler model \cite{Mielke:1991nn} which contains two more terms: A cosmological constant term and a Chern-Simons term built with $\omega^{ab}$. A general analysis of the surface charges for the Mielke-Baekler model is straightforward but for our purposes this torsional extra term is enough. In fact, an even more elegant perspective can be done by writing the full Mielke-Baekler model as a Chern-Simons theory and use the results of the Chern-Simons Section \ref{CSsection} below.

Because the theory depends only on vielbein and spin connection fields, the corresponding exact symmetry conditions are just the same as before, (\ref{symconde}) and (\ref{symcondomega}), which are valid for any dimension. The computation of the surface charge density is straightforward and is done in full detail in Appendix \ref{appECtorsion1}. It reads
\be
k_\epsilon= -\varepsilon_{abc}(\lambda^{ab}\delta e^c-\delta\omega^{ab} \xi\inter e^c)+2\beta\xi\inter e^a \delta e_a,
\ee
note that the first two terms have a similar structure than the four dimensional case \eqref{scEC}, this is not casual: In Section \ref{LC} we exhibit the general formula in arbitrary dimensions, for Lovelock-Cartan theories. 

As explained in the appendix we can go further and split $\omega^{ab}=\tilde\omega^{ab}+\bar\omega^{ab}$, such that the torsionless part of the connection satisfies $d_{\tilde \omega}e^a=de^a+\tilde \omega\indices{^a_b}e^b=0$, and we can use the equations of motion to solve algebraically the contorsion $\bar\omega^{ab}$. With this, the surface charge formula simplifies to
\be
\boxed{k_\epsilon= -\varepsilon_{abc}(\tilde \lambda^{ab}\delta e^c-\delta\tilde \omega^{ab} \xi\inter e^c)}
\label{SCtorsion1}
\ee
where $\tilde \lambda^{ab}=e^a\inter (d_{\tilde\omega} ( \xi\inter e^b))$ is the parameter that solves the exact symmetry condition with just the torsionless part of the connection, (\ref{splitlambda}). Then, (\ref{SCtorsion1}) shows that the contorsion is absent from the seed formula for the charges, that is, even before computing it for any specific solution or symmetry. As we see in a moment this is also true in a completely different theory.

%----------------------------------------------
\subsubsection{Einstein-Cartan-Dirac}
%----------------------------------------------

The action for the massless Einstein-Cartan-Dirac theory in four dimensions is
\be
S[e^a,\omega^{ab},\psi]=\int_{\scriptscriptstyle \cal M}\varepsilon_{abcd} e^a e^b\left[\kappa' R^{cd}-\frac{i}{3} \alpha_\psi\, e^c \left(\bar\psi \gamma^d\gamma_5 d_\omega \psi+\overline{d_\omega\psi}\gamma^d\gamma_5\psi \right)\right],
\ee
with $\alpha_\psi$ the coupling parameter and the $\gamma$-matrices satisfying $\{\gamma_a,\gamma_b\}=\gamma_a\gamma_b+\gamma_b\gamma_a=2\eta_{ab}$. The covariant derivative acting on spinors is $d_\omega\psi=d\psi+\frac{1}{2}\omega_{ab}\gamma^{ab}\psi$, with $\gamma_{ab}\equiv \frac{1}{4}[\gamma_a,\gamma_b]$. The special matrix $\gamma_5\equiv \gamma_0\gamma_1\gamma_2\gamma_3$ satisfies $\gamma_5\gamma_a=-\gamma_a\gamma_5$, and we use the bar to denote the complex conjugate.

Besides the exact symmetry condition on $e^a$ and $\omega^{ab}$ given by (\ref{symconde}) and (\ref{symcondomega}), we need to impose the exact symmetry condition directly on the spinor field
\ba
\delta_\epsilon \psi=\Lie_\xi\psi+\lambda' \psi=\xi\inter d_\omega \psi+\lambda \psi=0,
\ea
where again we used and improved version with the algebra valued parameter $\lambda'=\frac{1}{2}\lambda'^{ab}\gamma_{ab}=\frac{1}{2}(\lambda^{ab}+\xi\inter\omega^{ab})\gamma_{ab}$. Remember that the gauge symmetry for spinors is the local Lorentz symmetry, then $\epsilon=(\xi,\lambda^{ab})$. Spinors change under an infinitesimal Lorentz transformation as $\delta_{\lambda'}\psi=\lambda' \psi$, with the algebra valued parameter $\lambda'=\frac{1}{2}\lambda'^{ab}\gamma_{ab}$. Hence, in this section $\lambda'$ without indices is a matrix.  

For the surface charge density the calculations are long, details are in Appendix \ref{appECD}. The result is
\be
k_\epsilon=\mathring k_\epsilon-i\alpha_\psi\,\varepsilon_{abcd} \xi\inter e^ae^be^c \delta\left(\bar \psi\gamma^d\gamma_5 \psi\right),
\ee
which is again a simple modification of the Einstein-Cartan surface charge density produced by the spinor field. The new term comes directly from the spinor contribution to the boundary term in the varied action.

As we saw in the previous section we can go further if we consider the splitting $\omega^{ab}=\tilde \omega^{ab}+\bar\omega^{ab}$ such that $\tilde \omega^{ab}$ is the torsionless part of the connection and the contorsion field $\bar \omega^{ab}$ is solved from the equations of motion. Replacing this back we find a cancellation to simply get (see Appendix \ref{appECD})  
\be
\boxed{k_\epsilon=-\kappa'\varepsilon_{abcd}\left(\tilde\lambda^{ab}\delta(e^ce^d)-\delta\tilde\omega^{ab}\xi\inter(e^ce^d)\right)}
\ee
where again $\tilde\lambda^{ab}=e^a\inter (d_{\tilde\omega}(\xi\inter e^b))$. This surface charge density is exactly $\mathring k_\epsilon$ but using on it the Levi-Civita connection, $\tilde\omega^{ab}(e)$, instead of the general connection $\omega^{ab}$. Therefore, the conclusion for the Einstein-Cartan-Dirac theory is the same, contorsion leaves no trace on charges.
%----------------------------------------------
\subsection{Lovelock-Cartan action}
\label{LC}
%----------------------------------------------

For completeness we also consider in the differential forms language the generalization of the Einstein-Cartan action in arbitrary dimensions: The Lovelock-Cartan action (see for instance \cite{Zanelli:provisory}). Similarly than in the case of the metric formalism, as we increase the number of dimensions more and more terms are allowed into the action. A nice difference here is that in the differential forms language the allowed terms are precisely those that one is able to write down with two restrictions: First, that each of them is a $D$-form, and second, that indices are contracted using the $D$-dimensional Levi-Civita tensor. This automatically guarantees that the obtained equations of motion are first order, as expected. This is to be contrasted with the case of the metric language where to have the expected second order equations of motion we need to precisely select the coefficients of the specific contracted Riemann curvature terms (\emph{e.g.} for $D=5$ the Gauss-Bonnet term is  $[R^{\alpha\beta\gamma\delta}R_{\alpha\beta\gamma\delta}-4R^{\alpha\beta}R_{\alpha\beta}+R^2]$). Thus, in this sense the form language, with vielbein and connection variables, imposes the natural restriction to specify the theory in arbitrary dimensions.

The Lovelock-Cartan action reads

\be
S[e^a,\omega^{ab}]=\int_{\scriptscriptstyle \cal M}\sum_{p=0}^{\scriptscriptstyle [D/2]}L^{\scriptscriptstyle D}_p,
\ee
with $[\cdot]$ denoting the integer part function, and where $L^{\scriptscriptstyle D}_p$ is a $D$-form given by
\be 
L_p^{\scriptscriptstyle D}=\kappa^{\scriptscriptstyle D}_p\varepsilon_{a_1\cdots a_{\scriptscriptstyle D}}R^{a_1a_2}\cdots R^{a_{2p-1}a_{2p}}e^{a_{2p+1}}\cdots e^{a_{\scriptscriptstyle D}}.
\ee
Here $a_1,a_2,\dots = 0,\dots,D-1$. Notice that for $D=4$ to have (\ref{EC4D}) we just need to set $\kappa^{\scriptscriptstyle D=4}_0=\pm\frac{1}{2\ell^2}\kappa'$ to recover the cosmological constant term, $\kappa^{\scriptscriptstyle D=4}_1=\kappa'$ for the Einstein-Cartan term, and set $\kappa^{\scriptscriptstyle D=4}_2=0$ to avoid the Euler topological density (also called Gauss-Bonnet). This is a pure gravitational action and the only fields present are the vielbein and the connection, then, the exact symmetry conditions are just $\delta_\epsilon e^a=0$ and $\delta_\epsilon \omega^{ab}=0$ as given by (\ref{symconde}) and (\ref{symcondomega}).

With the theory and the expressions for the exact symmetries at hand we can compute the surface charge density. For the derivation the steps are similar to those shown in Appendix \ref{appendixECL} for the Einstein-Cartan-$\Lambda$ theory. The final formula is
\be
\label{LCSC}
k^{\scriptscriptstyle LC}_\epsilon=-\sum_{p=1}^{\scriptscriptstyle{\left[D/2\right]}}p\kappa^{\scriptscriptstyle D}_p\,\varepsilon_{a_1\cdots a_{\scriptscriptstyle D}}\left(\lambda^{a_1a_2}\delta-\delta\omega^{a_1a_2}\xi\inter\right)R^{a_3a_4}\cdots R^{a_{2p-1}a_{2p}} e^{a_{2p+1}}\cdots e^{a_{\scriptscriptstyle D}},
\ee
where both the $\delta$ and $\xi\inter$ operations have been factorized and should be understood as acting on all terms on the right. Notice that, as expected, there is no $p=0$ term which means that the cosmological constant term does not affect the formula for the charges in any dimension.

As noted in \cite{Frodden:2017qwh}, the previous formula can be rewritten by noticing that each contributing term in the sum of \eqref{LCSC} can be expressed as the sum of three terms, we suppress indices and write it schematically\footnote{
With all the indices on we have
\begin{multline}
\varepsilon_{a_1\cdots a_D}\left(\lambda^{a_1a_2}\delta-\delta\omega^{a_1a_2}\xi\inter\right)R^{a_3a_4}\cdots R^{a_{2p-1}a_{2p}} e^{a_{2p+1}}\cdots e^{a_D}= \\
(D-2p)\varepsilon_{a_1\cdots a_D}\left(\lambda^{a_1a_2}\delta e^{a_{3}}-\delta\omega^{a_1a_2}\xi\inter e^{a_{3}}\right)R^{a_4a_5}\cdots R^{a_{2p}a_{2p+1}} e^{a_{2p+2}}\cdots e^{a_D} \\
+(p-1)\varepsilon_{a_1\cdots a_D}\{d(\lambda^{a_1a_2}\delta\omega^{a_3a_4} R^{a_5a_6}\cdots R^{a_{2p-1}a_{2p}}e^{a_{2p+1}}\cdots e^{a_D})\\
+\delta_\epsilon \omega^{a_1a_2}\delta\omega^{a_3a_4} R^{a_5a_6}\cdots R^{a_{2p-1}a_{2p}} e^{a_{2p+1}}\cdots e^{a_D} \}.
\end{multline}
}
\ba
&\varepsilon\left(\lambda\delta-\delta\omega\xi\inter\right)R\cdots Re\cdots e= \n \\ 
&(D-2p)\varepsilon\left(\lambda\delta e-\delta\omega\xi\inter e\right)R\cdots R e\cdots e +(p-1)\varepsilon\left\{d(\lambda\delta\omega R\cdots Re\cdots e)+\delta_\epsilon \omega\delta\omega R\cdots R e\cdots e\right\}.\n
\ea
The terms inside the curly brackets do not contribute to the surface charges: The first one is an exact form, and the second one is proportional to the exact symmetry condition. Therefore, an equivalent surface charge density is 
\be
\label{LCSC2}
\boxed{k'^{\scriptscriptstyle LC}_\epsilon=-\sum_{p=1}^{\scriptscriptstyle{\left[\frac{D-1}{2}\right]}}p(D-2p)\kappa^{\scriptscriptstyle D}_p\,\varepsilon_{a_1\cdots a_{\scriptscriptstyle D}}\left(\lambda^{a_1a_2}\delta e^{a_{3}}-\delta\omega^{a_1a_2}\xi\inter e^{a_{3}}\right)R^{a_4a_5}\cdots R^{a_{2p}a_{2p+1}} e^{a_{2p+2}}\cdots e^{a_{\scriptscriptstyle D}}}\n
\ee
For even dimensions this formula has one less term than $k_\epsilon$ because in fact Lovelock-Cartan action contains topological terms: The generalization of the Euler/Gauss-Bonnet density. Their contributions to the surface charge density are cleaned out by the previous observation. Surprisingly, this second version of the formula is also obtained with the contracting homotopy operator method, which happens to be a direct and efficient method to get rid of spurious boundary terms appearing in the Lagrangian. 

%----------------------------------------------
\subsection{Chern-Simons action}
\label{CSsection}
%----------------------------------------------

In this section we discuss the surface charge formula for the Chern-Simons (CS) theory in $2+1$ dimensions. Because the calculations are simple this example is pedagogical, therefore, we put all the details here. However, in the context of Chern-Simons theories it is also interesting to consider general CS theories for an arbitrary odd dimension. The calculations for those theories are also worked out in full detail but relegated to the Appendix \ref{DCSapp}. 

Consider the Chern-Simons (CS) action in $D=3$ dimensions

\be \label{CSAction}
S[A]= \kappa_{\scriptscriptstyle CS} \int_{\scriptscriptstyle \cal M}\big<A\wedge dA+\frac{2}{3}A\wedge A\wedge A\big>, 
\ee
the one-form gauge connection $A$ is valued on the Lie algebra defining the theory, $\left<\cdot\right>$ denotes a group invariant symmetric polynomial here of rank $r=2$, in this dimension is also named the bilinear form; and $\kappa_{\scriptscriptstyle CS}$ the {\it level} of the theory which is not relevant for the classical analysis. Under a gauge transformation the CS action \eqref{CSAction} is not invariant but quasi-invariant because it produces a boundary term, this is of course still a gauge symmetry of the theory as far as the equations of motion are concerned. In the following we consider diffeomorphisms and gauge symmetries and group them in $\epsilon= (\xi, \lambda)$, with $\xi$ a vector field and $\lambda$ a Lie algebra valued gauge parameter. The general infinitesimal symmetry transformation reads
\be
\delta_\epsilon A= \Lie_\xi A - d_A \lambda'=\xi\inter F-d_A \lambda,
\label{symCS2}
\ee
notice that we use the exterior covariant derivative $d_A(\cdot) \equiv d(\cdot)+[A,(\cdot)]$ and define a displaced parameter as $\lambda=\lambda'-\xi\inter A$ to work directly with the improved general transformation, as is our custom through these notes.

Now, the variation of the action produces the equation of motion $F=dA+A\wedge A=0$ which holds only if one gets rid of the boundary term given by the potential $\Theta(\delta A)=\kappa_{\scriptscriptstyle CS}\langle A\wedge \delta A\rangle$. The symplectic structure density is simply $\Omega(\delta_1,\delta_2)=2\kappa_{\scriptscriptstyle CS}\langle \delta_1 A\wedge\delta_2 A\rangle$, which we evaluate with one of its entries on the gauge symmetry transformation (\ref{symCS2})
\ba
\Omega(\delta,\delta_\epsilon)&=&2\kappa_{\scriptscriptstyle CS}\, \langle \delta A \wedge (\xi\inter F-d_A \lambda)\rangle\\
&=&2\kappa_{\scriptscriptstyle CS}\, d \langle \delta A \lambda\rangle,
\ea 
where to get second line we used the equation of motion, $F=0$ and the linearized equation of motion $\delta F=d_A\delta A=0$. Hence, for the CS theory in $D=3$ we have the surface charge density 
\be \label{3DCsc}
\boxed{k^{\scriptscriptstyle CS}_\epsilon=2\kappa_{\scriptscriptstyle CS} \left<\lambda \delta A\right>} 
\ee 
This simple formula covers all CS theories in $3$-dimensions in the sense that the algebra of the theory is not specified yet\footnote{This formula coincides with \cite{Bergshoeff:2019rdb} when diffeomorphisms are considered only, i.e., setting $\lambda=-\xi\inter A$ in Eq. \eqref{3DCsc}.}. In particular we can choose the Poincar\'e  or (anti-)de Sitter group to obtain the surface charge formula for general relativity in $(2+1)$-dimensions (as we do in the first torsion example). We leave it as an exercise but, of course, the result is exactly what we got in the Lovelock-Cartan theory for $D=3$. As an explicit example, we derive the mass/energy of the non-rotating BTZ solution in the Subsection \ref{CSexample}.

The previous derivation is a particular case of the more general derivation for CS theory in $D=2n+1$ dimensions. The details of the general calculation are  explained in Appendix \ref{DCSapp}. The general result for the surface charge density is
\begin{equation}
\boxed{ k^{(2n+1)}_\epsilon = n(n+1)\kappa_{\scriptscriptstyle CS} \big< \lambda \delta A F^{n-1} \big> }
\end{equation}
 which could probably had been guessed, in fact we note there is also a very direct computation to get this result by using the contracting homotopy operator, see Appendix \ref{DCShomotopy}.
The infinitesimal symmetry transformation for the connection are the same \eqref{symCS2} and the actions for these theories are compactly written in equation (\ref{CSform}).  

%----------------------------------------------
\subsection{BF action}
%----------------------------------------------

The BF theories are a set of first order alternative formulation of General Relativity where the one-form frame field (vielbein) is traded by a two-form field usually named $B$. A possible explicit relation among them is $B^{ab}\sim\varepsilon\indices{^a^b_c_d}e^ce^d$ but as the $B^{ab}$ field is an independent variable this relation should be obtained as a consequence of the equations of motion. The $F$ letter in the name of the BF theory comes from the curvature two-form that originally was denoted by $F^{ab}$, although here we keep our notation $R^{ab}(\omega)= d\omega^{ab}+\omega\indices{^a_c}\omega^{cb}$. All BF theories contain a kinetic term in the action of the form $B^{ab}R_{ab}$, originally the BF term. If this is the only term in the action the theory is pure topological: There are non-propagating degrees of freedom present. To recover true General Relativity, with two propagating degrees of freedom, very specific additional constraints and their corresponding Lagrange multipliers should be incorporated into the action, see (\ref{bfaction0}) below. 

There are at least two good motivations to consider BF theories. The first one is due to its close relation with pure topological theories, this property makes BF actions a very interesting starting point to perform direct quantizations within the path integral formalism. Usually those approaches are based on manifold discretizations, and require several prescriptions to define the fields over the simplicial objects. The models for gravity quantization based on BF theories are called {\it Spin Foams} \cite{Perez:2012wv}. A second motivation arises because of the very different geometric nature of BF theories. In particular it suggests new ways to generalize General Relativity action, and consider it as a special case of a completely different family of theories that would be impossible to grasp within the metric or frame field formalism \cite{Krasnov:2017epi} (we suggest also the talks \cite{Krasnov:2019talk}). In this sense, within the perspective of GR as an effective field theory, it is worth exploring new windows that allows us to frame the theory in wider contexts.

In addition, another interesting technical thing of BF theories is that they can be seen as intermediate formulations of GR, that further allow for the integration of the Lagrange multipliers and even the $B$ field, to finally get a {\it pure connection formulation of GR} \cite{Krasnov:2011pp}. These are very special actions for GR whose applications have not been fully explored yet (see recent advances in \cite{Mitsou:2019nlt}). In contrast with the BF, that are first order formulations, this pure connection is a second order formulation, as the metric one, and that spoils the simplicity of the formula for surface charge. Hence, here we focus on a BF example and we leave the longer but straightforward surface charge analysis of pure connection formulations for a future work.

To have taste of the BF theories here we pick a particular BF formulation known as {\it Chiral} BF theory which have the nice feature of being one of the simplest first order formulations of GR. We use the fact that the Lorentz algebra can be decomposed as $\mathfrak{so}(3,1,\mathbb{C})=\mathfrak{sl}(2,\mathbb{C})\oplus \overline{\mathfrak{sl}(2,\mathbb{C})}$. Then, instead of the $B^{ab}$ and $\omega^{ab}$ variables we use a projector to define their self-dual and anti-self-dual components (under this projector). The interesting observation is that GR is redundant in this splitting and to pick only one of the components is enough to describe the full theory. By choosing one of them we obtain a chiral BF theory, which is more economic in components that the full BF. All the details of the equivalence  of chiral BF theory and the Einstein-Cartan-$\Lambda$ formulation are in the Appendix \ref{BFappendix}.
  
The action for a chiral BF formulation of GR is	 
\be
\label{bfaction0}
S[B^i,A^i,\chi^{ij}]=\int_{\scriptscriptstyle \cal M}\left(B^i F^i\pm\frac{1}{2\ell^2}B^iB^i+\frac{1}{2}\chi^{ij}B^iB^j\right),
\ee
where repeated indices implies summation and the internal metric is just $\delta^{ij}$ so we do not make any difference among upstairs or downstairs indices. The $B^i$ field is a self-dual two-form and with the self-dual spin connection one-form $A^i$ we define the curvature $F^i=dA^i+\frac{1}{2}\varepsilon^{ijk} A^jA^k$. As before the $\pm$ sign stands for both possible signs of the cosmological constant. The field $\chi^{ij}$ is a zero-form Lagrange multiplier demanded to be traceless $\chi^{ij}\delta_{ij}=0$. This theory is equivalent to gravity

To obtain the surface charge density for chiral BF we need to establish the exact symmetry condition for all fields involved, however, as the fields $\chi^{ij}$ is non-dynamical it does not appear in the symplectic structure. Thus, it is enough to impose the improved exact symmetry conditions
\ba
\delta_{\epsilon}A&=&\xi\inter F-d_A\lambda=0\\
\delta_\epsilon B&=&\xi\inter d_AB+d_A(\xi\inter B)+[\lambda,B]=0.
\ea

Then, the surface charge density, see Appendix \ref{BFappendix} for details, is simply  

\be \label{BF-surfacecharge}
\boxed{k^{\scriptscriptstyle ChBF}_\epsilon=-\delta B^i \lambda^i+\delta A^i \xi\inter B^i}
\ee
This expression is the most compact surface charge density formula for pure GR we have found so far, and it is certainly like this because of the easiness in components of the chiral BF formulation of GR. 

To see this formula at work we explicitly compute the surface charge for an (Euclidean) self-dual (anti-) de Sitter Taub-NUT solution in Subsection \ref{BFexample}.

As  a final remark,  we note there is a curious connection between the previous surface charge density formula for the chiral BF theory in $(3+1)$-dimensions and the formula one gets in the case of the BF-like formulation of the $(1+1)$-gravitational Jackiw-Teitelboim model. First note that because $B$, $\lambda$ and $A$ are valued in the algebra we can also write the previous surface charge density as 
\be \label{BF-surfacecharge2}
k^{\scriptscriptstyle ChBF}_\epsilon=-\big<\delta B, \lambda \big>+ \big<\delta A ,\xi\inter B \big>,
\ee
where $\big<\cdot,\cdot \big>$ is the trace on the algebra generators, therefore if we call $\tilde X_i$ the generators we have that $\big<\tilde X_i,\tilde X_j\big>$  reduces to a Kronecker delta, $\delta_{ij}$, and previous formula is exactly \eqref{BF-surfacecharge}.  As explained in the Appendix \ref{appendixJB}, the BF-like formulation of the Jackiw-Teitelboim model has a surface charge density given by 
\be
k^{\scriptscriptstyle JT}_\epsilon = -\big<\delta B , \lambda\big>=-\delta B^i \lambda^j \big<X_i,X_j \big>=-\delta \tilde B \lambda - \delta B^a \lambda_a, 
\ee
which is the formula one could deduce from the $3+1$ case remembering that in the $1+1$ theory the $B$ is a zero-form and therefore the second term in (\ref{BF-surfacecharge2}) must vanish. All the details of this statement are explicit in Appendix \ref{BFappendix}\footnote{A particular difference is that for the 1+1 BF-like gravity the $X_i$ are $\mathfrak{so}(2,1)$ algebra generators and thus the meaning of $\big<\cdot,\cdot \big>$ changes accordingly.}. In the same spirit of Lovelock-Cartan actions, the previous link certainly suggest a general formula of surface charge density for BF theories in even dimensions, we leave it for a future work.

%--------------------------------------------------------
\section{Examples: Surface Charges in Action}
\label{examples}
%--------------------------------------------------------

In this section we exhibit the surface charge method to compute charges in three different examples: (i) The rotating charged BTZ black hole; (ii) The non-rotating BTZ black hole from a Chern-Simons surface charge, and (iii) The (Euclidean) self-dual (anti-) de Sitter Taub-NUT solution in four dimensions. For each of them, we compute step-by-step the charges associated to the exact symmetries of the solution and provide explicit replicable calculation in {\it Mathematica} notebooks available at \href{https://sites.google.com/view/surfacechargetoolkit/home}{[sites.google.com/view/surfacechargetoolkit/home]}.   

%--------------------------------------------------------
\subsection{$2+1$ Black Hole with Rotation and Electric Charge}
\label{exBTZ1}
%--------------------------------------------------------

In $(2+1)$-dimensions the equations of motion for gravity coupled to electromagnetism \eqref{eomEHMAX} obtained from the action \eqref{SCem} admits a black hole solution with rotation and electric charge \cite{Martinez:1999qi}. The solution is a generalization of the black hole solution known as BTZ  \cite{Banados:1992wn}. 
Part of the solution is given by the metric field $g_{\mu\nu}$. The line element is
\be
\label{BTZds}
ds^2 = - \frac{r^2}{R^2} F^2 dt^2 + \frac{dr^2}{F^2} + R^2 (N^\phi dt + d \phi)^2,
\ee
with $r$-dependent functions \cite{Perez:2015kea}
\ba 
\label{BTZrr} R^2 & = & r^2 + \left(  \frac{\omega^2}{1-\omega^2} \right) r^{2}_{+} +\frac{2}{\pi} (q \omega \ell)^2 \ln \left( \frac{r}{r_{+}} \right), \\
 F^2 & = & \frac{r^2}{\ell^2} - \frac{r^{2}_{+}}{ \ell^2} - \frac{2}{ \pi} q^{2} (1- \omega^2) \ln \left(  \frac{r}{r_{+}}  \right),\\
 N^\phi & = & -\frac{\ell}{R^2} \left(  \frac{\omega}{1-\omega^2} \right)\left(  \frac{r^2}{\ell^2} - F^2 \right),
\ea
where the negative cosmological constant $\Lambda=-1/\ell^2$; and the arbitrary constants $r_+, \omega^2<1$, and $q$. Another part of the solution is the electromagnetic field
\ba
\label{BTZA}
A_\mu=\frac{q}{2\pi} \ln \left(\frac{r}{\ell}\right)\, (-1,0,\ell\omega ).
\ea
A curiosity of this solution is the logarithmic dependence on the radius brought in by the electric charge of the black hole. This particular dependence, not present in other dimensions, has put some troubles to the usual asymptotic analysis because standard asymptotic tailings do not admit it. The charges formulas for the standard asymptotia blow up when is computed for this black hole. In \cite{Perez:2015kea,Perez:2015jxn} the asymptotic analysis has been widened to work out the asymptotic charges correctly.  

Here, we take a different perspective and use the surface charge formula corresponding to the full theory. Because we have an explicit solution with exact symmetries defined everywhere we do not need to rely on asymptotic analysis and perform the calculation in a quasilocal manner. 

To illustrate this example we choose the formula for surface charge density $k_{\xi}^{\mu\nu}$ in metric variables \eqref{SCem}. Certainly the same can be done by using the surface charge formula in differential form language for Einstein-Cartan-Maxwell theory in $(2+1)$-dimensions\footnote{A Mathematica notebook with the surface charge formula implemented for gravity coupled to electromagnetism in $(2+1)$-dimensions can be found at \href{https://sites.google.com/view/surfacechargetoolkit/home}{[sites.google.com/view/surfacechargetoolkit/home]}.}. 

To compute charges we first identify the exact symmetries of the solution and the parameters generating them, $(\xi, \lambda)$. They should solve \eqref{symmEM0} and \eqref{symmEM}. Then, we replace the parameters and the field solutions into the corresponding surface charge formula. This has to be done for each independent exact symmetry, obtaining thus a surface charge for each of them.

For the solution \eqref{BTZds}-\eqref{BTZA} we have three exact symmetries (remember the use of improved prescription $\lambda=\lambda'-\xi^\mu A_\mu$)

\be
\begin{array}{lccccccl}
\text{Time symmetry:} & \xi_{(t)}^{\mu}&=&(1,0,0) & \quad \text{and}\quad\quad \quad  &\lambda_t&=&\frac{q}{2\pi}\ln\left(\frac{r}{\ell}\right) \\
\text{Axial symmetry:} & \xi_{(\phi)}^{\mu}&=&(0,0,-1) &\quad \text{and}\quad\quad \quad   &\lambda_\phi&=&\frac{\ell\omega q}{2\pi}\ln\left(\frac{r}{\ell}\right) \\
\text{$U(1)$-rigid symmetry:}   \quad\quad\quad & \xi_{(e)}^{\mu}&=&(0,0,0) & \quad \text{and}\quad \quad \quad   &\lambda_e&=&-1. 
\end{array}
\ee
There is a surface charge density for each of them: $k^{\mu\nu}_{(t)}$, $k^{\mu\nu}_{(\phi)}$, and $k^{\mu\nu}_{(e)}$. For the time symmetry the non-vanishing components are $k^{tr}_{(t)}=-k^{rt}_{(t)}$ and $k^{r\phi}_{(t)}=-k^{\phi r}_{(t)}$. Explicitly  
\begin{multline}
k^{tr}_{(t)}=\frac{1}{8 \pi ^2 \ell^2 r_+ \left(1-\omega^2\right)^2}\left[ (1-\omega^4)(\pi r_+^2-\ell^2q^2(1-\omega^2))\delta r_+ \right.\\
\quad\quad\quad\quad\quad-2\ell^2qr_+(1-\omega^2)^2\left(\omega^2+(1+\omega^2)\ln\left(\frac{r_+}{\ell}\right)\right)\delta q\\
\left. +2r_+\omega \left(\pi r_+^2-\ell^2 q^2(1-\omega^2)^2\left(1+\ln\left(\frac{r_+}{\ell}\right)\right)\right)\delta\omega \right], 
\label{ktime1}
\end{multline}
and
\begin{multline}
k^{r\phi}_{(t)}=\frac{1}{8 \pi ^2 \ell^3 r_+ \left(1-\omega^2\right)^2}\left[ -2\omega(1-\omega^2)(\pi r_+^2-\ell^2q^2(1-\omega^2))\delta r_+ \right.\\
\quad\quad\quad\quad\quad+2\ell^2qr_+\omega (1-\omega^2)^2\left(1+2\ln\left(\frac{r_+}{\ell}\right)\right)\delta q\\
\left. -r_+\left(\pi r_+^2(1+\omega^2)-\ell^2 q^2(1-\omega^2)^2\left(1+2\ln\left(\frac{r_+}{\ell}\right)\right)\right)\delta\omega \right].
\label{ktime2}
\end{multline}
It is not always the case, but for this example all surface charge densities $k_{\xi}^{\mu\nu}$ are coordinate independent. Usually coordinate independence is achieve only after spacetime integration. In particular, notice that there is no risk in evaluating them at asymptotic regions because the value is simply independent of $r$. By construction ($\partial_\mu k^{\mu\nu}_\xi=0$) the integration can be done over any loop around the origin with $t=cte$ and because coordinate independence happens it is trivially performed 
\be
\slashed \delta M\equiv \frac{1}{2}\oint k^{\mu\nu}_{(t)}\varepsilon_{\mu\nu\rho}dx^\rho=\int_0^{2\pi} k^{tr}_{(t)}d\phi=2\pi k^{tr}_{(t)}.
\ee
Note that  $k^{r\phi}_{(t)}$ does not play a role. 

A non-trivial check is the integrability on the reduced phase space $\delta(\slashed\delta M)=2\pi \delta  k^{tr}_{(t)}=0$, where the variation $\delta$ acts on $r_+$, $q$, and $\omega$. This guarantees the existence of the finite charge $M(r_+,q,\omega)$. If the integrability condition was not satisfied one may try to define a combination of Killing symmetries that produce integrable charges, this is the case for the Kerr-Newman-de Sitter black hole solution \cite{Frodden:2017qwh}. Here, by integrating the differential surface charge and setting the integration constant to zero we get the finite charge
\be
M= \frac{r_{+}^{2}}{8\ell^2} \left(   \frac{1+ \omega^2}{1-\omega^2} \right) -  \frac{q^2}{4\pi} \left( \omega^2      +(1+ \omega^2) \ln \left( \frac{r_+}{\ell} \right) \right).    
\label{massBTZ2}
\ee
Notice that leaving implicit the integration on time as $\Delta_t\equiv \int_{-\infty}^\infty dt$, we can compute a second differential charge as 
\be
\slashed \delta \widetilde M=\frac{1}{2}\oint k^{\mu\nu}_{(t)}\varepsilon_{\mu\nu\rho}dx^\rho=\int_{-\infty}^{\infty} k^{r\phi}_{(t)}dt=\Delta_t k^{r\phi}_{(t)},
\label{tildeM0}
\ee
where we identified future with past asymptotic regions in order to have a closed integration path. The integrated version of this charge is
\be
\widetilde M= -\frac{\omega}{\ell}\left(\frac{r_+^2}{4\ell^2(1-\omega^2)}-\frac{q^2}{4\pi}\left(1+2\ln\left(\frac{r_+}{\ell}\right)\right)\right)\frac{\Delta_t}{2\pi},
\label{tildeM}
\ee
the interpretation of this charge is left for curious readers, but we advance that it may has a precise meaning for the Euclidean BTZ black hole (for an Euclidean time $\Delta_t$ becomes finite). As far as the authors known this charge has not been reported.  

For the axial symmetry we find that the surface charge density is related with the previous one, \eqref{ktime1} and \eqref{ktime2}, as 
\ba
k^{tr}_{(\phi)}&=&-\ell^2 k^{r\phi}_{(t)},\\
k^{r\phi}_{(\phi)}&=&- k^{tr}_{(t)}+\delta\left( \frac{q^2}{8\pi^2}(1-\omega^2)\right).
\label{relationn}
\ea
After integration on any loop of constant time the associated surface charge differential is 
\be
\slashed \delta J\equiv \frac{1}{2}\oint k^{\mu\nu}_{(\phi)}\varepsilon_{\mu\nu\rho}dx^\rho=\int_0^{2\pi} k^{tr}_{(\phi)}d\phi=2\pi k^{tr}_{(\phi)},
\ee
this quantity can be integrated on phase space and produces a charge identified with the angular momentum\footnote{Notice the interesting relation $J=-\ell^2 \frac{2\pi}{\Delta_t} \widetilde M$. A similar relation is obtained from \eqref{relationn}. The relations among  tilded and untilded quantities is a clear signal of a duality for the Euclidean BTZ charges not reported so far.}
\be
J=\frac{\omega}{\ell}\left(\frac{r_+^2}{4(1-\omega^2)}-\frac{q^2\ell^2}{4\pi}\left(1+2\ln\left(\frac{r_+}{\ell}\right)\right)\right).
\ee
%\ba
%\widetilde J &=&-\frac{r_+^2}{8\ell^2}\left(\frac{1+\omega^2}{1-\omega^2}\right)+\frac{q^2}{4\pi}\left(1+(1+\omega^2)\ln\left(\frac{r_+}{\ell}\right)\right)\\
%&=&-M+\frac{q}{4\pi}(1-\omega^2)
%\ea
Finally,  for the $U(1)$-rigid symmetry
\be
k^{tr}_{(e)}=\frac{\delta q}{2\pi},\quad\quad\text{and}\quad\quad  k^{r\phi}_{(e)}=-\frac{1}{2\pi\ell}\left(q\delta\omega+\omega\delta q\right),
\ee
by integrating both in spacetime (loop around the origin) and in phase space  we obtain the (electric) charge\footnote{There is also the second charge, $\widetilde Q_e=-q\frac{\omega}{\ell}\frac{\Delta_t}{2\pi}$, we can obtain identifying past and future and integrating over an (infinite) loop in time, as we did for $\widetilde M$ in \eqref{tildeM0}.}
\be
Q_e=q.
\ee

Now, we can establish the first law of black hole mechanics for this family of black holes solutions spanned by the parameter $r_+$,  $\omega$, and $q$. With all the charges at hand and defining the entropy $S=\frac{L}{4}=\frac{\pi}{2}R(r_{+})=\frac{\pi r_{+}}{2\sqrt{1-\omega^2}}$ (with $L$ the perimeter of the black hole horizon and $R(r_+)$ is given by \eqref{BTZrr}), we have  
\be
\delta M = T\delta S+\Omega \delta J + \Phi \delta Q_e.
\ee
The previous first law, expressed in terms of $r_+$, $\omega$, and $q$; is a linear system exactly solved by
\ba
T&=&\frac{\sqrt{1-\omega^2}}{2\pi \ell^2}\left(1-\frac{q^2\ell^2}{\pi r_+^2}(1-\omega^2)\right),\\
\Omega&=&\frac{\omega}{\ell},\\
\Phi&=&-\frac{q}{2\pi}(1-\omega^2)\ln\left(\frac{r_+}{\ell}\right),
\ea
these quantities might be further identified with standard physical quantities: A temperature (Hawking thermal radiation), angular velocity of the horizon, and electrostatic potential, respectively.

However, and as a final comment, notice that by increasing the electric charge one may get a non-physical negative temperature. This problem is related to the possibility of having a negative mass \eqref{massBTZ2}. This is a non-desirable possibility if one expects to interpret the mass as the energy of the system. The problem of the negativity of this mass expression has been addressed from a Hamiltonian analysis of charges and an asymptotic perspective in \cite{Perez:2015kea,Perez:2015jxn}. There the authors introduce an {\it improper gauge transformation} for the electromagnetic field and then impose the so-called holographic asymptotic boundary conditions on the fields in order to fix it. The result within this frame is an always positive mass: A better candidate for the energy of the system. It would be interesting to frame that proposal in the present language of surface charges with a quasi-local perspective. We leave it as an open problem for interested readers.

%%%%%%%%%%%%%%%%%%%%%%%%%%%%%%%%%%%%%%%%%%%%%%%%%
\subsection{BTZ Energy from Chern-Simons formulation}\label{CSexample}
%%%%%%%%%%%%%%%%%%%%%%%%%%%%%%%%%%%%%%%%%%%%%%%%%
In this section we show the applicability of the Chern-Simons (CS) surface charge density formula by computing the energy of the BTZ black hole solution \cite{Banados:1992wn}. This can be done because the well-known relation between the theory of $2+1$ gravity with negative cosmological constant and the CS theory built with the anti-de Sitter (adS) group, $SO(2,2)$, as a gauge group \cite{Witten:1988hf, Achucarro:1987vz}. In Subsection \ref{CSsection} we obtained the three-dimensional CS surface charge density
\be\label{kCSex}
k_{\epsilon}^{\scriptscriptstyle (CS)} = 2\kappa_{\scriptscriptstyle CS} \left< \delta A \lambda  \right>,
\ee
with the CS connection $A$ and the gauge parameter $\lambda$, both, valued on an arbitrary algebra. Here we use the adS algebra
\be\label{ads3algebra}
[P_a, P_b] = \varepsilon_{abc} J^c, \quad [J_a, P_b] = \varepsilon_{abc} P^c, \quad [J_a, J_b] = \varepsilon_{abc} J^c, \quad
\ee
where the indices $a,b,c=0,1,2$ are raised and lowered with the three-dimensional Minkowski metric $\eta_{ab}=diag(-1,1,1)$, $P_a$ are the translation generators and for the Lorentz generators, $J^{ab}$, we use the dual form $J_a=\frac{1}{2}\varepsilon_{abc}J^{bc}$. The adS algebra admits a non-degenerate invariant bilinear form
\be\label{adspairing}
\left< J_a P_b \right> = \eta_{ab}. 
\ee
Then, the CS connection $A$ and the gauge parameter $\lambda$ are algebra valued as
\ba
 A  & = &\frac{1}{\ell} e^a P_a + \omega^a J_a \\
 \lambda & =&  \rho^a P_a + \lambda^a J_a,
\ea
with $\ell $ the adS radius, $e^a={e^a}_\mu dx^\mu$ the one-form dreibein field and $\omega^a = {\omega^a}_\mu dx^\mu$ the dual spin connection ($\omega_a = \frac{1}{2}\varepsilon_{abc} \omega^{bc}$). With this, the CS action quasi-invariant under $SO(2,2)$ group is
\be
S[e,\omega] = \frac{ \kappa_{\scriptscriptstyle CS} }{  \ell }\int_{\scriptscriptstyle \mathcal{M}} \left( 2e^a R_a(\omega) + \frac{1}{3\ell^2} \varepsilon_{abc}e^a e^b e^c	\right),
\ee
where the cosmological constant is identified as $\Lambda=-1/\ell^2$, the torsion and the curvature are $T^a =de^a - \varepsilon^{abc}\omega_b e_c$, and $R^a (\omega) = d\omega^a - \frac{1}{2}\varepsilon^{abc}\omega_b \omega_c$, respectively. The exact match between the CS action and the Einstein-Cartan action implies that their coefficients in front of the action are related as  $\kappa_{\scriptscriptstyle CS} =\frac{\ell}{8 \pi G}$, with $G=1$ the Newton's constant.

Now, for $k_{\epsilon}^{\scriptscriptstyle (CS)} $ to be conserved  the exact symmetry condition must hold, {\it i.e.} $\xi\inter F -d_{\scriptscriptstyle A}\lambda=0$ (vanishing of \eqref{symCS2}). By expanding on the generators the condition reduces to the following two equations
\ba
\xi\inter T^a -  d_\omega \rho^a + \varepsilon^{abc} e_b \lambda_c & = &0\label{symm1} \\
\xi \inter \left( R^a (\omega) -\frac{1}{2\ell^2} \varepsilon^{abc} e_be_c\right) - d_\omega \lambda^a +  \frac{1}{\ell} \varepsilon^{abc}e_b \rho_c & = &0.\label{symm2}
\ea 

If $\xi$ is a Killing vector, for the choice $\rho^a = \xi \inter e^a$ the Eqs. \eqref{symm1} and \eqref{symm2} reduce to the usual exact symmetry conditions of the Einstein-Cartan theory \eqref{symconde} and \eqref{symcondomega}, respectively. Then, from Eq.\eqref{symm1} we can solve the parameter (together with the on-shell condition $T^a=0$)
\be\label{lambdaBTZ}
\lambda^a = \frac{1}{2}\varepsilon^{abc}e_b \inter \left(d_\omega \xi \inter e_c \right). 
\ee
	
As expected, the CS surface charge \eqref{kCSex} valued on the adS algebra matches with Eq. \eqref{SCtorsion1} obtained from three-dimensional gravity (using the dual parameter $\lambda_{ab}=\varepsilon_{abc}\lambda^c$).

Now, consider the (non-rotating) BTZ black hole metric 
\be\label{BTZsolution}
ds^2 = -f(r)dt^2 + \frac{1}{f(r)} dr^2 + r^2 d\phi^2, \quad f(r) = -4  M+\frac{r^2}{\ell^2}, 
\ee
where the integration constant $M$, as we are going to check now, is the charge associated to time invariance (mass/energy).

To use the CS surface charge formula we need the CS connection expressed in terms of the tetrad and spin connection. The BTZ tetrad can be obtained from the relation $g_{\mu \nu} = {e^a}_\mu {e^b}_\nu \eta_{ab}$, while the spin connection is solved from $T^a =d_\omega e^a =0$,  we get
\ba \label{BTZformssolution}
&& e^{0}  =  \sqrt{f(r)}dt , \quad \quad   e^{1}  =  \frac{1}{\sqrt{f(r)}}dr, \quad \quad e^{2}  =  r d\phi\n \\
&&  \omega^{0}  =  - \sqrt{f(r)}d\phi, \quad \quad \omega^1 =0, \quad \quad  \quad \quad \omega^{2}  = -\frac{r}{\ell^2 } dt.
\ea
Now, consider a time translation timelike Killing vector as $\xi=\xi^\mu \partial_\mu$, $\xi^\mu = (1,0,0)$, the Eq. \eqref{lambdaBTZ} is solvable by\footnote{Note that $\lambda^2$ is the third component of $\lambda^a=(\lambda^0,\lambda^1,\lambda^2)$.}
\ba  \label{BTZparameter}
\lambda^{2} & = & \frac{r}{ \ell^2} .
\ea
 With the bilinear form \eqref{adspairing} the surface charge \eqref{kCSex} reduces to\footnote{The calculation is made explicit using the tetrad-connection variables in a Mathematica notebook at \href{https://sites.google.com/view/surfacechargetoolkit/home}{[sites.google.com/view/surfacechargetoolkit/home].}}
\ba
 \n k_{\epsilon}^{\scriptscriptstyle (CS)} & = &   \frac{\ell}{4 \pi } \left< \left(\frac{1}{\ell} \delta e^0 P_0 + \frac{1 }{\ell} \delta e^1 P_1 + \frac{1 }{\ell} \delta e^2 P_2 + \delta \omega^0 J_0 + \delta \omega^2  J_2\right)  \left(\frac{1 }{\ell} \xi \inter e^0 P_0+\lambda^2 J_2\right)\right> \\
\n & = & \frac{1}{4 \pi }  \left(- \delta \omega^0 \xi \inter e^0 + \delta e^2 \lambda^2 \right)\\
 & = &  \frac{1}{2 \pi }\delta M d\phi .\label{BTZkfinal} 
\ea
Notice that \eqref{BTZkfinal} does not depend on $r$, then $d k_{\epsilon}^{\scriptscriptstyle (CS)}=0$. The variation of the charge is
\be 
\slashed{\delta} Q_\epsilon =\oint k_{\epsilon}^{\scriptscriptstyle (CS)}=\frac{1}{2\pi} \oint_{S}  \delta M d \phi.
\ee
For this case, the integration is easily carried out on a circle $S: t=cte$ and  $r=cte$, we get
\be
\slashed{\delta} Q_\epsilon =\delta M.
\ee
The integration on phase space is trivially satisfied (with the integration constant set to zero), then $Q_\epsilon=M$. This charge is the mass of the non-rotating BTZ black hole.

%----------------------------------------------
\subsection{A Self-dual Taub-NUT with BF Variables: Vanishing of Charges}\label{BFexample}
%----------------------------------------------

The BF language is another alternative way to study pure General Relativity. One interesting feature of the formalism is its economy of components. Here we apply the surface charges BF formula to a simple Taub-NUT solution and show the power of having a compact formulation of gravity. 

For simplicity we use the Euclidean signature but the result is the same for the Lorentzian signature.  Consider the (Euclidean) Taub-NUT solution with cosmological constant \cite{Gibbons:1978zy}. For a particular value of the mass parameter, the solutions is self-dual. In (chiral) BF variables self-duality is simply expressed as\footnote{The BF formalism works by itself. For instance, by imposing self-dual condition one can solve the equation of motion and study the solution without making any contact with metric or tetrad-connection variables (check Section 6 in \cite{Herfray:2016azk}). However, for completeness and as a check, here we write down the explicit relation with the other variables. Let us start with the metric \cite{Gibbons:1978zy}
\be
ds^2=\frac{f}{\Delta}dr^2+\frac{4n^2\Delta}{f}(d\psi^2+\cos\theta d\phi)^2+f(d\theta^2+\sin^2\theta d\psi^2),
\ee
where $f=r^2-n^2$ and $\Delta=r^2-2mr+n^2+\frac{3}{\ell^2}\left(n^4+2n^2r^2-\frac{1}{3}r^4\right)$. The constant $m$ is the mass and $n$ the NUT parameter. The solution is (anti-)self-dual for $m=\pm n\left(1+\frac{4n^2}{\ell^2}\right)$. Consider the tetrad one-form compatible with the metric
\ba
e^0&=&\sqrt{\frac{f}{\Delta}}dr,\quad\quad\quad\quad\quad\quad\quad\quad\quad\quad e^1\ =\ 2n\sqrt{\frac{\Delta}{f}}\left(d\psi+\cos\theta d\phi\right), \n \\
e^2&=&\sqrt{f}\left(\cos\psi d\theta+\sin\psi\sin\theta d\phi\right),\quad e^3\ =\ \sqrt{f}\left(\sin\psi d\theta-\cos\psi \sin\theta d\phi\right),
\ea
the connection $\omega^{ab}(e)$ is obtained as usual by solving $de^a+\omega^a_{\ b}\wedge e^b=0$. Having the pair $(e^a,\omega^{ab})$ we can use the relation with the Euclidean BF variables
\be
B^i=e^0\wedge e^i-\frac{1}{2}\epsilon^{ijk}e^j\wedge e^k,\quad\quad A^i=\omega^{0i}-\frac{1}{2}\epsilon^{ijk}\omega^{jk},
\ee
compute curvature $F^i\equiv dA^i+\frac{1}{2}\epsilon^{ijk}A^j\wedge A^k$ and finally check the self-dual condition. The previous sketch can be explicitly checked using the Mathematical notebook at \href{https://sites.google.com/view/surfacechargetoolkit/home}{[sites.google.com/view/surfacechargetoolkit/home]}.
}
\be
F^i=\pm\frac{1}{\ell^2}B^i,
\ee
with $\pm$ the sign of the cosmological constant. The surface charge computed with \eqref{BF-surfacecharge}, for a solution satisfying this condition is simply 
\be
\slashed\delta Q_\epsilon=\pm\ell^2\oint\left(\delta F^i \lambda^i-\delta A^i \xi\inter F^i\right)=\pm\ell^2\oint\left( d_A\delta A ^i \lambda^i-\delta A^i  d_A \lambda^i\right)=\pm\ell^2\oint d\left(\lambda^i\delta A^i\right)=0,
\ee
where we used $\delta F^i=d_A\delta A^i$ and the exact symmetry condition $\delta_\epsilon A^i=\xi\inter F^i-d_A\lambda^i=0$.

This simple calculation shows that the surface charges for the self-dual Taub-NUT are trivial in the sense that its variations vanish.  

We compare this result with the analysis, in \cite{Araneda:2018orn}, using a Noether-like procedure to compute charges at the asymptotic region. There the authors, in the same spirit that \cite{Aros:1999id}, conclude the vanishing of charges with the help of adding an extra topological term into the Einstein-Cartan-$\Lambda$ action (the Pontriagyn invariant) with a specific coefficient in order to satisfy {\it self-dual asymptotic boundary conditions}. 

We stress two points. First, as explained in Section \ref{sec:introduction} because gravity is a gauge theory Noether First Theorem does not apply, therefore, Noether-like attemps might be misleading in this sense. Second, in the surface charge method the addition of a boundary term (a topological term here) to the Lagrangian action does not affect the surface charge $\slashed \delta Q_\epsilon=\oint k_\epsilon$. The boundary term always contribute to the surface charge density formula by an exact form, $k_\epsilon\to k_\epsilon+d\alpha$,  which integrated on a closed surface does not have any effect \cite{Frodden:2017qwh}.

%--------------------------------------------------------
\section{Discussion}
\label{}
%--------------------------------------------------------

In this toolkit we have presented the method, known as the surface charge method, to compute charges for gauge theories. As explained at the beginning this method extends the old Noether program to theories with gauge redundancy: (Exact) symmetries implies charges {\it also} for gauge theories. Here we have reviewed precisely how this happens and extensively applied the method to different gravity theories.

The main difference with the First Noether Theorem is that in the surface charge method the construction is over a one-dimension lower submanifold. That is, for theories without gauge symmetries, the usual Noether current $J_\epsilon^\mu$ corresponds to a $(D-1)$-form and defines a charge over a $(D-1)$-dimensional spacetime slice $\Sigma$ as $Q_\epsilon=\int_{\scriptscriptstyle \Sigma} J_\epsilon^\mu d\Sigma_\mu$. In contrast, for gauge theories, we have the surface charge density $k_\epsilon^{\mu\nu}$, playing the role of a {\it current}, that corresponds to a $(D-2)$-form and defines a (differential) charge over a closed $(D-2)$-dimensional surface $S$ as $\slashed \delta Q_\epsilon = \oint_{\scriptscriptstyle S} k_\epsilon^{\mu\nu}dS_{\mu\nu}$. Thus, a second difference is that in the surface charge method, instead of finite charges, one defines differential charges on phase space. They require further phase space integration. 

The integrability of $\slashed \delta Q$, to obtain a $Q$,  is usually assumed in the literature, but as we saw in detail is not ensured by the method. Still for exact solutions the integration on phase space of charges is achieved easily in most cases. 

In the context of asymptotic charges, which are a consequence of asymptotic symmetries, a lot of work is focused on providing the conditions to have integrable asymptotic charges. In contrast insufficient effort is put in providing an explanation of what those non-conserved asymptotic charges mean. Through these notes it is clear that to have a conservation law an exact symmetry equation is needed: An equation that by definition asymptotic symmetries do not satisfy.

It is on the purposes of these notes to bring clarity on the surface charge method in order to focus its applicability on physical systems in the context of current research. But also this toolkit is written in the hope to easy the work of gravity physicist, specially to those who decide to use alternative variables, and need to analyze the field equation solutions in the light of gauge invariant quantities. It is in this spirit that we systematically wrote the surface charge density formulas for many gravity theories, presented all of them in a Table, and furthermore we also performed all the step-by-step calculation in the appendices.

A key field where the results here are of direct use is in analyzing black holes, more specifically, their thermodynamic properties. The surface charge method allows us to have good control of the space of exact symmetries and the space of differential charges at once. In the case of black holes this analysis can lead to a well-posed first law of black hole mechanics\footnote{See \cite{Barnich:2003xg} for a general treatment or \cite{Astorino:2016hls} for a recent successful application on rotating magnetic black hole that has a non-usual asymptotic description.}.  In the first of the three examples, Subsection \ref{exBTZ1}, this was worked in detail for a complicated enough black hole family and showed that a consistent first law holds.

As a final comment, and to put this toolkit in a wider perspective, we should stress that all the methods to compute charges for gauge theories are defined in a classical regime. Then, it remains as a challenge to understand how the surface charge method works within a quantum gravity theory. It certainly depends on the quantization procedure one applies to the classical phase space structure and the different quantities defined there. Still, nowadays there is no clear path to achieve this quantization. We hope that future research will provide it.

%----------------------------------------------
\section*{Acknowledgments}

The authors thank Roberto Oliveri for collaboration and comments at several stages of this project.
The authors are also grateful to F. Canfora, K. Krasnov, M. Riquelme, S. Speziale, R. Troncoso, and J. Zanelli for enlightening discussions. EF wishes to thank the Centro de Estudios Cient\'ificos for the hospitality during a 2019 research visit. EF is partially funded by FONDECYT grant \#11150467 and DH is partially founded by CONICYT grant \#21160649. The Centro de Estudios Cient\'ificos (CECs) is funded by the Chilean Government through the Centers of Excellence Base Financing Program of CONICYT.
%----------------------------------------------

\appendix

%----------------------------------------------
\section{Notation and conventions}
\label{conventions}
%----------------------------------------------

For a $D$-dimensional spacetime, we use the signature convention for the flat metric $\eta_{ab} = \text{diag}(-1,1,\dots,1)$ and the Levi-Civita totally anti-symmetric symbol $\varepsilon_{\mu_1 \cdots \mu_D}$ such that $\varepsilon_{\scriptscriptstyle 01\cdots (D-1)}=1$, we also have
\ba
\varepsilon_{\mu_1 \cdots \mu_D}&=&\tilde \varepsilon^{\mu_1 \cdots \mu_D}\\
g^{\mu_1 \nu_1} \cdots g^{\mu_D \nu_D} \varepsilon_{\mu_1 \cdots \mu_D} & = & \frac{1}{g}\, \tilde\varepsilon^{\nu_1 \cdots 	\nu_D}=\varepsilon^{\nu_1 \cdots 	\nu_D}
\label{raiseg}
\\
g_{\mu_1 \nu_1} \cdots g_{\mu_D \nu_D} \tilde \varepsilon^{\mu_1 \cdots \mu_D} & = & g\, \varepsilon_{\nu_1 \cdots 	\nu_D},
\ea
with the spacetime metric determinant $g\equiv \det (g_{\mu\nu})$. We also introduced $\tilde \varepsilon^{\mu_1 \cdots \mu_D}$ such that  $\tilde\varepsilon^{\, 01\cdots (D-1)}=1$, this twiddle symbol is exactly the Levi-Civita symbol but with indices written upstairs. In contrast, $\varepsilon^{\mu_1\cdots \mu_D}$ is a spacetime function, not the Levi-Civita symbol, its indices are raised with the spacetime metric. Note that we use Greek letters for spacetime indices and Latin letters for internal indices. Thus, similarly to \eqref{raiseg}, to raise and lower indices with the internal flat metric, yields
\ba
 \eta^{a_1 b_1} \cdots \eta^{a_D b_D} \varepsilon_{a_1 \cdots a_D} & = & \det (\eta^{-1}) \tilde\varepsilon^{b_1 \cdots 	b_D}= -\tilde\varepsilon^{b_1 \cdots 	b_D}
=\varepsilon^{b_1 \cdots 	b_D}.
\label{leviflat}
\ea
The introduction of the object $\tilde \varepsilon$ is highly recommended as a way to keep consistent the Einstein notation of index contraction and thus to avoid some usual confusions on the computations. 
%----------------------------------------------
\subsection{Three differential forms operations}
\label{formsdefinitions}
%----------------------------------------------
Let us consider a $p$-form expressed in a coordinate basis 
\be
\alpha = \frac{1}{p!} \alpha_{\mu_1 ... \mu_p} dx^{\mu_1} \wedge \cdots \wedge dx^{\mu_p}. 
\ee
The operator exterior derivative on differential forms,  $d: \Omega^p \longrightarrow \Omega^{p+1}$, has the explicit action 
\ba
d \alpha = \frac{1}{p!} \partial_{\mu_0} \alpha_{\mu_1 ...\mu_p} dx^{\mu_0} \wedge  dx^{\mu_1} \wedge \cdots \wedge dx^{\mu_p}.
\ea
This operator satisfies the following Leibniz rule $d(\alpha \wedge \beta) = d\alpha \wedge \beta + (-1)^p \alpha \wedge d \beta$, and it is nilpotent, $d^2=0$. 

The Hodge dual operator $\star : \Omega^{\scriptscriptstyle p} \longrightarrow \Omega^{\scriptscriptstyle D-p}$, acts on a $p$-form as
\be
\star \alpha = \frac{1}{(D-p)! p!} \alpha^{a_1 ...a_p} \varepsilon_{a_1 ...a_pb_1 ... b_{D-p}} e^{b_1}\wedge \cdots \wedge e^{b_{D-p}}, 
\label{Hodgevielbein}
\ee
or with the differential form expressed in a coordinate basis
\be
\star \alpha = \frac{1}{(D-p)!p!} \sqrt{ |g|} \alpha^{\mu_1 ...\mu_p} \varepsilon_{\mu_1 ...\mu_p \nu_1...\nu_{\scriptscriptstyle D-p}} dx^{\nu_1} \wedge \cdots \wedge dx^{\nu_{D-p}}.  
\label{Hodgespacetime}
\ee

For a vector field $\xi=\vec{\xi}=\xi^a e_a = \xi^\mu \partial_\mu$, the interior product on forms is either denoted  by $i_\xi$ or also with the alternative notation  $\xi\inter$. This operation lows by one the form degree $i_\xi\equiv\xi\inter: \Omega^{\scriptscriptstyle p} \longrightarrow \Omega^{\scriptscriptstyle p-1}$. Explicitly, on a $p$-form it has the action
\be
\xi\inter \alpha = \frac{1}{(p-1)!} \xi^\nu \alpha_{\nu \mu_2 ....\mu_p}dx^{\mu_2} \wedge \cdots \wedge dx^{\mu_p}.  
\ee

%----------------------------------------------
\section{Einstein-Hilbert-$\Lambda$ action}
\label{gr}
%----------------------------------------------
Consider the Einstein-Hilbert-$\Lambda$ action
\be
S[g_{\mu\nu}]=\frac{\kappa}{2}\int_{\scriptscriptstyle \cal M}dx^{\scriptscriptstyle D}\sqrt{-g}\left(R-2\Lambda\right).
\ee
We first need to compute the variation of the Lagrangian generated by an arbitrary vector field $\xi=\xi^\mu\partial_\mu$. One has, from Eq.~\eqref{variationL}, that
\ba
\partial_\mu(\xi^\mu L) &=&E_{\mu\nu}\delta_{\xi}g^{\mu\nu}+\nabla_\mu \Theta^\mu(g, \delta_\xi g), \nonumber\\
&=&-2E_{\mu\nu}\nabla^{(\mu}\xi^{\nu)}+\nabla_\mu \Theta^\mu(g, \delta_\xi g), \label{hola0} \nonumber\\
&=& -\nabla^{(\mu}\left(2 E_{\mu\nu} \xi^{\nu)}\right) + 2\xi^{(\nu}\nabla^{\mu)}E_{\mu\nu}  + \nabla_{\mu}\Theta^\mu(g, \delta_\xi g), \nonumber\\
&=& -\nabla^{\mu}\left(2 E_{\mu\nu} \xi^{\nu}\right) + \nabla_{\mu} \Theta^{\mu}(g, \delta_\xi g),\nonumber\\ 
&=& \nabla^{\mu}\left[ -2 E_{\mu\nu} \xi^{\nu} + \Theta_{\mu}(g, \delta_\xi g)\right]. \label{hola}
\ea
In the second line we replaced\footnote{Notice that from the variation of the metric, $\delta_\xi g_{\mu\nu}=\nabla_\mu\xi_\nu+\nabla_\nu\xi_\mu$, and from the identity $0=\delta(\delta^\mu_\nu)=\delta g^{\mu\alpha}g_{\alpha\nu}+ g^{\mu\alpha}\delta g_{\alpha\nu}$, the variation of the inverse metric gets a minus sign.} $\delta_\xi g^{\mu\nu}=-\nabla^\mu\xi^\nu-\nabla^\nu\xi^\mu= -2\nabla^{(\mu}\xi^{\nu)}$, in the third line we made use of the Leibniz rule, and in the fourth line we used the symmetry of $E_{\mu\nu}$ and the Noether (Bianchi) identity $\nabla^{\mu}E_{\mu\nu} = 0$. 
%\footnote{Remember that Noether identities are proven from $\partial_\mu J^\mu(\xi)=\xi_\mu N^\mu$ for an arbitrary vector field $\xi$ implies $N^\mu=0$ identically (off-shell).}.

The explicit expressions of the quantities in Eq.~(\ref{hola}) are
\ba 
L&=& \frac{\kappa}{2}\sqrt{-g}\left(R-2\Lambda\right), \label{exprGR1}\\
E^{\mu\nu}&=& \frac{\kappa}{2}\sqrt{-g}\left(R^{\mu\nu}-\frac{1}{2}g^{\mu\nu}R+\Lambda g^{\mu\nu}\right), \label{exprGR2}\\
\Theta^\mu(g, \delta_\xi g) &=&\kappa \sqrt{-g}\nabla^{[\alpha}(g^{\mu]\beta}\delta_\xi g_{\alpha\beta}), \n\\
&=& \kappa \sqrt{-g}\left(\nabla_\alpha\nabla^{(\alpha} \xi^{\mu)}-\nabla^\mu\nabla^\alpha \xi_\alpha\right),\n\\
&=&\kappa \sqrt{-g}\left(\nabla_\alpha \nabla^{[\alpha}\xi^{\mu]}+[\nabla_\alpha,\nabla^\mu]\xi^\alpha\right),\n\\
&=&\kappa\sqrt{-g}\left(\nabla_\alpha \nabla^{[\alpha}\xi^{\mu]}+R^{\mu\alpha}\xi_\alpha\right), \label{exprGR3}
\ea
where we used the identity $[\nabla_\alpha,\nabla^\mu]\xi^\alpha=R^{\mu\nu}\xi_\nu$.
As seen in the general case, Eq.~\eqref{totd}, and using Eqs.~\eqref{exprGR1}-\eqref{exprGR2}-\eqref{exprGR3}, one has the trivially conserved current
\ba
J_{\xi}^{\mu} &=& \Theta^{\mu}(g,\delta_{\xi} g)- \xi^{\mu} L - 2\xi_{\nu} E^{\mu\nu}, \n\\
&=& \kappa \sqrt{-g} \nabla_{\nu} \nabla^{[\nu}\xi^{\mu]},\n \\
&=&\kappa\, \partial_{\nu} \left(\sqrt{-g}\nabla^{[\nu}\xi^{\mu]}\right)=\kappa\, \partial_{\nu} \left(-\sqrt{-g}\nabla^{[\mu}\xi^{\nu]}\right).
\ea
Then, the Noether potential
\be
\widetilde Q_\xi^{\mu\nu}= -\kappa\sqrt{-g}\nabla^{[\mu}\xi^{\nu]},
\ee
because its anti-symmetry the current is trivially conserved, $\partial_{\mu} \partial_{\nu}\widetilde Q_\xi^{\mu\nu} = 0$, without using the equations of motion.

Finally, the surface charge density is given by
\ba
k_\xi^{\mu\nu}&=&\delta \widetilde Q_\xi^{\mu\nu} + 2\xi^{[\mu}\Theta^{\nu]}(g, \delta g), \n\\
&=&-\kappa \delta\left(\sqrt{-g}\nabla^{[\mu}\xi^{\nu]}\right)+2\kappa\sqrt{-g}\xi^{{\color{blue}\dot [}\mu}\nabla^{[\alpha}\left(g^{\nu]{\color{blue}\dot ]}\beta}\delta g_{\alpha\beta}\right).
\label{pekin}
\ea
To compute the variation of the first term we use 
 \be
\delta (\nabla_\alpha \xi^\nu)=\delta \Gamma^\nu_{\ \alpha\gamma}\xi^\gamma=\frac{1}{2}g^{\nu\lambda}\left(\nabla_\gamma \delta g_{\alpha\lambda}+\nabla_\alpha\delta g_{\gamma\lambda}-\nabla_\lambda \delta g_{\alpha \gamma}\right)\xi^\gamma,
\ee
we insert this in the first term of \eqref{pekin} as
\ba
\delta\left(\sqrt{-g}\nabla^{[\mu}\xi^{\nu]}\right)&=&\delta\left(\sqrt{-g}g^{[\mu|\alpha}\nabla_\alpha \xi^{\nu]}\right),\\
&=&\sqrt{-g}\left(-\frac{1}{2}\delta g \nabla^{[\mu}\xi^{\nu]}+\delta g^{\sigma[\mu}\nabla_\sigma \xi^{\nu]}-\xi_\sigma\nabla^{[\mu}\delta g^{\nu]\sigma}\right),
\ea
with $\delta g\equiv g_{\mu\nu}\delta g^{\alpha\beta}$, then we replace the result in (\ref{pekin}) to get the final expression of the surface charge in Eq.~(\ref{metricGR}).

%----------------------------------------------
\section{Einstein-Hilbert-Maxwell action}
\label{appEHM}
%----------------------------------------------

The variation of the Lagrangian generated by an arbitrary vector field $\xi=\xi^\mu\partial_\mu$ and gauge transformation $\lambda'=\lambda+\xi^\mu A_\mu$  (with $\delta_\epsilon =\delta_\xi+\delta_{\lambda'}$)
\ba
\delta_\xi L&=&E_{\mu\nu}\delta_{\xi}g^{\mu\nu}+E_\mu \delta_\epsilon A^\mu+\partial_\mu \Theta^\mu(\delta_\epsilon),\\
\partial_\mu(\xi^\mu L)&=&-E_{\mu\nu}(\nabla^\mu\xi^\nu+\nabla^\nu\xi^\mu)+E_\mu( \xi_\nu F^{\nu\mu}-\nabla^\mu \lambda )+\partial_\mu \Theta^\mu(\delta_\epsilon),\\
\partial_\mu (\xi^\mu L)&=&\partial_\mu\left[-2(\xi_\nu E^{\mu\nu})-E^\mu\lambda+\Theta^\mu(\delta_\epsilon)\right],\label{holaelectro}
\ea
where we used the corresponding Noether identities $F\indices{^\mu_\nu} E^\nu-2\nabla_\nu E^{\mu\nu}=0$ and $\nabla_\mu E^\mu=0$.  The explicit functions in (\ref{holaelectro}) are
\ba
L&=&\sqrt{-g}\left(\frac{\kappa}{2}(R-2\Lambda)-\frac{1}{4}F_{\mu\nu}F^{\mu\nu}\right),\\
E^{\mu\nu}&=&\sqrt{-g}\left[\frac{\kappa}{2}\left(R^{\mu\nu}-\frac{1}{2}g^{\mu\nu}R+\Lambda g^{\mu\nu}\right)-\frac{1}{2}\left(g_{\alpha\beta}F^{\alpha\mu}F^{\beta\nu}-\frac{1}{4}g^{\mu\nu}F^{\alpha\beta}F_{\alpha\beta}\right)\right],\label{eomEHMAX}\\
E^\mu&=&-\sqrt{-g}\nabla_\nu F^{\mu\nu},\\
\Theta^\mu(\delta_\epsilon) &=&\sqrt{-g}\left(\kappa \nabla^{[\alpha}(g^{\mu]\beta}\delta_\xi g_{\alpha\beta})-\delta_\epsilon A_\nu F^{\mu\nu}\right),\\
&=& \sqrt{-g}\left[\kappa \left(\nabla_\alpha\nabla^{(\alpha} \xi^{\mu)}-\nabla^\mu\nabla^\alpha \xi_\alpha\right)-\left(\xi^\alpha F_{\alpha\nu}-\nabla_\nu \lambda\right)F^{\mu\nu}\right].\n
\ea
Replacing them we note that the three terms inside the total derivative in (\ref{holaelectro}) can be rewritten as a total derivative too
\ba
J_\epsilon^\mu\equiv\Theta^{\mu}(\delta_\epsilon)-\xi^\mu L-2\xi_\nu E^{\mu\nu}-\lambda E^\mu=\partial_\nu\left[-\sqrt{-g}\left(\kappa \nabla^{[\mu}\xi^{\nu]}-\lambda F^{\mu\nu}\right)\right],
\ea
where we used $[\nabla^\mu,\nabla^\nu]\xi_\mu=R^{\mu\nu}\xi_\mu$. The term inside the total derivative is $\widetilde Q_\xi^{\mu\nu}$. Then we obtain the surface charge density
\ba
k_\xi^{\mu\nu}&=&\delta \widetilde Q_\xi^{\mu\nu}+2\xi^{{\color{blue}[}\mu}\Theta^{\nu{\color{blue}]}}(\delta),\\
&=&-\delta \left[\sqrt{-g}\left(\kappa \nabla^{[\mu}\xi^{\nu]}-\lambda F^{\mu\nu}\right)\right]+2\sqrt{-g}\xi^{{\color{blue}{\dot[}}\mu}\left( \kappa\nabla^{[\alpha}(g^{\nu]{\color{blue}{\dot]}}\beta}\delta g_{\alpha\beta})-\delta A_\alpha F^{\nu{\color{blue}{\dot ]}}\alpha}\right).
\n
\ea
Note that when using improved transformations we have $\delta \lambda =\delta (\lambda'-\xi^\mu A_\mu)=-\xi^\mu \delta A_\mu$\footnote{ Equivalently we can forget this and note that in the symplectic structure density the variations do not commute, thus we should include a term $\Theta([\delta,\delta_\epsilon])$ which will contribute to the surface charge.  The specific non commutation in this case is $[\delta,\delta_\epsilon] A_\mu =\delta_{(\xi^\nu\delta A_\nu)}A_\mu=-\partial_\mu (\xi^\nu \delta A_\nu)$.}. Expanding all we obtain the surface charge density for this theory \eqref{SCem}.

%----------------------------------------------
\section{ Einstein-Hilbert action with a conformally coupled scalar field}
\label{appEHscalar}
%----------------------------------------------
The variation of the Lagrangian \eqref{scalarL} generated by an arbitrary vector field $\xi=\xi^\mu\partial_\mu$ is
\ba
\delta_\xi L&=&E_{\mu\nu}\delta_{\xi}g^{\mu\nu}+E_\Phi \delta_\xi \Phi+\partial_\mu \Theta^\mu(\delta_\xi),\n\\
\partial_\mu(\xi^\mu L)&=&-E_{\mu\nu}(\nabla^\mu\xi^\nu+\nabla^\nu\xi^\mu)+E_\Phi \xi_\mu\nabla^\mu\Phi+\partial_\mu \Theta^\mu(\delta_\xi),\n\\
\partial_\mu (\xi^\mu L)&=&\partial_\mu\left[-2(\xi_\nu E^{\mu\nu})+\Theta^\mu(\delta_\xi)\right]\label{holascalar},
\ea
where we used the corresponding Noether identity $\frac{1}{2}\nabla^\mu \Phi E_\Phi-\nabla_\nu E^{\mu\nu}=0$.  

The explicit expressions in (\ref{holascalar}) are
\ba
L&=&\frac{\kappa}{2}\sqrt{-g}\left(R-2\Lambda-\frac{1}{\kappa}\nabla^\mu\Phi\nabla_\mu \Phi +\frac{\zeta_{\scriptscriptstyle D}}{\kappa} R\Phi^2 \right)\\
E^{\mu\nu}&=&\frac{\kappa}{2}\sqrt{-g}\left[\left(R^{\mu\nu}-\frac{1}{2}g^{\mu\nu}R\right)\left(1+\frac{\zeta_{\scriptscriptstyle D}}{\kappa} \Phi^2\right)+\Lambda g^{\mu\nu}\right.\n \\
&&\quad\quad\left.+\frac{1}{\kappa} \left(-\nabla^\mu \Phi\nabla^\nu \Phi+\frac{1}{2}g^{\mu\nu}  \nabla^\alpha\Phi\nabla_\alpha \Phi+\zeta_{\scriptscriptstyle D}\left(g^{\mu\nu}\square \Phi^2-\nabla^{(\mu}\nabla^{\nu)} \Phi^2 \right)\right)\right]\\
\Theta^\mu(\delta_\xi) &=&\kappa\sqrt{-g}\left(\nabla^{[\alpha}(g^{\mu]\beta}\delta_\xi g_{\alpha\beta})\left(1+\frac{\zeta_{\scriptscriptstyle D}}{\kappa} \Phi^2\right)-\frac{1}{\kappa}\delta_\xi \Phi \nabla^\mu \Phi+\frac{\zeta_{\scriptscriptstyle D}}{\kappa} g^{\beta[\alpha}\delta_\xi g_{\alpha\beta}\nabla^{\mu]}\Phi^2\right)\n\\
&=& \kappa \sqrt{-g}\left[\left(\nabla_\alpha\nabla^{(\alpha} \xi^{\mu)}-\nabla^\mu\nabla^\alpha \xi_\alpha\right)\left(1+\frac{\zeta_{\scriptscriptstyle D}}{\kappa} \Phi^2\right)-\frac{1}{\kappa}\xi^\alpha\nabla_\alpha\Phi \nabla^\mu\Phi\right.\n\\
&&\quad\quad\quad\quad\quad\quad\quad\quad\quad\quad\quad\quad\quad\quad\left.+\frac{\zeta_{\scriptscriptstyle D}}{\kappa}\left(\nabla_\alpha \xi^\alpha \nabla^\mu \Phi^2-\nabla^{(\alpha}\xi^{\mu)}\nabla_\alpha \Phi^2\right)\right].
\ea
Replacing them we note that the three terms inside the total derivative in (\ref{holascalar}) can be rewritten as a total derivative too
\ba
J_\xi^\mu\equiv\Theta^{\mu}(\delta_\xi)-\xi^\mu L-2\xi_\nu E^{\mu\nu}=\partial_\nu\left[-\kappa\sqrt{-g}\left(\nabla^{[\mu}\xi^{\nu]}\left(1+\frac{\zeta_{\scriptscriptstyle D}}{\kappa} \Phi^2\right)+\frac{2\zeta_{\scriptscriptstyle D}}{\kappa} \xi^{[\mu}\nabla^{\nu]}\Phi^2\right)\right],\n
\ea
where we used $[\nabla^\mu,\nabla^\nu]\xi_\mu=R^{\mu\nu}\xi_\mu$. The term inside the total derivative is $\widetilde Q_\xi^{\mu\nu}$. Then we obtain the surface charge density
\ba
k_\xi^{\mu\nu}&=&\delta \widetilde Q_\xi^{\mu\nu}+2\xi^{{\color{blue}[}\mu}\Theta^{\nu{\color{blue}]}}(\delta)\\
&=&\delta\left[-\kappa\sqrt{-g}\left(\nabla^{[\mu}\xi^{\nu]}\left(1+\frac{\zeta_{\scriptscriptstyle D}}{\kappa} \Phi^2\right)+\frac{2\zeta_{\scriptscriptstyle D}}{\kappa} \xi^{[\mu}\nabla^{\nu]}\Phi^2\right)\right]\\
&&+2\kappa\sqrt{-g}\xi^{{\color{blue}{\dot [}}\mu}\left(\nabla^{[\alpha}(g^{\nu]{\color{blue}{\dot ]}}\beta}\delta g_{\alpha\beta})\left(1+\frac{\zeta_{\scriptscriptstyle D}}{\kappa}  \Phi^2\right)-\frac{1}{\kappa} \delta \Phi \nabla^{\nu{\color{blue}{\dot ]}}} \Phi+\frac{\zeta_{\scriptscriptstyle D}}{\kappa}  g^{\beta[\alpha}\delta g_{\alpha\beta}\nabla^{\nu]{\color{blue}{\dot ]}}}\Phi^2\right),
\n
\ea
the previous expression reduces to (\ref{metricGRPhi}).

%----------------------------------------------
\section{Einstein-Hilbert-Skyrme action} 
\label{appEHS}
%----------------------------------------------

The variation of the Lagrangian \eqref{skyrmeaction} generated by an arbitrary vector field $\xi=\xi^\mu\partial_\mu$ 
\ba
\delta_\xi L&=&E_{\mu\nu}\delta_{\xi}g^{\mu\nu}+\left<E_U \delta_\xi U\right>+\partial_\mu \Theta^\mu(\delta_\xi)\\
\partial_\mu(\xi^\mu L)&=&-E_{\mu\nu}(\nabla^\mu\xi^\nu+\nabla^\nu\xi^\mu)+\left<E_U \xi_\mu \nabla^\mu U\right>+\partial_\mu \Theta^\mu(\delta_\xi)\\
\partial_\mu (\xi^\mu L)&=&\partial_\mu\left[-2\xi_\nu E^{\mu\nu}+\Theta^\mu(\delta_\xi)\right],\label{holaskyrme}
\ea
where we used the Noether identity $2\nabla^\mu E\indices{_\mu_\nu}+\left<E_U\nabla_\nu U\right>=0$.  The explicit functions in (\ref{holaskyrme}) are
\ba
L&=& \sqrt{-g}\left[\frac{\kappa}{2}\left(R-2\Lambda\right)+\frac{K}{4}\left<R^\mu R_\mu+\frac{\lambda}{8}F_{\mu\nu}F^{\mu\nu}\right>\right]\\
E^{\mu\nu}&=&\sqrt{-g}\left[\frac{\kappa}{2}\left(R^{\mu\nu}-\frac{1}{2}g^{\mu\nu}R-\Lambda g^{\mu\nu}\right)\right.\n\\
&&\left.\quad\quad+\frac{K}{4}\left<R^\mu R^\nu-\frac{1}{2}g^{\mu\nu}R^\alpha R_\alpha+\frac{\lambda}{4}\left(F^{\mu\alpha}F\indices{^\nu_\alpha}-\frac{1}{4}g^{\mu\nu}F_{\alpha\beta}F^{\alpha\beta}\right)\right>\right]\\
E_{\scriptscriptstyle U}&=&-\frac{K}{2}\sqrt{-g}\left<\nabla_\mu\left(R^\mu+\frac{\lambda}{4}[R_\nu,F^{\mu\nu}]\right)U^{-1}\right>\\
\Theta^\mu(\delta_\xi) &=&\sqrt{-g} \left[\kappa\nabla^{[\alpha }(g^{\mu]\beta}\delta_\xi g_{\alpha\beta})+\frac{K}{2}\ \left<\left(R^\mu +\frac{\lambda}{4}[R_\nu, F^{\mu\nu}]\right)U^{-1}\delta_\xi U\right>\right].
\ea
and again, after cancellations, we simply have
\ba
J_\epsilon^\mu\equiv\Theta^{\mu}(\delta_\xi)-\xi^\mu L-2\xi_\nu E^{\mu\nu}=\partial_\nu\left[-\kappa\sqrt{-g}\nabla^{[\mu}\xi^{\nu]}\right],\n
\ea
where the cyclic property of the trace was used to show that $\left<[R_\nu,F^{\mu\nu}]\xi^\alpha R_\alpha-\xi_\nu F^{\mu\alpha}F\indices{^\nu_\alpha}\right>=0$. Therefore ${\widetilde Q}_\xi^{\mu\nu}=-\kappa\sqrt{-g}\nabla^{[\mu}\xi^{\nu]}$, and we use the surface charge formula $k^{\mu\nu}_\xi=\delta \widetilde Q^{\mu\nu}_\xi +2\xi^{[\mu} \Theta^{\nu]}(\delta)$ to get (\ref{SCskyrme}).

%----------------------------------------------
\section{Einstein-Cartan-$\Lambda$}
\label{appendixECL}
%----------------------------------------------
Consider four-dimensional General Relativity with a cosmological term in the differential form language
\be \label{EClambda}
S[e^a,\omega^{ab}]=\kappa'\int_{\scriptscriptstyle \cal M} \varepsilon_{abcd} \left( R^{ab}e^c e^d \pm \frac{1}{2\ell^2} e^a e^b e^c e^d      \right) , \quad \ell^2 = \frac{3}{|\Lambda|}.
\ee
The general variation of the Lagrangian density is
\ba
\delta L = E_a \delta e^a +E_{ab}\delta \omega^{ab} + d \Theta(\delta \omega).
\ea
The equations of motion and boundary term are given by
\ba
E_a &=& 2\kappa' \varepsilon_{abcd}\left(R^{bc} \pm \frac{1}{\ell^2} e^b e^c  \right) e^d  =0, \\
E_{ab}&=& 2\kappa' \varepsilon_{abcd}T^c e^d=0,\\
\Theta (\delta \omega )  & = & \kappa'\varepsilon_{abcd} \delta \omega^{ab} e^c e^d.
\ea
The action \eqref{EClambda} is invariant under diffeomorphisms and local Lorentz transformations. Infinitesimal generators of these symmetries are a vector field $\xi$ and the set of parameters $\lambda^{ab}$, we group them in $\epsilon=(\xi,\lambda^{ab})$. Both can be combined such that the dynamical fields transform infinitesimally as
\ba
\delta_\epsilon e^a &=& d_\omega (\xi\inter e^a)+\xi\inter (d_\omega e^a)+{\lambda^a}_{b} e^b,
\label{transftetrad}\\
\delta_\epsilon \omega^{ab} &=& \xi\inter R^{ab}-d_\omega\lambda^{ab}. 
\label{transfomega}
\ea
The symplectic structure density computed with these local symmetries as one of its entries is
\ba 
\Omega (\delta, \delta_\epsilon)& = & \delta \Theta (\delta_\epsilon \omega) - \delta_\epsilon \Theta (\delta \omega) - \Theta ([\delta, \delta_\epsilon] \omega) , \\
& = & \kappa' \varepsilon_{abcd} \left(  \delta_\epsilon \omega^{ab} \delta (e^c e^d) - \delta \omega^{ab} \delta_{\epsilon} (e^c e^d)     \right) , \\
& = & 2\kappa' \varepsilon_{abcd} \left(  [\xi\inter R^{ab}-d_\omega \lambda^{ab}] e^c \delta e^d\right. - \left.\delta \omega^{ab} [d_\omega \xi\inter e^c+\xi\inter d_\omega e^c+\lambda\indices{^c_f}e^f] e^d     \right) , \label{presympEC0}\\
& = & dk_\epsilon ,
\label{presympEC}
\ea 
note a subtle point, the term $\Theta ([\delta, \delta_\epsilon] \omega)$ should be formally included to guarantee the bilinearity on $\delta$ and $\delta_\epsilon$ because the variations do not commute in general. The surface charge density is
\ba \label{k-EC}
k_\epsilon & = & -\kappa'\varepsilon_{abcd}\left(\lambda^{ab}\delta (e^ce^d)-\delta \omega^{ab}\xi\inter (e^ce^d)\right).
\ea
In the last step of \eqref{presympEC} we used both, the equations of motion and the linearized equations of motion too. Knowing the result of this kind of calculation the strategy is always to rearrange the exterior derivatives on the parameters $\xi$ and $\lambda^{ab}$, second and third terms in \eqref{presympEC0}, to complete exact differential forms and then check that all the remaining terms vanish due to the equations of motion and the linearized equations of motion, for instance $T^a=de^a+\omega\indices{^a_b}e^b=0$ and $\delta T^a=d\delta e^a+\delta\omega\indices{^a_b}e^b+\omega\indices{^a_b}\delta e^b=0$, as a general rule all of them have to be explicitly used.

%----------------------------------------------
\section{Einstein-Cartan-Yang-Mills} 
\label{ECYM}
%----------------------------------------------

The four-dimensional Einstein-Cartan gravity coupled to a non-Abelian field reads 

\be\label{ECYMactionap}
S[e^a, \omega^{ab}, A]=  \int_{\scriptscriptstyle \cal M} \left( \kappa' \varepsilon_{abcd}R^{ab} e^ce^d   + \alpha_{\scriptscriptstyle YM} \left<  F\star F\right> \right) ,
\ee
where $\left< \cdot  \right> $ denotes an invariant bilinear form on the Lie algebra of the non-Abelian Lie group $SU(N)$, $\alpha_{ \scriptscriptstyle YM}$ is the coupling constant, the two-form $F$ is defined as $F=dA+A\wedge A= \frac{1}{2}F\indices{^i_a_b}e^a e^b \tau_i$, the Hodge operator acts as $\star F= \frac{1}{2} \varepsilon_{abcd} F^{iab} e^c e^d \tau_i$, with $\tau_i$ the $SU(N)$ generators.  

The variation of the Lagrangian is
\ba
\delta L =  E_a \delta e^a +E_{ab}\delta \omega^{ab} + \left<E_{A} 	\delta A\right> +d \Theta(\delta \omega, \delta A),
\ea
with the equations of motion and boundary term given by
\ba
E_a &=& \kappa' \varepsilon_{abcd}  R^{bc}e^d   -  \alpha_{\scriptscriptstyle YM}  \left<  e_a\inter  F  \star F  -Fe_a \inter \star F  \right>=0 \\
E_{ab}&=& 2\kappa' \varepsilon_{abcd}T^c e^d=0, \\
E_{A} & = & -2\alpha_{\scriptscriptstyle YM}\,  d_{\scriptscriptstyle A} \star F =0,\\
\Theta (\delta \omega, \delta A)  & = &\kappa' \varepsilon_{abcd} \delta \omega^{ab} e^c e^d+2 \alpha_{\scriptscriptstyle YM}    \left<   \delta A \star F \right> ,\label{btYM}
\ea
with the covariant derivative defined by $d_{\scriptscriptstyle A} (\cdot) = d (\cdot) + [A ,\cdot ]$. Remember the operation of the interior product $e_a\inter F=\frac{1}{2}F_{bc}e_a\inter(e^be^c)=\frac{1}{2}F_{bc}(\delta^b_a e^c-e^b\delta^c_a)=F_{ac}e^c$. 

The gauge symmetries are diffeomorphisms, local Lorentz transformations, and $SU(N)$ acting on $A$. The parameters of the infinitesimal symmetries are grouped in $\epsilon =(\xi, \lambda^{ab}, \lambda^i)$, with $\lambda^i$ the components of the algebra valued gauge parameter $\lambda=\lambda^i\tau_i$.  The improved exact symmetry conditions are 
\ba
\delta_\epsilon e^a &=& d_\omega (\xi\inter e^a)+\xi\inter (d_\omega e^a)+{\lambda^a}_{b} e^b=0 ,\\
\delta_\epsilon \omega^{ab} &=& \xi\inter R^{ab}-d_\omega\lambda^{ab}=0, \\
\delta_\epsilon A^i & =  &  \xi\inter F^i- d_A \lambda^i=0.
\ea

As showed in the general case for the differential form language, the surface charge density is the sum of three terms
\be
k_\epsilon=\delta \widetilde Q_\epsilon -\xi\inter \Theta(\delta)-B_{\delta \epsilon}.
\label{gralexp}
\ee
We already have the boundary term, \eqref{btYM}.  Evaluating on an infinitesimal gauge symmetry $E_a \delta_\epsilon e^a +E_{ab}\delta_\epsilon \omega^{ab} + \left<E_{\scriptscriptstyle A} 	\delta_\epsilon A\right>=dS_\epsilon+N_\epsilon$, with the Noether identities $N_\epsilon=0$, we obtain $S_\epsilon$.  Then, as usual, the would-be Noether charge $J_\epsilon=\Theta(\delta_\epsilon)-\xi\inter L+S_\epsilon=d\widetilde Q_\epsilon$ is an exact form with
\be
\widetilde Q_\epsilon = -\kappa' \varepsilon_{abcd}\lambda^{ab}e^ce^d-2\alpha_{\scriptscriptstyle YM}\left<\lambda \star F\right>.
\ee
Now, we use that $[\delta,\delta_{\epsilon}]=[\delta,\Lie_\xi+\delta_{\lambda^{ab}+\xi\inter \omega^{ab}}+\delta_{\lambda+\xi\inter A}]=\delta_{\delta\lambda^{ab}+\xi\inter \delta\omega^{ab}}+\delta_{\delta\lambda+\xi\inter\delta A}$ because the vector field $\xi$ is assumed fixed on the phase space. Then, we have $\Theta([\delta,\delta_\epsilon])=dB_{\delta\epsilon}+C_\epsilon$ such that on-shell $C_\epsilon\approx 0$. Thus, we obtain
\be
B_{\delta \epsilon}=-\kappa'\varepsilon_{abcd}(\delta \lambda^{ab}+\xi\inter \delta \omega^{ab})e^c e^d-2\alpha_{\scriptscriptstyle YM}\left<(\delta\lambda+\xi\inter \delta A)\star F\right>.
\ee
Replacing all back in the general expression \eqref{gralexp} we get
\be
k_\epsilon=-\kappa'\varepsilon_{abcd}\left(\lambda^{ab}\delta(e^ce^d)-\delta \omega^{ab}\xi\inter (e^ce^d)\right) -2 \alpha_{\scriptscriptstyle YM}  \left< \lambda \delta \star F - \delta A \xi \inter \star F  \right>,
\ee
with the first two terms just the surface charge density of pure gravity \eqref{k-EC}. Thus, roughly, to consider the extension to a general YM theory from a pure electromagnetic field one should include the $\left<\cdot \right>$ brackets to deal with the algebra valued fields.

\section{Einstein-Cartan in $(2+1)$-dimensions plus a Torsional Term}
\label{appECtorsion1}
%----------------------------------------------
In this appendix we compute the surface charge density explicitly. Because is faster and equivalent, we use the contracting homotopy operator method to do it. Consider the Lagrangian for $(2+1)$-spacetime dimensions
\be \label{eT-term}
L=\varepsilon_{abc}e^aR^{bc}+\beta e_aT^a,
\ee
where $T^a=de^a+\omega\indices{^a_b}e^b$, and the term proportional to $\beta$ is the one that would produce torsion. The variation of the Lagrangian is
\be
\delta L= \delta e^a(\varepsilon_{abc} R^{bc}+2\beta T_a)+\delta \omega^{ab}(\varepsilon_{abc} T^c-\beta e_ae_b)-d(\varepsilon_{abc} e^a\delta\omega^{bc}+\beta e^a\delta e_a),
\label{varLtorsion}
\ee
the second equation of motion tells us that torsion does not vanish, and because of the first one, we can advance that in fact it plays the role of a cosmological constant term.
 
With the improved infinitesimal gauge transformation for the fields, $\delta_\epsilon e^a$ and $\delta_\epsilon \omega^{ab}$ from (\ref{symconde}) and (\ref{symcondomega}) respectively, we rearrange the combination $E_a\delta_\epsilon e^a+E_{ab}\delta_\epsilon \omega^{ab}=dS_\epsilon+N_\epsilon$, such that
\be
S_\epsilon=\xi\inter e^a(\varepsilon_{abc} R^{bc}+2\beta T_a)-\lambda^{ab}(\varepsilon_{abc} T^c-\beta e_a e_b), 
\ee
and $N_\epsilon=0$ are the Noether identites. The $S_\epsilon$ is what we need to compute the surface charge density using the contracting homotopy operator. For the Einstein-Cartan theory the operator is (see Eq. (3.29) in \cite{Frodden:2017qwh})
\be
I_{\delta e,\delta\omega}\equiv\delta e^a\frac{\partial\ \ }{\partial T^a}+\delta \omega^{ab} \frac{\partial\ \ }{\partial R^{ab}},
\label{EChomo}
\ee
because apart from $e^a$ and $\omega^{ab}$ there are no extra fields in the phase space, this is the full operator for this theory.
Then, the surface charge density, $k_\epsilon \equiv I_{\delta e,\delta\omega} S_\epsilon$, becomes
\be
\boxed{k_\epsilon=-\varepsilon_{abc}(\lambda^{ab}\delta e^c-\delta\omega^{ab} \xi\inter e^c)+2\beta\xi\inter e^a \delta e_a}
\label{withhomotopy}
\ee
By following the prescription that uses the symplectic structure density, $\Omega(\delta,\delta_\epsilon)=dk_\epsilon$, we arrive at exactly the same expression.

We remember that in general, boundary terms (exact forms) at the level of the Lagrangian do not contribute to the surface charges, or at most they contribute as exact forms in the $k_\epsilon$ formula and therefore can be neglected. However, notice that the term used here can not be written as an exact form, the tentative term one would try vanishes identically $d (e^ae_a)=T^ae_a-e^aT_a=0$. Then, at this level $\beta e_aT^a$ is a genuine term that produces torsion. In particular, it contributes to the surface charge density as is explicit in (\ref{withhomotopy}) with the last term, and again, this contribution is not an exact form at this level either.

Now, we make a step further to analyse the surface charge density  by performing a split of the connection in torsionless and contorsion parts. That is,
\be
\omega^{ab}=\tilde\omega^{ab}+\bar \omega^{ab},
\label{omegasplit}
\ee 
with $\tilde\omega^{ab}(e)$ solving $de^a+\tilde\omega\indices{^a_b}e^b=0$. Then, the torsion is simply $T^a=\bar\omega\indices{^a_b}e^b$.
From the equation of motion, $\varepsilon_{abc} T^a=\beta e_be_c$, we solve 
\be
\bar \omega^{ab}=\frac{\beta}{2}\varepsilon\indices{^a^b^c}e_c,
\label{omegabar1}
\ee 
where we used $\varepsilon\indices{_a_b_c}\varepsilon^{dbc}=-\varepsilon\indices{_a_b_c}\tilde\varepsilon^{dbc}=- 2\delta^{d}_a$\footnote{Remember $\varepsilon^{dbc}=\eta^{dd'}\eta^{bb'}\eta^{cc'}\varepsilon_{d'b'c'}=-\varepsilon_{dbc}=-\tilde \varepsilon^{dbc}$ we put a twiddle to the Levi-Civita symbol that do not carry information about the flat metric even if it has upstairs indices. Check this prescription in (\ref{leviflat}).}.  
The split of the connection induces also a split of the parameter, $\lambda^{ab}$, that solves the exact symmetry condition. In general the condition $\delta_\epsilon e^a=0$ is solved by
\ba
\lambda^{ab}&=&e^a\inter (d_{\omega}(\xi\inter e^b)+\xi\inter d_\omega e^b)\\
&=&e^a\inter (d(\xi\inter e^b)+\tilde\omega\indices{^b_c}\xi\inter e^c+\bar\omega\indices{^b_c}\xi\inter e^c+\xi\inter(\bar\omega\indices{^b_c}e^c))\\
&=&e^a\inter (d_{\tilde\omega}(\xi\inter e^b))-\xi\inter\bar\omega\indices{^a^b}\\
&=&\tilde\lambda^{ab}-\xi\inter\bar\omega\indices{^a^b},
\label{splitlambda}
\ea 
where we used $d_{\tilde\omega}e^b=0$, $e^a\inter e^c=\eta^{ac}$, and introduced $\tilde\lambda^{ab}\equiv e^a\inter (d_{\tilde\omega}(\xi\inter e^b))$, the torsionless part of the parameter. An equivalent way to define this parameter is to use the exact symmetry condition but improving the transformation only with the torsionless connection
\be
\delta_\epsilon e^a=\xi\inter (d_{\tilde\omega} e^a)+d_{\tilde\omega}(\xi\inter e^a)+\tilde\lambda\indices{^a_b} e^b=0,
\ee 
note that the first term vanishes by construction.
Collecting all, we can now use the split of the connection and the $\lambda^{ab}$ parameter on the surface charge density formula to show that
\ba
k_\epsilon &=&-\varepsilon_{abc}(\tilde\lambda^{ab}\delta e^c-\delta\tilde\omega^{ab} \xi\inter e^c)+\varepsilon_{abc}(\xi\inter \bar\omega^{ab}\delta e^c+\delta\bar\omega^{ab} \xi\inter e^c)+2\beta\xi\inter e_a\delta e^a\\
&=&\tilde{k}_\epsilon+\frac{\beta}{2}\varepsilon_{abc}\varepsilon\indices{^a^b^d}(\xi\inter e_d \delta e^c+\delta e_d\xi\inter e^c)+2\beta\xi\inter e_a\delta e^a\\
&=&\tilde{k}_\epsilon,
\ea
in the second line we defined $\tilde k_\epsilon\equiv -\varepsilon_{abc}(\tilde\lambda^{ab}\delta e^c-\delta\tilde\omega^{ab} \xi\inter e^c)$ and used the explicit expression for $\bar\omega^{ab}$ computed at (\ref{omegabar1}). To reach the third line remember that $\varepsilon\indices{^a^b^d}=-\tilde \varepsilon\indices{^a^b^d}$, as in (\ref{leviflat}), then $\varepsilon_{abc}\varepsilon\indices{^a^b^d}=-2\delta^d_c$. The remarkable fact is the third line. At the level of surface charge density, the torsion contribution to the connection cancels exactly with the extra source term $2\beta\xi\inter e_a\delta e^a$. In other words, the surface charge can be computed with the usual expression if one uses the Levi-Civita, or torsionless, connection $\tilde \omega^{ab}(e)$. 

The equation $k_\epsilon=\tilde k_\epsilon$ tells us that contorsion will never contribute to the charges in this theory. For this simple theory this result could have been expected as from the beginning we knew that the torsional term in the action is equivalent to a cosmological constant term. And we already know, at least in four dimensions but for any dimensions is the same, that a cosmological term do not enter in the surface charge formula\footnote{As an extra comment, we contrast our results with the charges computed in \cite{Garcia:2003nm}. To compare, all contributions coming from the action term, $\omega d\omega+\frac{2}{3}\omega^3$, in \cite{Garcia:2003nm}, shall be set to zero. Still, due to the $e_a T^a$ term in the action, it is found that the theory admits a so-called {\it BTZ solution with torsion}. The formulas for the mass and angular momentum presented in \cite{Garcia:2003nm} have a direct torsion contribution, and not only through the {\it effective cosmological constant} parameter, as can be appreciated in Eqs. (20) and (21) there. This is in tension with our results because, as we just checked, torsion disappears from our charge formulas. Another curiosity is that in \cite{Garcia:2003nm} the proposed quasilocal charge expressions depend on the $r$ coordinate. This dependence is avoided there, in their final formula, by taking the usual $r\to\infty$. From the surface charges density perspective we adopt here this can not happen simply because of the conservation law, $dk_\epsilon=0$, guarantee independence of the radius.
}.

%----------------------------------------------
\subsection{From a Chern-Simons perspective }
%----------------------------------------------
The previous result can be understood from a Chern-Simons (CS) perspective too. In fact, it is well-known that three-dimensional General Relativity with negative cosmological constant can be written as a topological Chern-Simons theory of a one-form gauge connection $\widetilde A = e^a \widetilde P_a + \omega^a \widetilde J_a$ valued on the anti-de Sitter algebra in three dimensions ($AdS_3$), $\mathfrak{so}(2,1)$. For the spin connection we use $\omega_a=\frac{1}{2}\varepsilon_{abc}\omega^{cb}$. The algebra reads
\be \label{AdS3algebra}
[ \widetilde P_a, \widetilde P_b ] =  \Lambda \varepsilon_{abc} \widetilde J^c , \quad  
[\widetilde J_a , \widetilde P_b ]   =  \varepsilon_{abc}  \widetilde P^c, \quad 
[  \widetilde J_a,  \widetilde J_b ]   =  \varepsilon_{abc}  \widetilde J^c ,
\ee
with $\Lambda$ the cosmological constant. The algebra \eqref{AdS3algebra} admits a non-degenerate and invariant bilinear form
\be
 \big< \widetilde J_a , \widetilde P_b \big> =   \eta_{ab}.
\ee
These two ingredients provide a CS construction for three-dimensional General Relativity as follows
\ba
\nonumber L_{\scriptscriptstyle{CS}} & = & \left< \widetilde  A \wedge d \widetilde A + \frac{1}{3}   \widetilde A \wedge  [\widetilde A, \widetilde A]      \right> \\
& = &  2e^a R_a (\omega) + \frac{\Lambda}{3} \varepsilon_{abc} e^a e^b e^c+d(e^a\omega_a),
\label{EC-ads}
\ea
with the curvature $R_a (\omega)=d\omega_a + \frac{1}{2}\varepsilon_{abc} \omega^b \omega^c$. Note that the equivalence is up to a boundary term. Now, the physics does not depend on the chosen algebra basis. Let us introduce a different basis for the algebra generators
\ba
 P_a  &= & \widetilde P_a  +(\beta/2 )\widetilde J_a   \\
 J_a &=&  \widetilde J_a ,
\ea
with $\beta$ a constant. Then, the $AdS_3$ algebra commutators \eqref{AdS3algebra} in this basis are 
\ba 
 \left[ P_a, P_b \right]  &= & (\Lambda - (\beta^2/4)) \varepsilon_{abc}J^c + \beta \varepsilon_{abc}P^c \n \\
 \left[ J_a , P_b \right]  &= & \varepsilon_{abc} P^c\n \\
   \left[  J_a, J_b \right]   & =&  \varepsilon_{abc} J^c . 
\label{newpoincarealgebra}
\ea 
The invariant and non-degenerate bilinear form associated to \eqref{newpoincarealgebra} is now
\be \label{newpairingAdS3}
\big< J_a , P_b \big> =  \eta_{ab}, \quad \big< P_a , P_b \big> = \beta \eta_{ab} .
\ee
Thus, the equivalent Chern-Simons Lagrangian for the one-form gauge connection $A= e^a P_a + \omega^a J_a $ valued on the algebra \eqref{AdS3algebra} and associated to the bilinear form \eqref{newpairingAdS3} is
\ba
L_{\scriptscriptstyle{CS}} & = &   2 e^a R_a (\omega) +\beta e^a T_a  + \frac{\Lambda_{\scriptscriptstyle{eff}}}{3}\varepsilon_{abc} e^a e^b e^c +d(e^a\omega_a)  ,
\ea 
with the torsion $T_a = de_a + \varepsilon_{abc} \omega^b e^c$, and the effective cosmological constant $\Lambda_{\scriptscriptstyle{eff}}= \Lambda - \frac{3}{4}\beta^2$. We can choose the parameter $\beta^2=\frac{4}{3}\Lambda$ and the last Lagrangian becomes exactly the Lagrangian \eqref{eT-term} considered previously. Therefore, we conclude that the torsional Lagrangian \eqref{eT-term} is just equivalent to the Einstein-Cartan Lagrangian \eqref{EC-ads} with a specific value for the cosmological constant $\Lambda=\frac{3}{4}\beta^2$.

%----------------------------------------------
\section{Einstein-Cartan-Dirac}
\label{appECD}
%----------------------------------------------
Consider the gravity contribution to surface charge density in four spacetime dimensions
\be
\mathring{k}_\epsilon=-\kappa'\varepsilon_{abcd}\left[\lambda^{ab}\delta(e^ce^d)-\delta\omega^{ab}\xi\inter(e^ce^d)\right].
\ee
As a preliminary we will rearrange this formula. First note that the spin connection can have a torsion part, named the contorsion, we want to isolate its contribution into the formula. Exactly as in the previous appendix, (\ref{splitlambda}), we perform a split of the spin connection, $\omega^{ab}=\tilde\omega^{ab}(e)+\bar\omega^{ab}$, such that $d_{\tilde \omega} e^a=0$. The exact symmetry condition, $\delta_\epsilon e^a=0$, is solved by the parameter 
\be
\lambda^{ab}=e^{[a}\inter(\xi\inter d_\omega e^{b]})+e^{[a}\inter(d_\omega\xi\inter e^{b]})=e^{[a}\inter (d_{\tilde\omega}\xi\inter e^{b]})-\xi\inter\bar\omega^{ab},
\ee
thus, the split of the connection is translated in a split of the parameter $\lambda^{ab}=\tilde\lambda^{ab}+\bar\lambda^{ab}$. Then, the gravity contribution to the surface charge density has two parts
\be
\mathring{k}_\epsilon=\tilde k_\epsilon+\bar k_\epsilon,
\label{splitgr}
\ee
the torsionless part
\be
\tilde k_\epsilon=-\kappa'\varepsilon_{abcd}\left[\tilde\lambda^{ab}\delta(e^ce^d)-\delta\tilde\omega^{ab}\xi\inter(e^ce^d)\right],
\label{aqui}
\ee
where $\tilde\lambda^{ab}=e^a\inter (d_{\tilde\omega}(\xi\inter e^b))$, and the contorsion part that we wanted to isolate
\be
\bar k_\epsilon=-\kappa'\varepsilon_{abcd}\left[\bar\lambda^{ab}\delta(e^ce^d)-\delta\bar\omega^{ab}\xi\inter(e^ce^d)\right],
\ee
where $\bar\lambda^{ab}=-\xi\inter \bar\omega^{ab}$. If we further express the contorsion one-form in frame components, $\bar\omega^{ab}=\bar\omega\indices{^a^b_f}e^f$, we can write
\be
\bar k_\epsilon=2\kappa'\varepsilon_{abcd}\left[\bar\omega\indices{^a^b_f}e^c(\xi\inter e^f\delta e^d+\xi\inter e^d\delta e^f)-\delta\bar\omega\indices{^a^b_f}e^f e^c\xi\inter e^d\right].
\label{scini}
\ee

Now, we compute the whole surface charge density. Consider the Einstein-Cartan-Dirac action %(see for instance 0908.2755 (or in 2+1 1002.0958))
\be
S[e^a,\omega^{ab},\psi]=\int_{\scriptscriptstyle \cal M}\varepsilon_{abcd} e^a e^b\left[\kappa' R^{cd}-\frac{i}{3} \alpha_\psi\, e^c \left(\bar\psi \gamma^d\gamma_5 d_\omega \psi+\overline{d_\omega\psi}\gamma^d\gamma_5\psi \right)\right],
\ee
with $d_\omega\psi=d\psi+\frac{1}{2}\omega_{ab}\gamma^{ab}\psi$ and $\gamma_{ab}\equiv \frac{1}{4}[\gamma_a,\gamma_b]$ satisfying the Lorentz algebra. The special matrix $\gamma_5\equiv \gamma_0\gamma_1\gamma_2\gamma_3$ satisfies $\gamma_5\gamma_a=-\gamma_a\gamma_5$. The following computation of surface charge is very sensitive to the coefficients, therefore we make a {\it scriptsize} detour to be self-contained and to check the consistence of our conventions.

\vspace{0.2cm}

{\scriptsize The $\gamma$-matrices satisfy the Clifford algebra $\{\gamma_a,\gamma_b\}= \gamma_a\gamma_b+\gamma_b\gamma_a=2\eta_{ab}$. Then, we have also $[\gamma_a,(\gamma_b\gamma_c-\gamma_c\gamma_b)]=4(\eta_{ab}\gamma_c-\eta_{ac}\gamma_b)$. If we define $\gamma_{ab}\equiv \frac{1}{4}[\gamma_a,\gamma_b]$ we can check that $[\gamma_a,\gamma_{bc}]=\eta_{ab}\gamma_c-\eta_{ac}\gamma_b$, and the matrices $\gamma_{ab}$ satisfy the Lorentz algebra, namely
\be
[\gamma_{ab},\gamma_{cd}]=\eta_{bd}\gamma_{ca}-\eta_{ad}\gamma_{cb}+\eta_{bc}\gamma_{ad}-\eta_{ac}\gamma_{bd},
\ee
this fix the $1/4$ normalization of the $\gamma_{ab}$ definition. The coefficient multiplying the connection in the covariant derivative acting on a spinor is fixed by the defining equation of the covariant derivatives on spinors 
\be
d_{\omega'}(\Lambda\psi)=\Lambda d_\omega\psi. 
\label{covonspinors}
\ee
The Lorentz transformation is $\Lambda=\exp\left(\frac{1}{2}\lambda^{ab}\gamma_{ab}\right)$, thus we define the algebra valued coefficient $\lambda=\frac{1}{2}\lambda^{ab}\gamma_{ab}$. We check that the $1/2$ is consistent with (\ref{covonspinors}). If we expand to first order in the Lorentz transformation both sides of (\ref{covonspinors}), we get
\be
d\lambda\psi+\lambda d\psi-\frac{1}{2}(d_\omega\lambda_{ab})\gamma^{ab}\psi+\frac{1}{2}\omega_{ab}\gamma^{ab}\lambda\psi=\lambda d\psi +\frac{1}{2}\lambda \omega_{ab}\gamma^{ab} \psi,
\label{key}
\ee
where we used $\omega'^{ab}=\omega^{ab}+\delta_\lambda\omega^{ab}$ with $\delta_\lambda\omega_{ab}=-d_\omega\lambda_{ab}=-d\lambda_{ab}+\lambda_{ac}\omega\indices{^c_b}-\lambda\indices{^c_b}\omega_{ac}$. The last expression brings into the formula the convention for the infinitesimal transformation $\lambda_{ab}$ used in the other variables ($\delta_\lambda e^a=\lambda\indices{^a_b}e^b$ or equivalently $\delta_\lambda\omega_{ab}=-d_\omega\lambda_{ab}$). Then we check that (\ref{key}) is in fact and identity because our conventions are correct such that $d\lambda=\frac{1}{2}d\lambda_{ab}\gamma^{ab}$, and
\ba
\omega_{ab}\left[\lambda,\gamma^{ab}\right]&=&\frac{1}{2}\omega_{ab}\lambda_{cd}\left[\gamma^{cd},\gamma^{ab}\right]=-\frac{1}{2}\omega_{ab}\lambda_{cd}\left[\gamma^{ab},\gamma^{cd}\right]\\
&=&\left(\lambda_{ac}\omega\indices{^c_b}-\lambda\indices{^c_b}\omega_{ac}\right)\gamma^{ab}.
\ea
Those checks set correctly the three coefficients in $\gamma_{ab}=\frac{1}{4}[\gamma_a,\gamma_b]$, $\lambda=\frac{1}{2}\lambda_{ab}\gamma^{ab}$, and $d_\omega\psi=d\psi+\frac{1}{2}\omega_{ab}\gamma^{ab}\psi$.}

\vspace{0.2cm}

Now, besides the usual exact symmetry conditions on the gravity fields, (\ref{symconde}) and (\ref{symcondomega}), we should impose the exact symmetry  condition on the spinor field. Spinor field transform under an infinitesimal local Lorentz transformation as $\delta_{\lambda'}\psi=\lambda'\psi$.  Therefore, the correct exact symmetry condition is
\be
\delta_\epsilon \psi=\Lie_\xi\psi+\lambda' \psi=\xi\inter d_\omega \psi+\lambda \psi=0,
\label{spinorLT}
\ee
with, again, the improved prescription given by $\lambda=\frac{1}{2}\gamma_{ab}(\lambda'^{ab}-\xi\inter\omega^{ab})$ (remember $\lambda_{ab}=\lambda'_{ab}-\xi\inter\omega_{ab}$).

A general formula for the surface charge density in differential form language is (see Eq. (2.19) in \cite{Frodden:2017qwh})
\be
k_\epsilon=\delta \widetilde Q_\epsilon-\xi\inter\Theta(\delta)-B_{\delta\epsilon}.
\label{gralk}
\ee
The variation of the Lagrangian is
\be
\delta L=E_a\delta e^a+E_{ab}\omega^{ab}+E_{\psi}\delta \psi++E_{\bar\psi}\delta \bar\psi+d\Theta(\delta),
\ee
with the boundary term 
\ba
\Theta(\delta)=\varepsilon_{abcd}e^ae^b\left(\kappa'\delta\omega^{cd}+\frac{i}{3}\alpha_\psi\, e^c \delta \left(\bar\psi \gamma^d\gamma_5 \psi\right)\right).
\label{boundaryc}
\ea
This is the middle term we need in (\ref{gralk}). For the first term we compute the {\it trivial current} $J_\epsilon=\Theta(\delta_\epsilon)-\xi\inter L+S_\epsilon=d\widetilde Q_\epsilon$, and after cancellations we get the usual $\widetilde Q_\epsilon=-\kappa'\varepsilon_{abcd}e^ae^b \lambda^{cd}$ we find for pure Einstein-Cartan theory. For the third term in (\ref{gralk}), the prescription tells us that the symplectic potential term $\Theta([\delta,\delta_\epsilon])=dB_{\delta\epsilon}+C_{\delta\epsilon}$ with $C_{\delta\epsilon}\approx 0$, thus we use the commutation of variations $[\delta,\delta_{\epsilon}]=[\delta,\Lie_\xi +\delta_{\lambda+\xi\inter\omega}]=\delta_{\delta\lambda+\xi\inter\delta\omega}$, and we get $B_{\delta\epsilon}=-\kappa'\varepsilon_{abcd}e^ae^b(\delta\lambda^{cd}+\xi\inter \delta\omega^{cd})$. Thus the spinor field does not contribute through $B_{\delta\epsilon}$ nor $\widetilde Q_\epsilon$ in the general formula (\ref{gralk}), it only enters through the extra boundary term in (\ref{boundaryc}). The complete surface charge density for the Einstein-Cartan-Dirac theory is
\be
k_\epsilon=-\varepsilon_{abcd}\left(\kappa'\left(\lambda^{ab}\delta(e^c e^d)-\delta\omega^{ab}\xi\inter(e^c  e^d)\right)+i\alpha_\psi\, \xi\inter e^ae^be^c \delta\left(\bar \psi\gamma^d\gamma_5 \psi\right)\right).
\label{scECD}
\ee
We stress that the addition of a spinorial mass term in the action does not change the surface charge formula. Then, this result is already useful enough to compute charges for this theory, its massive spinor equivalent, or even with an additional cosmological constant term.  But we can go further. Let us split the gravity terms as we did at the beginning, (\ref{splitgr}), then  
\be
k_\epsilon=\tilde k_\epsilon+\bar k_\epsilon+k^{\psi}_\epsilon
\ee
with 
\be
k^{\psi}_\epsilon=-i \alpha_\psi\, \varepsilon_{abcd}\xi\inter e^ae^be^c \delta(\bar \psi\gamma^d\gamma_5 \psi).
\label{scpsi}
\ee
We can compute $\bar k_\epsilon$ explicitly, as given by (\ref{scini}), by solving the contorsion $\bar \omega^{ab}$ from the torsion equation of motion. We do it step by step. The equation we need is
\be
\varepsilon_{abcd}T^ce^d=\frac{i}{12}\alpha_\psi\, \varepsilon_{cdmn}e^ce^de^n\bar\psi \left(\delta^{m}_{a}\gamma_{b}-\delta^{m}_{b}\gamma_{a}\right)\gamma_5 \psi,
\ee 
with $T^c=d_\omega e^c=\bar\omega\indices{^c_f}e^f=\bar\omega\indices{^a_f_g}e^ge^f$, we have
\be
\varepsilon_{abcd}\bar\omega\indices{^a_f_g}e^ge^fe^d=\frac{i}{12}\alpha_\psi\,\varepsilon_{cdmn}e^ce^de^n\bar\psi \left(\delta^{m}_{a}\gamma_{b}-\delta^{m}_{b}\gamma_{a}\right)\gamma_5 \psi,
\ee 
or its dual equation
\be
\varepsilon_{abcd}\varepsilon^{gfdh}\bar\omega\indices{^c_f_g}=\frac{i}{12}\alpha_\psi\,\varepsilon_{cdmn}\varepsilon^{cdnh}\bar\psi \left(\delta^{m}_{a}\gamma_{b}-\delta^{m}_{b}\gamma_{a}\right)\gamma_5 \psi.
\ee
Note that in the l.h.s. $\varepsilon_{abcd}\varepsilon^{gfdh}=-(-3!\delta^{g}_{[a}\delta^{f}_{b}\delta^{h}_{c]})=3!\delta^{g}_{[a}\delta^{f}_{b}\delta^{h}_{c]}$, where we have to be careful with the extra minus sign because we raise indices with the flat metric $\eta^{ab}$. We also use in the r.h.s. $\varepsilon_{cdmn}\varepsilon^{cdnh}=-\varepsilon_{cdnm}\varepsilon^{cdnh}=-(-6\delta^h_m)=6\delta^h_m$. Therefore
\be
2\bar\omega\indices{^h_b_a}+\delta^h_a\bar\omega\indices{^c_c_b}-\delta^h_b\bar\omega\indices{^c_c_a}=\frac{i}{2}\alpha_\psi\, \bar\psi \left(\delta^{m}_{a}\gamma_{b}-\delta^{m}_{b}\gamma_{a}\right)\gamma_5 \psi.
\ee
To completely solve this equation we have to contract it and then replace the result in itself. Contracting $h=b$ we get $\bar\omega\indices{^c_c_a}=\frac{3i}{2}\alpha_\psi\, \bar\psi \gamma_{a}\gamma_5 \psi$. Then
\be
\bar\omega\indices{^h_b_a}=\frac{i}{2}\alpha_\psi\, \bar\psi \left(\delta^{h}_{b}\gamma_{a}-\delta^{h}_{a}\gamma_{b}\right)\gamma_5 \psi,
\ee
or equivalently
\be
\bar\omega\indices{^a^b_f}=\frac{i}{2}\alpha_\psi\, \bar\psi \left(\eta^{ab}\gamma_{f}-\delta^{a}_{f}\gamma^{b}\right)\gamma_5 \psi.
\ee
We are ready to replace this into the expression for $\bar k_\epsilon$, (\ref{scini}).  We do it by parts. First note that the following combination simply vanishes
\ba
\varepsilon_{abcd}\bar\omega\indices{^a^b_f}e^c(\xi\inter e^f\delta e^d+\xi\inter e^d\delta e^f)=\frac{i}{2}\alpha_\psi\, \varepsilon_{abcd}\bar\psi \left(\eta^{ab}\gamma_{f}-\delta^{a}_{f}\gamma^{b}\right)\gamma_5 \psi e^c(\xi\inter e^f\delta e^d+\xi\inter e^d\delta e^f)=0.\n
\ea
Then
\ba
\bar k_\epsilon=-2\varepsilon_{abcd}\delta\bar\omega\indices{^a^b_f}e^f e^c\xi\inter e^d=i \alpha_\psi\, \varepsilon_{abcd}e^ae^b\xi\inter e^c\delta(\bar\psi\gamma^d\gamma_5\psi),
\ea
this term is exactly $-k^{\psi}_\epsilon$ as in (\ref{scpsi}).  Therefore, as it might have been suspected, for the surface charge density, we have an exact cancellation of all the terms concerning the spinor field
\be
\bar k_\epsilon+k^{\psi}_\epsilon=0,
\ee 
or equivalently, this means that the full surface charge density for the Einstein-Cartan-Dirac theory is simply $k_\epsilon=\tilde k_\epsilon$ as in equation (\ref{aqui}). In particular, this implies that in a spacetime with a spinor field living on it, as far as exact symmetries are satisfied, it is not needed to have the explicit solution for the spinor field to compute charges.

\section{$D$-dimensional Chern-Simons form}
\label{DCSapp}
%----------------------------------------------

A Chern-Simons (CS) Lagrangian in $D=2n+1$ dimensions is a local function of a one-form gauge connection, $A$, valued on a Lie algebra. That is $A=A^i\tau_i=A^i_a\tau_i\, e^a$ with $e^a$ the one-form frame field, $\tau_i$ the generators of the algebra, $[\tau_i,\tau_j]=f\indices{_i_j^k}\tau_k$, and $f\indices{_i_j^k}$ the algebra structure constants. The full CS Lagrangian can be expressed in a very compact form using the trick of an integral over an auxiliary variable $t$ \cite{Azcarraga:book} 
\begin{equation}\label{CSform}
L^{(2n+1)} [A]   =  \kappa_n \int_{0}^{1} dt  \langle A F_{t}^{n} \rangle,
\end{equation}
where $F_{t} \equiv dA_t +  A_{t}\wedge A_t $,  $A_t =t A$, and $\kappa_n=\kappa_{\scriptscriptstyle  CS}(n+1)$ with $\kappa_{\scriptscriptstyle  CS}$ the CS level. Notice that the one-form nature of $A$ induces the algebra commutator on the $A_{t}^2$ term, explicitly $A_{t}^2=(t A)\wedge(t A)=t^2 A\wedge A=\frac{1}{2}t^2\, A^i\wedge A^j [\tau_i,\tau_j]$. The angled bracket $\big<\cdot \big >$ denotes the symmetric invariant polynomial on the algebra such that for any two algebra valued forms, says the $p$-form $P$ and the $q$-form $Q$, the usual commutation properties are respected, namely
\be
\big<\cdots P Q\cdots\big>=(-1)^{pq}\big<\cdots QP\cdots\big>.
\ee

The CS theory possesses two symmetries, in fact the Lagrangian is invariant under diffeomorphisms and also {\it quasi-invariant} (up to a boundary term/exact form) under gauge symmetries, these are the key gauge symmetries that push us to compute surface charges. The trick to have a compact expression for the Lagrangian allows us to perform all calculations directly, we show them in detail. Let us start with a general variation of \eqref{CSform}, we have
\begin{eqnarray}
\delta L^{(2n+1)} [A] % &  = & k(n+1) \int_{0}^{1} dt \langle \delta (AF_{t}^{n})   \rangle   \\
& = &\kappa_n \int_{0}^{1} dt \langle \delta A F^{n}_{t}  + n A \delta F_{t} F_{t}^{n-1}  \rangle \n \\
& = &  \kappa_n \int_{0}^{1} dt \langle \delta A F^{n}_{t}  + n A\, d_{\scriptscriptstyle A_t} (\delta A_t ) F_{t}^{n-1}  \rangle  \n \\
&  =&  \kappa_n \int_{0}^{1} dt \langle \delta A F^{n}_{t}   +\frac{d}{dt} \delta A_t F^{n}_{t} -\delta A F^{n}_{t} -nd(A \delta A_t F^{n-1}_{t})  \rangle\n  \\
& = & \kappa_n\langle \delta A F^n \rangle - d\Theta (\delta A) ,
\end{eqnarray}
where we used, $\delta F_t=d\delta A_t+[A_t,\delta A_t]=d_{\scriptscriptstyle A_t}\delta A_t$ with the notation $d_{\scriptscriptstyle A_t}$ for the exterior covariant derivative for the connection $A_t$, we used  also the Leibniz's rule for $d_{\scriptscriptstyle A_t}$, that $d_{A_t} F_t=0$, the identity $\frac{d\ }{dt} F_t = d_{\scriptscriptstyle A_t}  A$, integration by parts in the variable $t$, that $\frac{d\ }{dt}\delta A_t=\delta A$, and the invariance property of the symmetric polynomial $\big< \cdot \big>$. Then, the equations of motion and the boundary term are
\ba
\label{eomCS} \langle F^n \rangle & = & 0, \\
\Theta (\delta A )  & = &  -n\kappa_n \int_{0}^{1}dt  \langle  \delta A_t A  F_{t}^{n-1} \rangle .
\ea
Notice that we defined the boundary term with an overall minus sign, this convention save us of carrying a minus sing in the following calculations. This is conventional, remember that surface charge densities are defined up to overall factors.

Later we will also need the linearized equation of motion, namely
\be
\delta \langle F^n \rangle  = n \langle  (\delta F) F^{n-1} \rangle  = n \langle  d_{\scriptscriptstyle A}(\delta A) F^{n-1} \rangle =0,
\label{linearizedeomCS}
\ee
where $d_{\scriptscriptstyle A}(\cdot)\equiv d(\cdot)+[A,\cdot]$ denotes the covariant exterior derivative for the connection $A$. 

To obtain the surface charge density we first compute the symplectic structure density using the boundary term. With two independent general variations on the phase space, say $\delta_1$ and $\delta_2$, the symplectic structure density reads
\ba 
\nonumber \Omega (\delta_1 , \delta_2) &  =& \delta_1 \Theta (\delta_2A) - \delta_2 \Theta(\delta_1A) - \Theta ([\delta_1 , \delta_2]A)  \\
%& = & n(n+1)k\int_{0}^{1} dt \big<    \delta_2 A_t \delta_1 (AF_{t}^{n-1}) - \delta_1 A_t \delta_2 (A F_{t}^{n-1})   \big> \\
%& = & n(n+1)k\int_{0}^{1} dt \big<    \delta_2 A_t \delta_1 A F_{t}^{n-1} + \delta_2 A_t A \delta_1 F_{t}^{n-1} - \delta_1 A_t \delta_2 A F_{t}^{n-1} - \delta_1 A_t A \delta_2 F_{t}^{n-1}  \big> , \\
 & = &  -n\kappa_n \int_{0}^{1}dt \big<    2\delta_2 A_t \delta_1 A F_{t}^{n-1} + \delta_2 A_t A \delta_1 F_{t}^{n-1}  - \delta_1 A_t A \delta_2 F_{t}^{n-1}  \big> .
\label{CSsymplectic2}
\ea
Now, the key to get a more tractable expression is to rewrite the second term as
\begin{eqnarray}
\nonumber 
&& \hspace{-0.8cm}\big< \delta_2 A_t  A \delta_1 F_{t}^{n-1}\big>
\nonumber  \\
& =& (n-1) \big< \delta_2 A_t A\, d_{\scriptscriptstyle A_t} (\delta_1  A_t) F_{t}^{n-2}\big> \n\\
\nonumber & = &   (n-1)\big < d\left(  \delta_2 A_t A \delta_1 A_t F_{t}^{n-2}   \right) -   d_{A_t} (\delta_2 A_t) A \delta_1 A_t F_{t}^{n-2}  +\delta_2 A_t\, d_{\scriptscriptstyle A_t} A \delta_1 A_t F^{n-2}_{t}  \big>  \\
\nonumber & =  &  (n-1)\big< d ( \delta_2 A_t A \delta_1 A_t F_{t}^{n-2}) -   \delta_2 F_t A \delta_1 A_t F_{t}^{n-2} +\delta_2 A_t \frac{d}{dt}(F_t) \delta_1 A_t F^{n-2}_{t} \big>\\
\nonumber & = &  (n-1) \big< d ( \delta_2 A_t A \delta_1 A_t F_{t}^{n-2}) -   \delta_2 F_t A \delta_1 A_t F_{t}^{n-2} \big> +\frac{d}{dt} \big<\delta_2 A_t \delta_1 A_t F^{n-1}_{t}\big>- 2\big<\delta_2 A_t \delta_1 A F^{n-1}_{t}\big>\\
\label{key22}
\end{eqnarray}
where we used in the second line the Leibniz's rule for the covariant derivative $d_{\scriptscriptstyle A_t}$ and the identity $d_{\scriptscriptstyle A_t} F_{t}^{n-2}=0$. In the third line, $d_{\scriptscriptstyle A_t} (\delta_2 A_t)=\delta_2 F_t$ and $\frac{d\ }{dt}F_t=d_{\scriptscriptstyle A_t} A$. In the fourth line, we introduce a total derivative in $t$, we use that all the expression is inside the bracket $\langle\cdot \rangle$ to perform commutations of the algebra valued forms, and used also that $\frac{d\ }{dt}\delta A_t=\delta A$. 

Now, replacing back, the second and fourth terms of (\ref{key22}) cancel exactly the first and third terms of (\ref{CSsymplectic2}), respectively. We are left with a total derivative in $t$ which we can integrate trivially, and also an exact form. Then, the result is a symplectic structure density composed by a piece that could have been expected plus another piece which is an exact form
\ba
\Omega^{(2n+1)} (\delta_1 , \delta_2) & =&n\kappa_n\big<\delta_1 A \delta_2 A F^{n-1}\big>-n(n+1)\kappa_n\, d\left(\int_{0}^{1}dt \big< \delta_2 A_t A \delta_1 A_t F_{t}^{n-2} \big>\right).
\label{CSsymplectic3}
\ea
We observe that unlike other theories, the symplectic structure density for CS is not gauge invariant due to the last term, this is expected because the Lagrangian as well as the boundary term $\Theta(\delta A)$ are not gauge invariant forms. Remember that the theory is just quasi-invariant.

Now, we combine infinitesimal diffeomorphisms and gauge transformations for the connection to write an improved general infinitesimal symmetry transformation as
\be
\delta_\epsilon A =\Lie_\xi A-d_A \lambda'= \xi \inter F -d_A \lambda , 
\label{connectiont}
\ee
where as usual we select the parameter as $\lambda' = \lambda +\xi \inter A$ in order to define an overall homogeneous infinitesimal transformation. Remember that $\lambda=\lambda^i\tau_i=(\lambda'^i-\xi^\mu A_\mu^i)\tau_i$ is an algebra valued gauge parameter which is also field dependent. 

Then, we evaluate the symplectic structure density, (\ref{CSsymplectic3}), such that one of its entries is an improved symmetry transformation, $\delta_2A\to \delta_\epsilon A$ as in (\ref{connectiont}) (and $\delta_1 A\to \delta A$). Using the equation of motion (\ref{eomCS}), and also the linearized equation of motion (\ref{linearizedeomCS}) (varied e.o.m. on phase space), it is straightforward to show that the first term becomes also an exact form
\ba
\Omega^{(2n+1)} (\delta, \delta_\epsilon) & =&n\kappa_n d\big<\lambda \delta A F^{n-1}\big>-n(n+1)\kappa_n d\left(\int_{0}^{1}dt \big< \delta_\epsilon A A_t \delta A_t F_{t}^{n-2} \big>\right).
\label{sympl444}
\ea
When the exact symmetry condition is satisfied: $\delta_\epsilon A=0$ the symplectic structure density simply vanishes and it also vanishes the integral second term of the last expression. Therefore, we conclude that the surface charge density for a $D$-dimensional Chern-Simons theory, that satisfies the conservation law (\emph{i.e.} is closed $dk_\epsilon=0$), is 
\begin{equation}
k^{(2n+1)}_\epsilon = n\kappa_n \big< \lambda \delta A F^{n-1} \big> . 
\label{CSSCN}
\end{equation}
This is the main result of this appendix and it could have been expected by symmetry considerations. In fact, with only a connection at disposal there are no other ways to write a $(D-2)$-form which is also a variation (or one-form in field space) and at the same time a gauge invariant expression.

 On ther other hand, in contrast to \cite{Izaurieta:2005vp} and \cite{Mora:2006ka} we observe the advantage to group the improved gauge parameter $\lambda$ and the Killing vector $\xi$ in the symmetry parameter $\epsilon$ which allows us to define an unique charge.

Note that for $n=1$ we recover the standard $D=3$ dimensional surface charge for a CS theory computed in the main text.  

We  remark that in the last step we needed to invoke the exact symmetry condition to get rid of the integral term in the symplectic structure density, (\ref{sympl444}). This is not usually the case. For all other theories worked out through these notes the surface charge density is read directly once we replace the symmetry transformation as one of the entries of the symplectic structure. Instead, here there is this extra exact form, expressed as an integral in $t$. As stressed before this is related with the quasi-invariance of the CS theory and that our prescription to define the symplectic structure relies on the Lagrangian\footnote{In the method based on the contracting homotopy operator this is not the case and the symplectic structure is defined directly form the equations of motion. Because of this reason this alternative prescription is sometimes called {\it invariant} symplectic structure \cite{Barnich:2007bf}.}. Having said that, at this stage it should be already clear through our discussions that it is only for those cases, when the exact symmetry condition holds, that the surface charge density is closed and therefore it becomes a meaningful formula to compute true charges. 

%----------------------------------------------
\subsection{$D$-CS surface charge from the contracting homotopy operator}
\label{DCShomotopy}
%----------------------------------------------

As a final remark of this appendix we note that the surface charge density formula for CS in $D=2n+1$ could had been easily obtained using the corresponding contracting homotopy operator. For a CS theory we can sketch the operator as
\be
I_{\scriptscriptstyle \delta A}\equiv\delta A\frac{\partial\ \ }{\partial F}.
\label{CShomo}
\ee   
Now the Noether identity implies that $-\langle \kappa_n\delta_\epsilon A F^n \rangle=dS_\epsilon+\cancelto{0}{N_\epsilon}$ with $S_\epsilon=\kappa_n \langle \lambda F^n\rangle$. Thus, for the surface charge density
\be
k^{(2n+1)}_\epsilon\equiv I_{\scriptscriptstyle \delta A} S_\epsilon=n\kappa_n \big< \lambda \delta A F^{n-1} \big>.
\ee   
and we directly recover \eqref{CSSCN}. This short calculation shows the power of the contracting homotopy operator approach. Of course the procedure to obtain (\ref{CShomo}) as a rigorous expression for the operator is the missing part here but, as we checked, its naive application it is certainly powerful enough.  

%----------------------------------------------
\section{BF Theory}
\label{BFappendix}
%----------------------------------------------

\subsection{From Einstein-Cartan-$\Lambda$ to Chiral BF}

In the following we show how the Einstein-Cartan-$\Lambda$ action in four dimensions can be written as a chiral theory (for instructive talks see \cite{Krasnov:2019talk}). Afterwards, we go further and rewrite it as a {\it chiral BF theory}. 

The first observation is that the Einstein-Cartan-$\Lambda$ Lagrangian in \eqref{EC4D} can be supplemented with two particular extra terms of the form $e_ae_b R^{ab}$ and $e_ae_be^ae^b$ that do not modify the equations of motion. The later term is trivially zero, but the former term is part of the Nieh-Yan topological density \cite{Nieh:1981ww}: $d(e_aT^a)=T^aT_a-e_ae_bR^{ab}$, therefore $e_ae_b R^{ab}$ is not a boundary term but can be traded, up to a boundary term, by $T^aT_a$. Still, if we add this term to the action, at the level of the equations of motion it can be checked that they are exactly equivalent to the Einstein-Cartan-$\Lambda$ equations of motion. This observation is also the basis of the so-called Holst action \cite{Holst:1995pc}.  

Consider the modified Einstein-Cartan-$\Lambda$ action 
\ba
S[e^a,\omega^{ab}]&=&\kappa'\int_{\scriptscriptstyle \cal M}\varepsilon_{abcd}e^ae^b\left(R^{cd}\pm\frac{1}{2\ell^2} e^ce^d\right)+2i\eta_{a[c}\eta_{b|d]}e^ae^b\left(R^{cd}\pm\frac{1}{2\ell^2}e^ce^d\right)\\
&=&4i\kappa'\int_{\scriptscriptstyle \cal M}e_ae_b P\indices{_+^a^b_c_d}\bar R^{cd}\\
&=&4i\kappa'\int_{\scriptscriptstyle \cal M}\left(e_ae_b R^{ab}(\omega^+)\pm\frac{1}{12i\ell^2}\varepsilon_{abcd}e^ae^be^ce^d\right),
\ea
in the first line we added both mentioned terms with a specific imaginary coefficient to the action. The compact notation in the second line uses the (anti)-de Sitter curvature $\bar R^{ab}=R^{ab}\pm\frac{1}{2\ell^2}e^ae^b$ and the definition of the projectors $P\indices{_\pm^a^b_c_d}=\frac{1}{2}\left(\delta^a_{[c}\delta^b_{d]}\pm\frac{1}{2i}\varepsilon\indices{^a^b_c_d}\right)$. In the third line we use the projector to define $\omega^{ab}_\pm=P\indices{_\pm^a^b_c_d}\omega^{cd}$ and a short calculation shows that the curvature splits $R(\omega)=R(\omega_++\omega_-)=R(\omega_+)+R(\omega_-)$ such that $P_+ R(\omega)=R(\omega_+)$. Thus, we have written an equivalent action that depends only on half of the connection components: The $\omega_-^{ab}$ are absent. This is a chiral Einstein-Cartan action that in fact still encodes the full GR dynamics.

With the action in this form we select an internal time direction and make an explicit split of the internal Lorentz group to define the Plebanski variables \cite{Plebanski:77} as
\ba
\Sigma^l(e)&=&\frac{1}{2i}P\indices{_+^0^l_c_d}e^ce^d=ie^0\wedge e^l-\frac{1}{2}\varepsilon^{ljk}e^j\wedge e^k \\
A^l(\omega)&=&\frac{1}{2i}P\indices{_+^0^l_c_d}\omega^{cd}=i\omega^{0l}-\frac{1}{2}\varepsilon^{ljk}\omega^{jk}, 
\ea
where the summation rule on the internal indices $i,j,k=1,2,3$ is assumed, and because the Euclidean metric for the space indices $\eta_{ij}=\delta_{ij}$ no distinction between upstairs and downstairs indices is needed in the following $e^i=e_i$. With these definitions at hand the chiral action is rewritten as
\be
S[e^a,\omega^{ab}]=4i\kappa'\int_{\scriptscriptstyle \cal M}\left(\Sigma^i F^i(A)\pm \frac{1}{2\ell^2}\Sigma^i\Sigma^i\right),
\ee
where the field strength is $F^i=dA^i+\frac{1}{2}\varepsilon^{ijk}A^j\wedge A^k$ (remember $\varepsilon_{0ijk}=\varepsilon_{ijk}=\varepsilon^{ijk}$). 

To get a BF formulation, here we follow \cite{Freidel:2012np}, we note that $\Sigma^i$ is a basis for self-dual two-forms in four dimensions. Therefore, if we have a collection of three arbitrary two-form self-dual fields named $B^i$ we have
\be
B^i=\sigma b\indices{^i_j}\Sigma^j,
\ee 
where the $b\indices{^i_j}$ are the components on the basis, and $\sigma=\pm$ is a sign. As explained in \cite{Freidel:2012np} because of the conformal invariance of the self-dual property we can always choose the $3\times 3$ matrix as unimodular: $\det b\indices{^i_j}=1$. In order to write the action in terms of $B^i$ fields as true variables and still have a GR theory we have to impose a constraint on the $B$ fields to fix $b\indices{^i_j}$, such that the $b\indices{^i_j}$ matrices are just the identity or at most a 3D rotation, that is $b\indices{^i_k}b\indices{^j_l}\delta^{kl}=\delta^{ij}$. 
Now, note that the Plebanski variables satisfy
\be
\Sigma^i\wedge \Sigma^j\sim \delta^{ij},
\label{sigmasigma}
\ee
inspired by previous equation, a constraint that does the job is $B^i\wedge B^j\sim\delta^{ij}$, or more explicitly
\be
B^i\wedge B^j=\frac{1}{3}\delta^{ij}B_k\wedge B^k.
\ee  
To implement the constraint into the action we introduce a term with a Lagrange multiplier $\chi^{ij}B^i\wedge B^j$ that has a traceless condition $\chi^{ij}\delta_{ij}=0$. Then, we have the following chiral BF action
\be
S[B^i,A^i,\chi^{ij}]=\int_{\scriptscriptstyle \cal M}\left(B^i F^i\pm\frac{1}{2\ell^2}B^iB^i+\frac{1}{2}\chi^{ij}B^iB^j\right).
\label{BFaction0}
\ee
The overall factor $4i\kappa'$ is dropped because it does not affect the equations of motion and here we will not consider the coupling with non-gravitational fields. The constraint ensures us that the $B^i$ has a Plebanski variable form in terms of the vielbein and therefore a GR metric is implicit. However, it is interesting to note that the metric components can be obtained at once through the Urbantke's formula \cite{Urbantke:1984eb}
\be
\sqrt{g}g_{\mu\nu}=\frac{1}{12}\varepsilon_{ijk}\varepsilon^{\alpha\beta\gamma\delta}B^i_{\mu\alpha}B^j_{\beta\gamma}B^k_{\delta\nu},
\ee
this is the equivalent of the equation $g_{\mu\nu}=e^a_\mu e^b_\nu \eta_{ab}$ for the $B^i$ field. 

Here we have explained the equivalence of Einstein-Cartan-$\Lambda$ and chiral BF theories at the level of the action. It is also instructive to recover the Einstein field equations in its metric form directly from the chiral BF equations of motion, for the interested reader we suggest \cite{Freidel:2012np} and \cite{Krasnov:2009pu}.

%---------------------------------------------- 
\subsection{Surface Charges for a Chiral BF Theory}
%----------------------------------------------
Consider the chiral BF action \eqref{BFaction0}. It should be noted that the connection $A^i$ defines, as usual, a covariant exterior derivative $d_A(\cdot)=d(\cdot)+[A,(\cdot)]$. The variation of the Lagrangian is straightforward
\be
\delta L=\delta B^i E^i_{\scriptscriptstyle (B)}+\delta A^i E^i_{\scriptscriptstyle (A)}+\delta \chi^{ij}E^{ij}_{\scriptscriptstyle (\chi)}+d\Theta(\delta A),
\ee
with the three equations of motion and the boundary term given by
\ba
E^i_{\scriptscriptstyle (B)}&=&F^i\pm\frac{1}{\ell^2}B^i+\chi^{ij}B^j=0\\
E^i_{\scriptscriptstyle (A)}&=& d_AB^i=0\\
E^{ij}_{\scriptscriptstyle (\chi)}&=&0\to B^i\wedge B^j=\frac{1}{3}\delta^{ij}B_k\wedge B^k\\
\Theta(\delta A)&=& B^i\delta A^i.
\ea
The boundary term is used to compute the symplectic structure density
\be
\Omega(\delta_1,\delta_2)=\delta_1B^i\wedge \delta_2 A^i-\delta_2 B^i\wedge \delta_1A^i.
\ee
The improved infinitesimal gauge plus diffeomorphism transformations are
\ba
\delta_\epsilon B^i &=&\xi\inter d_A B^i+d_A\xi\inter B^i+[\lambda,B]^i
\label{exactB}\\
\delta_\epsilon A^i &=&\xi\inter F^i-d_A\lambda^i,
\label{exactA}
\ea
then, as usual, we evaluate the symplectic structure density on those transformations 
\ba
\Omega(\delta,\delta_\epsilon)&=& \delta B^i\left(\xi\inter F^i-d_A\lambda^i\right)-\left(d_A\xi\inter B^i+\xi\inter d_AB^i+[\lambda, B]^i\right)\delta A^i\\
&=&- d\left(\delta B^i \lambda^i+\xi\inter B^i\delta A^i\right)+\delta B^i\xi\inter F^i+d_A\delta B^i \lambda^i-\xi\inter d_A B^i\delta A^i\\
&&-\xi\inter B^i d_A \delta A^i-[\lambda,B]^i \delta A^i\\
&=&- d\left(\delta B^i \lambda^i+\xi\inter B^i\delta A^i\right),
\ea
where to get the second line we use the Leibniz's rule for the exterior covariant derivative to express the second and third terms of the first line as exact forms (plus the rest). Then, we use the equations of motion and the linearized equations of motions. With this, all the five non-exact forms on the third line cancel. In particular note $\delta F^{i}=d_{\scriptscriptstyle A}\delta A^i$ and the linearized equation of motion: $\delta F^i\pm\frac{1}{\ell^2}\delta B^i-\chi^{ij}\delta B^j=0$.
Hence, the surface charge density formula for the chiral BF theory is simply 
\be
k_\epsilon=-\delta B^i\lambda^i+\delta A^i\xi\inter B^i.
\ee
For the surface charge, built as a closed integral of the previous formula, to be conserved we require, as usual, the exact symmetry condition to hold: $\delta_\epsilon B^i=0$ and $\delta_\epsilon A^i=0$ as in \eqref{exactB} and \eqref{exactA}, respectively. We stress that still from the condition for $B^i$ it is possible to directly solve the parameter $\lambda^i$, we leave it as an exercise.

%----------------------------------------------
\subsection{Jackiw-Teitelboim model as a BF theory}
\label{appendixJB}
%----------------------------------------------
Here we show how pure gravity in $1+1$ dimensions admits a BF-like formulation \cite{Chamseddine:91} and then we compute the corresponding surface charge density.

Analogous to the BF formulation of General Relativity in four dimensions, it is possible to have a similar construction for a gravity theory on a line, \emph{i.e.}, {\it lineal gravity}. Recently, this topic has active interest due to its use as a $1+1$ model in the context of the $AdS_2/CFT_1$ conjecture \cite{Polchinski,Astorino:2002bj} and also as a low-energy limits model of near-extremal black holes \cite{sarosi-review}; in both cases this is a toy model with the attractiveness of being simple enough such that many computation are doable even in a quantum regime. 

The Einstein-Hilbert action in $1+1$ is an uninteresting theory of gravity because the action is just a boundary term\footnote{The Einstein-Hilbert action $\int d^2x \sqrt{-g}R$ is a surface term, it is in fact the topological Euler invariant and therefore does not lead to sensitive equations of motion.}. In \cite{Teitelboim:83, Jackiw:84} Jackiw and Teitelboim established a new model -the Jackiw-Teitelboim (JT) model-to describe the dynamics of gravity in $(1+1)$-dimensions. They solved the problem by simply introducing an additional variable, a Lagrange multiplier field $\eta$, as follows
\be \label{JTaction}
 S[g_{\mu\nu},\eta]= \int_{\scriptscriptstyle \mathcal{M}} d^2 x \sqrt{-g}\, \eta\, (R-\Lambda),
\ee
where $\Lambda$ is the cosmological constant. 

Here we are interested in the fact that this JT action admits a BF-like formulation \cite{Chamseddine:91}, where now the $B$ field is a zero-form that plays the role of a Lagrange multiplier. The first order action to consider is 
\be \label{JT}
S [B^i,A^i]=  \int_{\scriptscriptstyle \mathcal{M}} \big<B, F(A)\big>, 
\ee
with the field strength $F=dA + A\wedge A$, and $A$ an algebra valued gauge connection $A=A^iX_i$, with $X_i$ the algebra generators. The $B$ field is also algebra valued $B=B^iX_i$ and the $\big<\cdot, \cdot \big>$ denotes the non-degenerate bilinear form compatible with the algebra. 

In order to get a gravitational model we choose our variables valued on the anti-de Sitter algebra $\mathfrak{so}(2,1)$ given by
\be
\left[ P_a, P_b \right] =  \varepsilon_{ab} J, \quad \quad 
\left[ J, P_a \right] = {\varepsilon_a}^b P_b ,
\ee
where we have hidden the cosmological constant re-defining the generator as $P_a = \displaystyle \frac{1}{\sqrt{\Lambda} } \widetilde{P}_a$.  This algebra admits a non-degenerate invariant bilinear form (in this subsection the indices run as $a,b,c,...=0,1$)
\be
\big< P_a, P_b \big> =   \eta_{ab} ,\quad  \quad \big< J,J \big> = 1 .
\ee
Explicitly $A=A^iX_i=e^aP_a+\omega J$ and $B=B^iX_i=B^aP_a+\tilde B J$, with $P_a$ the translations and $J$ the rotation algebra generators. The $e^a$ field corresponds to the one-form zweibein (related with a metric field $g_{\mu\nu}=e\indices{^a_\mu} e\indices{^b_\nu}\eta_{ab}$), and $\omega\equiv \frac{1}{2}\varepsilon_{ab}\omega^{ab}$ is the dual spin connection ($\varepsilon_{01}=-\varepsilon_{10}=1$). With the bilinear form of the anti-de Sitter algebra the (\ref{JT}) action expands as
\be
\label{JT2}
S [B,e^a,\omega]= \int_{\mathcal{M}} \big< B ,F \big> = \int_{\mathcal{M}} B^i F^j \big< X_i, X_j \big>=\int_{\mathcal{M}} \left(    B_a (de^a +{\varepsilon^a}_b \omega e^b  ) +\tilde B (d\omega + \varepsilon_{ab} e^a e^b)    \right),
\ee
where the $B^i$ field decomposes in the Lagrange multipliers: $B^a$ enforcing the vanishing of torsion and $\tilde B$, equivalent to the field $\eta$, enforcing the vanishing of the (anti-de Sitter) curvature.  The torsionless equation allows us to solve the spin connection in terms of the geometric field $e^a$, by replacing it back into the action \eqref{JT2} it is possible to obtain a second order formulation of the action which is nothing but the JT model \eqref{JTaction}.

Now we compute the surface charge density. We start by varying the action \eqref{JT} to read the symplectic potential
\ba
\Theta (\delta A) & = &  \big<B,\delta A\big>= B^i \delta A ^j \big< X_i, X_j \big>= B^a \delta e_a + \tilde B \delta \omega .
\ea

Then, the symplectic structure density yields
\ba 
\nonumber \Omega(\delta_1,\delta_2)& = & \big<\delta_1B,\delta_2 A \big>- \big<\delta_2B,\delta_1 A\big>=  \delta_1 B^a \delta_2 e_a - \delta_2 B^a \delta_1 e_a+ \delta_1 \tilde B \delta_2 \omega - \delta_2 \tilde B \delta_1 \omega. \\ 
\label{JTsymplectic2}
%\nonumber & = & \big< \delta_1 B^a P_a + \delta_1 B J , \delta_2 e^b P_b + \delta_2 \omega J\big> - \big< \delta_2 B^a P_a + \delta_2 B J , \delta_1 e^b P_b + \delta_1 \omega J\big>, \\
%& = & \Lambda \delta_1 B^a \delta_2 A_a + \delta_1 B \delta_2 A -\Lambda \delta_2 B^a \delta_1 A_a - \delta_2 B \delta_1 A .
\ea 
 The action \eqref{JT2} is invariant under diffeomorphism and gauge transformations. The infinitesimal transformation of the field, in their improved version, are
\ba\label{JTsy}
\nonumber \delta_\epsilon A & = & \xi \inter F - d_A \lambda, \\
\delta_\epsilon B& = & \xi\inter d_A B+d_A\xi\inter B+[\lambda, B ],
\ea
with $\xi$ a vector field and $\lambda = \lambda^i X_i=\lambda^aP_a + \lambda J$ a gauge parameter valued in the $\mathfrak{so}(2,1)$ algebra. Then, plugging \eqref{JTsy} into \eqref{JTsymplectic2}, and using the equations of motion and the linearized equations of motions we have
\be
\Omega (\delta, \delta_\epsilon) = dk_\epsilon ,
\ee
with the surface charge density
\be
k_\epsilon = -\big<\delta B , \lambda \big>= - \delta B^i \lambda^j \big< X_i, X_j \big>=-\delta \tilde B \lambda - \delta B^a \lambda_a.
\ee
Notice that, as it might have guessed, this expression corresponds to the formula \eqref{BF-surfacecharge}, where $\xi\inter B^i=0$ because in this context it is a zero form.

%\bibliographystyle{JHEP}
%\bibliographystyle{fullsort}
%\bibliography{Bibliography}

\end{document}